\documentclass{article}
\usepackage[margin=1.0in]{geometry}
\usepackage{cite}
\usepackage{amsmath,amssymb,amsfonts}
\usepackage{algorithm}
\usepackage{algpseudocode}
\usepackage{algorithmicx}
\usepackage{graphicx}
\usepackage{textcomp}
\usepackage[dvipsnames]{xcolor}

\usepackage[utf8]{inputenc} 
\usepackage[T1]{fontenc}    
\usepackage{hyperref}       
\usepackage{url}            
\usepackage{booktabs}       
\usepackage{nicefrac}       
\usepackage{microtype}      
\usepackage{multirow}
\usepackage{subcaption}

\usepackage[export]{adjustbox}
\usepackage{graphbox}
\newcommand{\eps}{\varepsilon}
\newcommand{\T}{\top}
\DeclareMathOperator*{\argmin}{\arg\,\min}
\DeclareMathOperator*{\DC}{DC}
\newcommand{\R}{\mathbb{R}}
\newcommand{\ie}{\emph{i.e.},~}
\newcommand{\eg}{\emph{e.g.},~}
\newcommand{\pnp}{P\&P}

\newcommand{\rnr}{R\&R}

\newcommand{\xhat}{\widehat{x}}

\newcommand{\change}[1]{{\color{black}#1}}

\title{Model Adaptation for Inverse Problems in Imaging}

\author{Davis Gilton, Gregory Ongie, and Rebecca Willett.
\thanks{D.\,Gilton is with the Department of Electrical and Computer Engineering at the University of Wisconsin-Madison, 1415 Engineering Dr, Madison, WI 53706 USA. G.\,Ongie is with the Department of Mathematical and Statistical Sciences at Marquette University, 1250 W Wisconsin Ave, Milwaukee, WI 53233 USA. R.\,Willett is with the Departments of Computer Science and Statistics at the University of Chicago, 5747 S Ellis Ave, Chicago, IL 60637 USA.}}

\begin{document}

\maketitle

\begin{abstract}
Deep neural networks have been applied successfully to a wide variety of inverse problems arising in computational imaging. These networks are typically trained using a forward model that describes the measurement process to be inverted, which is often incorporated directly into the network itself. However, these approaches are sensitive to changes in the forward model: if at test time the forward model varies (even slightly) from the one the network was trained for, the reconstruction performance can degrade substantially.
Given a network trained to solve an initial inverse problem with a known forward model, we propose two novel procedures that adapt the network to a change in the forward model, even without full knowledge of the change. Our approaches do not require access to more labeled data (i.e., ground truth images). We show these simple model adaptation approaches achieve empirical success in a variety of inverse problems, including deblurring, super-resolution, and undersampled image reconstruction in magnetic resonance imaging.
\end{abstract}

\section{Introduction}

Repeated studies have illustrated that neural networks can be trained to solve inverse problems in imaging, including problems such as image reconstruction in MRI, inpainting, superresolution, deblurring, and more. Recent reviews and tutorials on this topic \cite{arridge2019solving,ongie2020deep} have described various approaches to this problem. \change{For concreteness, we focus on the case of \emph{linear} inverse problems in imaging.} In the general framework of interest, an unknown
$n$-pixel image (in vectorized form) $x \in \R^n$ (or $\mathbb{C}^n$) is observed via $m$ noisy \change{linear} measurements 
$y\in\mathbb{R}^m$ (or $\mathbb{C}^m$) according to the model
\begin{equation}\label{eq:invprob0}
y=A_0x+\varepsilon,
\end{equation}
where the matrix $A_0 \in \R^{m\times n}$ (or $\mathbb{C}^{m\times n})$ is the {\em forward model} and $\varepsilon$ represents a vector of noise. The goal is to recover $x$ from $y$.

\begin{figure*}[tb]
\centering
\setlength{\tabcolsep}{0pt}
\begin{tabular}{cccc}
\subfloat{\includegraphics[width = 0.245\textwidth]{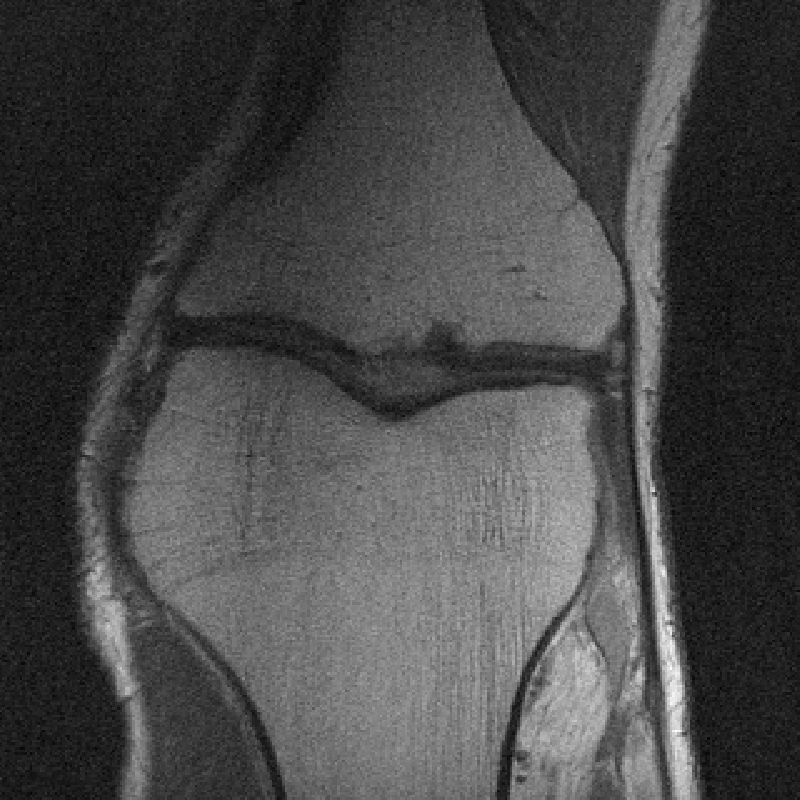}}\hspace{0.1em} &
\subfloat{\includegraphics[width = 0.245\textwidth]{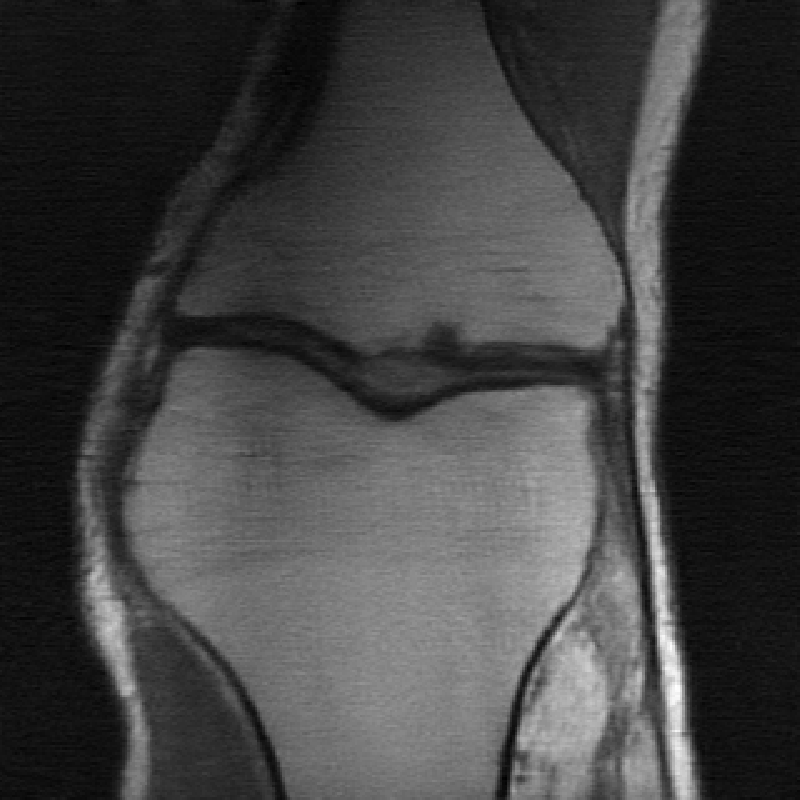}}\hspace{0.1em}  &
\subfloat{\includegraphics[width = 0.245\textwidth]{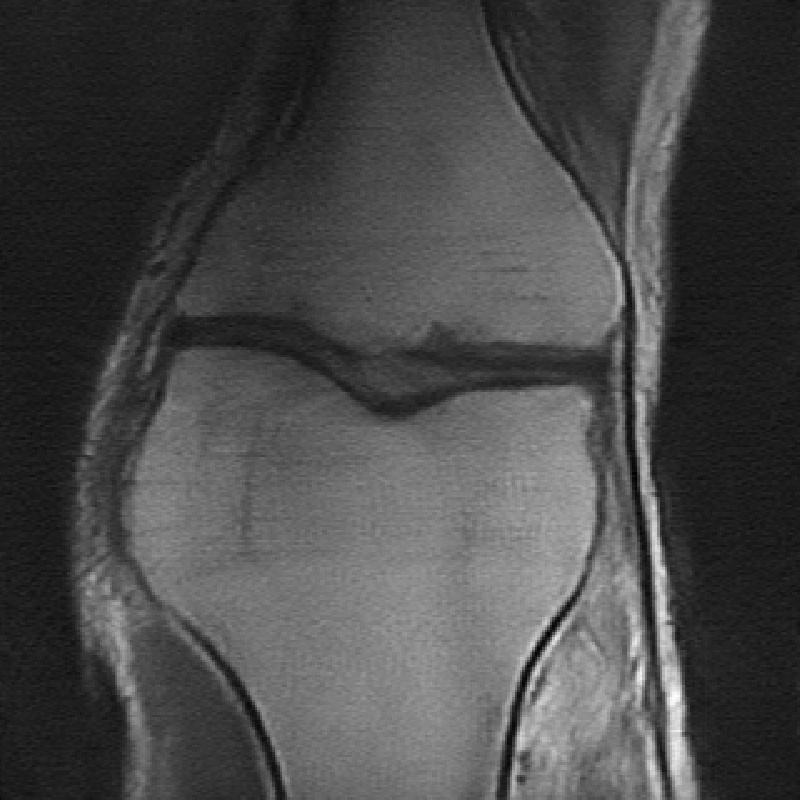}} \hspace{0.1em} &
\subfloat{\includegraphics[width = 0.245\textwidth]{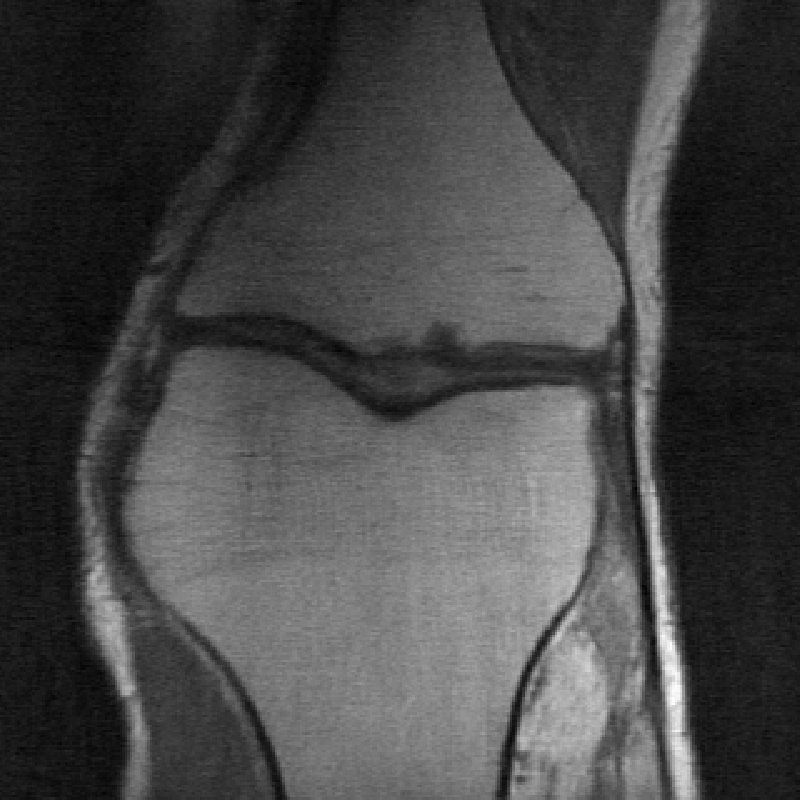}} \\
\subfloat{\includegraphics[width = 0.245\textwidth]{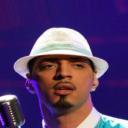}}\hspace{0.1em}  &
\subfloat{\includegraphics[width = 0.245\textwidth]{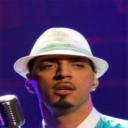}}\hspace{0.1em}  &
\subfloat{\includegraphics[width = 0.245\textwidth]{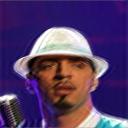}}\hspace{0.1em}  &
\subfloat{\includegraphics[width = 0.245\textwidth]{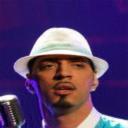}} \\
\small (a) Ground truth & \small (b) No model drift & \small (c) Model drift  & \begin{minipage}{0.25\linewidth}\small (d) Model drift w/model adaptation \end{minipage}
\end{tabular} \\
\caption{Small perturbations in measurements for deep learning-based image reconstruction operators can lead to both subtle and obvious artifacts in reconstructions across problems and domains. In the top row, we present results for undersampled MRI reconstruction of knee images, and the second row illustrates deblurring images of human faces.
(a) Ground truth image. (b) No model drift. Training and test data correspond to same model, $A_0$, yielding accurate reconstruction  via learned model. (c) Model drift but no model adaptation. Training assumes model $A_0$ but at test time we have model $A_1$. Reconstruction using trained network {\em without model adaptation} gives significant distortions. (d) Model drift and model adaptation. Training assumes model $A_0$ but at test time we have model $A_1$. Reconstruction using model adaptation prevents distortions and compares favorably to the setting without model drift. The MRI example demonstrates our Reuse and Regularize method (Alg. \ref{alg:rnr}), and the deblurring example demonstrates our Parameterize and Perturb method (Alg. \ref{alg:pnp}). Experimental details are in Section~\ref{sec:exp}.}
\label{fig:fig1}
\end{figure*}

In this paper, we focus on the setting in which the forward model $A_0$ is known and used during training. Past work has illustrated that leveraging knowledge of $A_0$ during training can reduce the sample complexity \cite{gilton2019neumann}. This paradigm is particularly common in applications such as medical imaging, where $A_0$ represents a model of the imaging system. For instance, in magnetic resonance imaging (MRI), $A_0$ reflects which k-space measurements are collected. 



Unfortunately, these methods can be surprisingly  fragile in the face of {\em model drift}, which occurs when, at test time, we are provided samples of the form 
\begin{equation}
\label{eq:invprob1}
y = A_1 x+\varepsilon'
\end{equation}
for some new forward model $A_1 \neq A_0$  \change{and/or a change in the noise distribution (\ie the noise $\varepsilon'$ is distributed differently than $\varepsilon$).} That is, 
assume we have trained a solver that is a function of both the original forward model $A_0$ and a learned neural network. 
One might try to reconstruct $x$
from $y$ using this solver, but it will perform poorly because it is using a misspecified model ($A_0$ instead of $A_1$). Alternatively, we might attempt to use the same general solver where we replace $A_0$ with $A_1$ but leave the learned component intact. In this case, the estimate $x$ computed from $y$ may also be poor, as illustrated in \cite{antun2020instabilities} and \cite{hussein2020correction}. The situation is complicated even further if we do not have a precise model of $A_1$ at test time. 

These are real challenges in practice. For example, in MRI reconstruction there is substantial variation in the forward model depending on the type of acquisition -- e.g., Cartesian versus non-Cartesian k-space sampling trajectories, different undersampling factors, different number of coils and coil sensitivity maps, magnetic field inhomogeneity maps, and other calibration parameters \cite{fessler} -- all which need to be accounted for during training and testing. A network trained for one of these forward models may need to be retrained from scratch in order to perform well on even a slightly different setting (e.g., from six-fold to four-fold undersampling of k-space). Furthermore, training a new network from scratch may not always be feasible after deployment due to a lack of access to ground truth images. This could be either due to privacy concerns of sharing patient data between hospitals and researchers, or because acquiring ground truth images is difficult for the new inverse problem.

This leads us to formulate the problem of {\em model adaptation}: given a reconstruction network trained on measurements from one forward model adapt/retrain/modify the network to reconstruct images from measurements reflecting a new forward model. We consider a few  variants of this problem: (a) the new forward model $A_1$ is known, along with one or more unlabeled training samples $y_i$ reflecting $A_1$, and (b) $A_1$ is unknown or only partially known, and we only have one or more unlabeled training samples reflecting $A_1$. These training samples are unlabeled in the sense that they are not paired with ``ground truth'' images used to generate the $y_i$'s. 
Our proposed model adaptation methods allow 
a reconstruction network to be trained for a known forward model and then adapted to a related forward model without access to ground truth images, and without knowing the exact parameters of the new forward model.

Model drift as stated above is a particular form of \emph{distribution drift}, in which the distribution of $Y|X=x$ changes between training and deployment and we know $Y$ has a linear dependence on $X$ before and after the drift (even if we do not know the parameters of those linear relationships, represented as $A_0$ and $A_1$). \change{That is, if we assume $\varepsilon \sim {\cal N}(0,\sigma^2 I)$, then the training distribution is $Y|X=x \sim {\cal N}(A_0 x,\sigma^2 I)$ and the distribution at deployment is $Y|X=x \sim {\cal N}(A_1 x,\sigma^2 I)$. In general, distribution drift challenges may be addressed using transfer learning \cite{pan2009survey,weiss2016survey,tan2018survey} and domain adaptation \cite{muandet2013domain,li2017deeper,wang2018deep}. One of the methods we explore in the body of the paper, Parameterize and Perturb, shares several features with transfer learning methodology. However, since in our setting we have a specific form of distribution drift, it is possible to design more targeted methods with better performance, as illustrated by
existing specialized methods for image inpainting \cite{fawzi2016image}, as well as our general-purpose Reuse and Regularize method (detailed below).}
\subsection{Related Work}

A broad collection of recent works, as surveyed by \cite{arridge2019solving} and \cite{ongie2020deep}, have explored using machine learning methods to help solve inverse problems in imaging. The current paper is motivated in part by experiments presented in \cite{antun2020instabilities}, which show that deep neural networks trained to solve inverse problems are prone to several types of instabilities. 
Specifically, they 
showed that model drift in the form of slight changes in the forward model (even ``beneficial'' ones, like increasing the number of k-space samples in MRI) often have detrimental impacts on reconstruction performance. While \cite{antun2020instabilities}
is mostly empirical in nature, a follow-up mathematical study \cite{gottschling2020troublesome} provides theoretical support to this finding, implying that instability arises naturally from training standard deep learning inverse solvers. However, recent work also shows that the instabilities observed in in \cite{antun2020instabilities} can be mitigated to some extent by adding noise to measurements during training, though such techniques are not sufficient to resolve artifacts arising from substantial model drift.
To address a subset of these issues, \cite{raj2020improving} and \cite{lunz2018adversarial} propose  adversarial training frameworks that increases the robustness of inverse problem solvers. However, \cite{raj2020improving} and \cite{lunz2018adversarial} focus on robustness to adversarial perturbations in the measurements for a fixed forward model, and do not address \change{a global change in the forward model}, which is the focus of this work.

Similar to this work, a recent paper \cite{jong} has proposed domain adaptation techniques to transfer a reconstruction network from one inverse problem setting to another, e.g., adapting a network trained for CT reconstruction to perform MRI reconstruction. However, the focus of that approach is on adapting to {\em changes in the image distribution}, whereas our approaches focus on {\em changes to the forward model} assuming the image distribution is unchanged. Additionally, to our knowledge, no existing domain adaptation approaches consider the scenario where the new forward model depends on unknown calibration parameters, as we do in this work. 

Another line of work explores learned methods for image reconstruction with automatic parameter tuning; see \cite{wei2020tuning} and references therein. However, this work focuses on learning regularization and optimization parameters, not parameters of a drifting forward model. Also, \cite{wu2019learning} describes a unrolling approach to learning a forward model in an imaging context, but with the goal of designing a forward model that optimizes reconstruction quality, rather than estimating a correction to the forward model from measurements. Some recent studies have used pre-trained generative models to solve inverse problems with unknown calibration parameters \cite{anirudh2018unsupervised}; this line of work can be viewed as an extension of compressed sensing with general models framework introduced in \cite{bora2017compressed}.

\section{Problem Formulation}
Here we formalize the problem of \emph{model adaptation} as introduced above.

Suppose we have access to an estimator $\xhat  = f_0(y)$ that has been designed/trained to solve the inverse problem
\begin{equation}\label{eq:invprob0a}
    y = A_0x + \varepsilon,\quad x \sim P_X,~\eps\sim P_{N_0}\tag{P0}
\end{equation}
where $A_0$ is a known (linear) forward model, $P_X$ denotes the distribution of images $x$ and $P_{N_0}$ denotes the distribution of the noise $\eps$. We assume the trained estimator ``solves'' the inverse problem in the sense that it produces an estimate $\hat x = f_0(y)$ such that the mean-squared error (MSE) $\mathbb{E}_{x,\eps}[\|\hat x - x\|^2]$ is small. 

Now assume that the forward model has changed from $A_0$ to a new model $A_1$ and/or the noise distribution has changed from $P_{N_0}$ to a new noise distribution $P_{N_1}$, resulting in the new inverse problem
\begin{equation}\label{eq:invprob1a}
    y = A_1 x + \varepsilon',\quad x \sim P_X,~\varepsilon'\sim P_{N_1}.\tag{P1}
\end{equation}
We consider both the case where $A_1$ is \emph{known} (\ie we have an accurate estimate of $A_1$) and the case \change{where $A_1$ is partially unknown, in the sense that it belongs to a class of parameterized forward models, \ie $A_1 \in \{A(\sigma): \sigma \in \R^q\}$, where the parameters $\sigma \in \R^q$ are unknown.}

The goal of \emph{model adaptation} is to adapt/retrain/modify the estimator $\xhat = f_0(y)$ that was designed to solve the original inverse problem (P0) to solve the new inverse problem (P1). We will consider two variants of this problem:
\change{
\begin{itemize}
    \item \emph{Model adaptation without calibration data:} In this setting, we assume access to only one measurement vector $y$ generated according to \eqref{eq:invprob1a}.
    \item \emph{Model adaptation with calibration data:} In this setting we assume access to a new set of measurement vectors $\{y_i\}_{i=1}^N$ generated according to \eqref{eq:invprob1a}, but without access to the paired ground truth images (i.e., the corresponding $x_i$'s).
\end{itemize}
}

While the above discussion centers around a general estimator $\xhat = f_0(y)$, we are particularly interested in estimators that combine a trained trained deep neural network component depending on a vector of weights/parameters $\theta_0$, along with the original forward model $A_0$; we will call such an estimator a \emph{reconstruction network}. Specifically, we assume that the forward model $A_0$ (or other derived quantities, such as its transpose $A_0^
\T$, psuedo-inverse $A_0^\dagger$, etc.) is embedded in the reconstruction network, either in an initialization layer and/or in multiple subsequent layers. This is the case for networks based on unrolling of iterative algorithms (see, for example, \cite{gregor2010learning, sun2016deep, mardani2018neural,ongie2020deep,arridge2019solving}, and references therein), in which $A_0$ appears repeatedly in the network in ``data-consistency'' layers that approximately re-project the intermediate outputs of the network onto the set of data constraints $\{x\in \R^n : A_0 x = y\}$.
In general, we will
assume the reconstruction network can be parametrized as $f_0(\cdot) =  f(\cdot;\theta_0,A_0)$ where  $\theta_0\in\R^p$ is the vector of pre-trained neural network weights/parameters and $A_0$ is the original forward model.

\change{Finally, to simplify the presentation, we will assume an additive white Gaussian noise model for both (P0) and (P1), \ie $P_{N_0} = \mathcal{N}(0,\sigma_0^2 I)$ and $P_{N_1} = \mathcal{N}(0,\sigma_1^2 I)$ with known variances $\sigma_0^2$ and $\sigma_1^2$. In this case the negative log-likelihood of $x$ given $y$ under measurement model (P1) is $\frac{1}{2\sigma_1^2}\|A_1x-y\|_2^2$, which justifies our use of quadratic data-consistency terms in the development below.}

\begin{figure}[ht]
\centering
\begin{subfigure}{.8\linewidth}

    \caption{Original reconstruction network.}
    \centering
  \includegraphics[width=.6\linewidth]{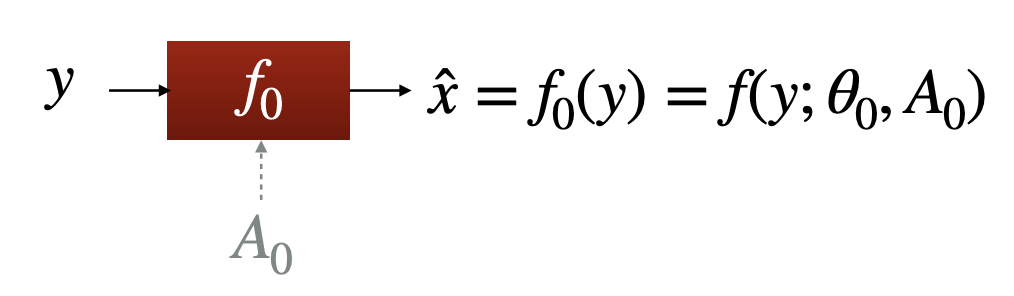}  
\end{subfigure} \\[1em]
\begin{subfigure}{.8\linewidth}
    \caption{Model adaptation by Parametrize and Perturb (\pnp).}
    \centering
  \includegraphics[width=.6\linewidth]{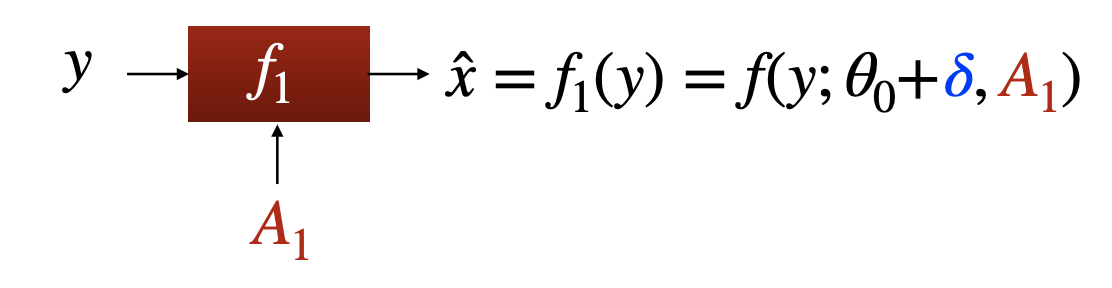}  
\end{subfigure} \\[1em]
\begin{subfigure}{0.8\linewidth}
  \caption{Model adaptation by Reuse \& Regularize (\rnr).}
  \centering
  \includegraphics[width=0.8\linewidth]{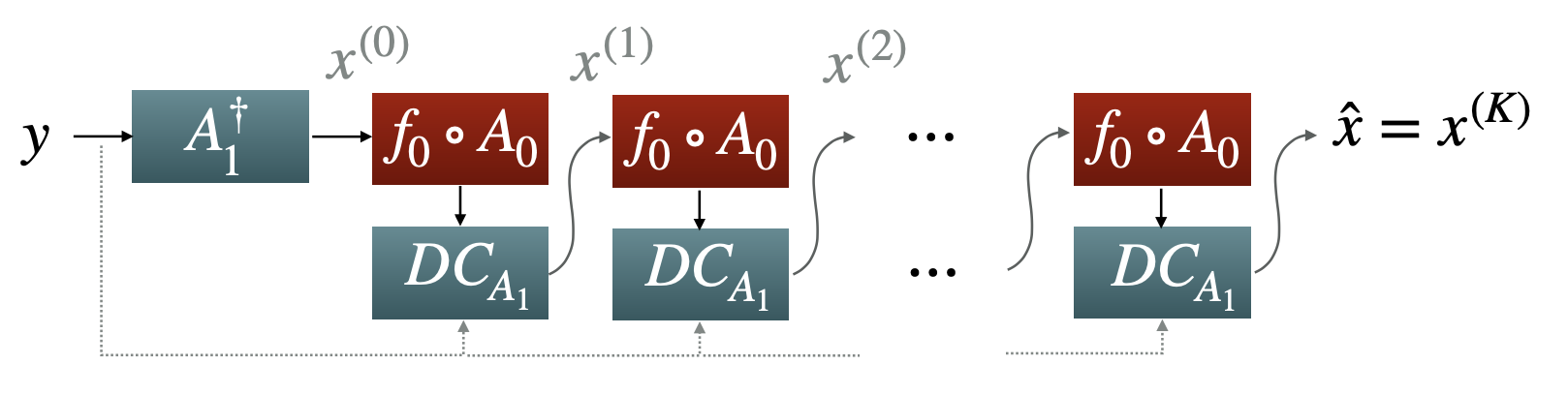}  
\end{subfigure}
\caption{Three basic paradigms of reconstruction under ``model drift''. (a) If the training data is generated using the model $y = A_0 x + \varepsilon$, this can be used to learn a reconstruction network $f(y;\theta_0,A_0)$ which is parameterized by weights or parameters $\theta_0$ and may also explicitly depend on forward model $A_0$. (b) {\bf Parametrize and Perturb (P\&P):} If at test time we are presented with data corresponding to the model $y = A_1 x + \varepsilon'$, we may not only use the new forward model $A_1$ but also learn a perturbation $\delta$ to the original network parameters $\theta_0$ to compensate for the model drift.
(c) {\bf Reuse and Regularize (\rnr):} Alternatively to \pnp, we may reuse the pre-trained network $f_0$ as an implicit regularizer in an iterative model-based reconstruction scheme. The proposed scheme alternates between applying $f_0\circ A_0$, which denoises and/or removes artifacts, and a data-consistency step (denoted by $\DC_{A_1}$ above) that enforces the estimated image $\xhat$ satisfies $A_1 \xhat \approx y$.
}
\label{fig:proposedmethods}
\end{figure}

\subsection{The feasibility of model adaptation}

To compute an accurate reconstruction under the original forward model, $A_0$, the learned solver must reconstruct components of the image that lie in the null space $N(A_0)$: for superresolution, these are high-frequency details lost during downsampling, and in inpainting, these are the pixels removed by $A_0$. 
\change{The neural network trained as a component of $f_0$ implicitly represents a mapping from image components in the range of $A_0$ to components in $N(A_0)$.} 

\change{Reconstructing under a different forward model, $A_1$, requires reconstructing different components of the image in the null space $N(A_1)$. The general intuition behind model adaptation is that if $A_0$ and $A_1$ are similar, then the mapping represented by $f_0$ can inform the new mapping that we need to learn from image components in the range of $A_1$ to components in $N(A_1)$. For example, in an inpainting setting, the learned network not only represents the missing pixels, but it also represents some function of the observed pixels that are relevant to filling in the missing pixels. Thus if $A_1$ has a similar null space (\eg an offset in the collection of missing pixels), it is reasonable to expect that the original network has learned to represent some information about image components in the null space of $A_1$ but not in the null space of $A_0$. As the null spaces of $A_0$ and $A_1$ get further apart, model adaptation becomes less effective. This is similar to the widely-noted behavior of transfer learning, where transfer learning efficacy depends on the similarity of the training and target distributions. This intuition is supported by our empirical results, which illustrate that when $A_0$ and $A_1$ correspond to different blur kernels or perturbed k-space sampling patterns in MRI, the learned mapping $f_0$ does contain information about image components in the null space of $A_1$ that can be leveraged to improve reconstruction accuracy, even without additional training samples drawn using the model $A_1$.}

\section{Proposed Approaches}
\label{sec:proposed}
We propose two distinct model adaptation approaches, {\em Parameterize \& Perturb (\pnp)} and {\em Reuse \& Regularize (\rnr)}, as detailed below.

\subsection{Parametrize and Perturb: A transfer learning approach}

Let $f_0$ be a reconstruction network trained to solve inverse problem (P0). Suppose we can explicitly parameterize $f_0$ both in terms of the trained weights/parameters $\theta_0$ and the original forward model $A_0$, \ie we may write $f_0(\cdot) = f(\cdot\,;\theta,A_0)$. Given a new measurement vector $y$ under the measurement model (P1), a ``naive'' approach to model adaptation is to simply to replace substitute the new forward model $A_1$ for $A_0$ in this parametrization, and estimate the image as $f(y;\theta_0,A_1)$. However, as illustrated in Figure \ref{fig:fig1}, this can lead to artifacts in the reconstruction due to model mismatch.

Instead, we propose estimating the image as $ f(y;\theta_1,A_1)$ where $\theta_1$ is a perturbed set of of network parameters obtained by solving the optimization problem:
\begin{equation}\label{eq:A1}
\min_{\theta} \|y - A_1 f(y;\theta,A_1)\|_2^2 + \mu \|\theta-\theta_0\|_2^2.
\end{equation}
where $\mu > 0$ is a tunable parameter.
The first term enforces data consistency, i.e., the estimated image $\xhat$ should satisfy $y \approx A_1 \xhat$, while the second term $\|\theta-\theta_0\|_2^2$ ensures the retrained parameters $\theta$ stay close to the original network parameters $\theta_0$. This term is necessary to avoid degenerate solutions, \change{which we demonstrate in the Supplement. Our use of a proximity term of this form is also inspired in part by its success in other transfer learning applications (see, \eg \cite{xuhong2018explicit}).}

If the forward model $A_1$ is also unknown, we propose optimizing for it as well in the above formulation, which gives:
\begin{equation}\label{eq:A2}
\min_{\theta,A \in {\cal A}} \|y - A f(y;\theta,A)\|_2^2 + \mu \|\theta-\theta_0\|_2^2.
\end{equation}
where $\mathcal{A}$ denotes a constraint set.
We assume the forward model is parameterized such that the constraint set is given by $\mathcal{A} = \{A(\sigma) : \sigma \in \mathbb{R}^q\}$, where $A(\sigma)$ denotes a class of of forward models parametrized by a vector $\sigma \in \R^q$ with $q \ll m\cdot n$ (e.g., in a blind deconvolution setting, $A(\sigma)$ corresponds convolution with an unknown kernel $\sigma$). In particular, we propose optimizing over the parameters $\sigma$, which is possible with first-order methods such as stochastic gradient descent, provided the map $\sigma \mapsto A_\sigma$ is first-order differentiable.

{\centering
\begin{minipage}{.6\linewidth}
\centering
\begin{algorithm}[H]
    \caption{Parameterize \& Perturb (\pnp)}
    \label{alg:pnp}
    \centering
    \begin{algorithmic}[1]
        \Require{Original forward model $A_0$, new forward model $A_1$, pre-trained reconstruction network $f_0(\cdot) = f(\cdot; \theta_0,A_0)$,  regularization parameter $\mu$, new measurements $y$.}
            \State Modify the reconstruction network $f_0$ by internally changing $A_0$ to $A_1$, to obtain the estimate $f(y;\theta_0, A_1)$
            \State Fine-tune the network weights as  $\theta_1 = \theta_0+\delta$ where $\delta$ is a perturbation learned by solving \eqref{eq:A1}
    \end{algorithmic}    
\end{algorithm}
\end{minipage}
\par
}\vspace{1em}

The preceding discussion focused on the case of reconstructing single measurement vector $y$ at test time, \ie model adaptation without calibration data. Additionally, we consider a \pnp\ approach in the case where we have access to calibration data $y_1,...,y_N$ generated according to (P1). In this case we propose retraining the network
by minimizing the sum of data-consistency terms over the calibration set:
\begin{equation}\label{eq:A3}
(\theta_1,A_1) = \argmin_{\theta,A \in {\cal A}} \frac{1}{N}\sum_{i=1}^N\|y_i - A f(y_i;\theta,A)\|_2^2 + \lambda \|\theta-\theta_0\|_2^2.
\end{equation}
At deployment, we propose using the retrained  network $\xhat = f(y;\theta_1,A_1)$ as our estimator.

\change{
It is worth noting that the \pnp\ model adaptation technique presented above bears similarities to the deep image prior (DIP) approach to solving inverse problems as introduced in \cite{dip}.  However, \pnp\ differs from DIP in two key aspects: First, in DIP the reconstruction network is initialized with random weights, whereas in \pnp\ we start with a network whose initial weights $\theta_0$ are obtained by training to solve the initial inverse problem (P0). Second, we explicitly enforce proximity to the initial weights to prevent overfitting to the data, and do not rely on early stopping heuristics as is the case DIP. The \pnp\ approach also shares similarities to the ``fine-tuning'' step proposed in the $\Sigma$-net MRI reconstruction framework \cite{hammernik2019sigmanet}, where a loss similar to \eqref{eq:A1} is minimized to enforce data consistency at test time. However, different from \pnp, the fine-tuning approach in \cite{hammernik2019sigmanet} regularizes the reconstruction by minimizing the loss between initial reconstruction and the new network output in the SSIM metric. As demonstrated in Figure \ref{fig:fig1}, this initial reconstruction can have severe artifacts in certain settings due to model mismatch, in which case enforcing proximity in image space to an initial reconstruction is less justified.
}

\subsection{Reuse \& Regularize: Model adaptation without retraining} 
\change{
One drawback of the \pnp\ approach is that it requires fine-tuning the network for each input $y$, which is computationally expensive relative to a feed-forward reconstruction approach. Additionally, the \pnp\ approach is somewhat indirect, relying only on the inductive bias of the network architecture and its original parameter configuration to impart a regularization effect for the new inverse problem (P1). Here we propose a different model adaptation approach that does not require retraining the original reconstruction network, and explicitly makes use of the fact that the original network is designed to solve (P0).

Suppose we are given a reconstruction network $f_0(y)$
trained to solve (P0). The key idea we exploit is that the composition of $f_0$  with the original forward model $A_0$, should act as an \emph{auto-encoder}, \ie if we define the map $g:\mathbb{R}^d\rightarrow\mathbb{R}^d$ by $g(x) = f_0(A_0x)$ then by design we should have $g(x)\approx x$ for any image $x$ sampled from the image distribution $P_X$. See Figure \ref{fig:autoencoder} for an illustration in the case of undersampled MRI reconstruction.

Given this fact, one simple approach to reconstructing a measurement vector $y$ under (P1) is to start from an initial guess, \eg the least squares solution $x ^{(0)}= A_1^\dagger y$, and attempt to find a fixed-point of $g(\cdot)$ by iterating:
\begin{equation}\label{eq:fp}
x^{(k+1)} = g(x^{(k)}),~~k=0,1,2,...
\end{equation}
However, this approach only uses knowledge of the new forward model $A_1$ in the initialization step. Also, unless we can guarantee the map $g(\cdot)$ is non-expansive (i.e., its Jacobian is 1-Lipschitz), these iterations could diverge.

\begin{figure}
    \centering
    \includegraphics[width=0.7\columnwidth]{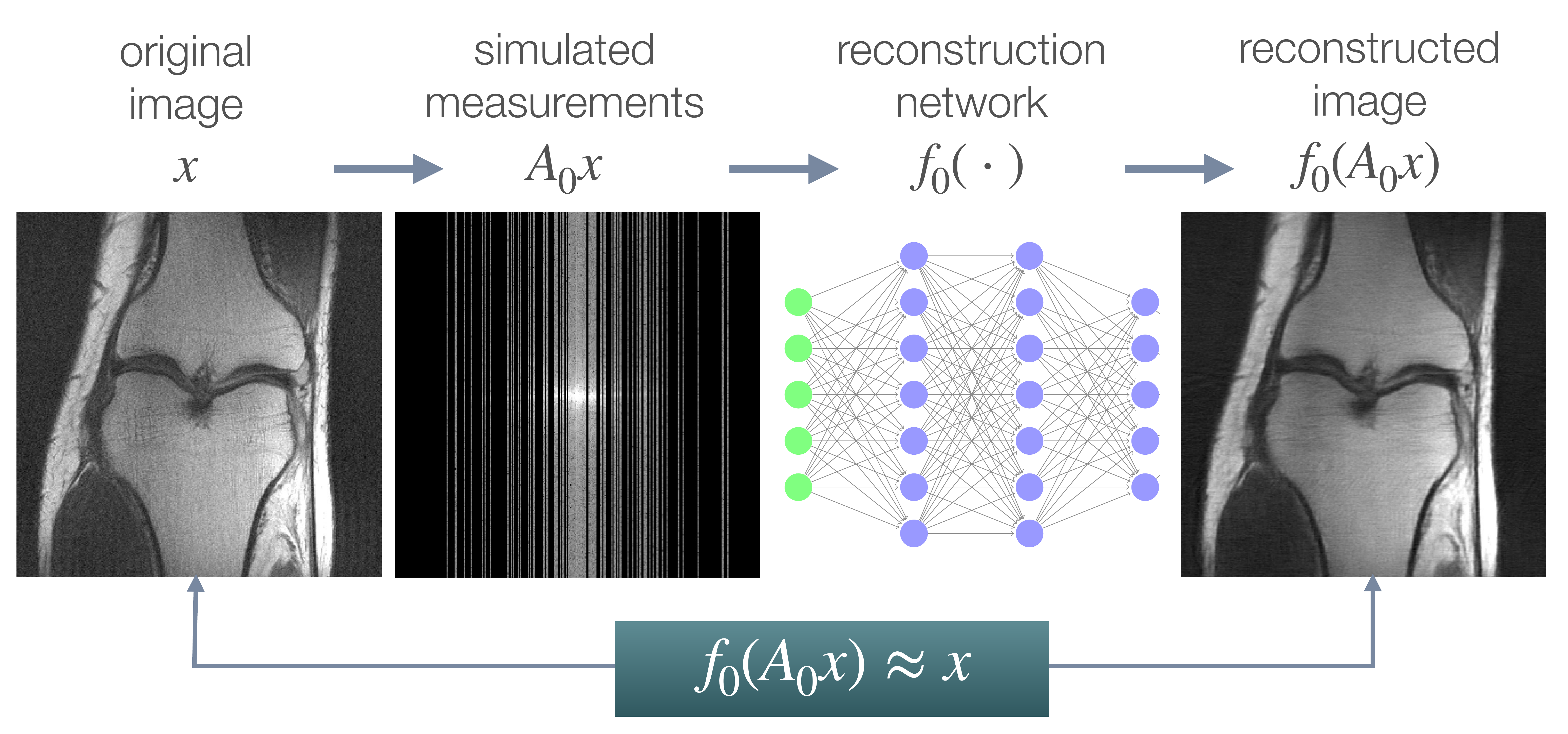}
    \caption{Illustration of the auto-encoding property of the map $f_0 \circ A_0$ as used in the proposed \rnr\ model adaptation approach, illustrated in an undersampled MRI reconstruction setting.}
    \label{fig:autoencoder}
\end{figure}

Instead, building off the intuition that $g$ acts as an auto-encoder, we propose using $g$ as a regularizer in an iterative model-based reconstruction scheme. In particular, we adopt a regularization-by-denoising (RED) approach, which allows one to convert an arbitrary denoiser/de-artifacting map into a regularizer  \cite{RED}. The RED approach is motivated by the following cost function:
\begin{equation}\label{eq:RED}
    \min_{x} \frac{1}{2}\|A_1x - y\|_2^2 + \lambda\rho(x)
\end{equation}
where the function $\rho(x) := x^\T(x-g(x))$ can be interpreted as a regularizer induced by the map $g(x)$ and $\lambda > 0$ is a regularization parameter. Under appropriate conditions on the function $g$, one can show $\nabla \rho(x) = x-g(x)$. This fact is used in \cite{RED} to derive a variety of iterative algorithms based on first-order methods (see also \cite{reehorst2018regularization} for further analysis of RED, including convergence guarantees).

{\centering
\begin{minipage}{.6\linewidth}
\centering
\begin{algorithm}[H]
    \caption{Reuse \& Regularize (R\&R)}
    \label{alg:rnr}
    \centering
    \begin{algorithmic}[1]
        \Require{Pre-trained reconstruction network $f_0(\cdot)$, original forward model $A_0$, new forward model $A_1$, regularization parameter $\lambda>0$, max iterations $K$, new measurements $y$.}
            \State $x\gets A_1^\dagger y$ \Comment{\emph{least-squares initialization}}
            \For{$k=1,2,...,K$}
                \State $z \gets f_0(A_0x)$ \Comment{\emph{regularize using pre-trained network}}
                \State $x \gets (A_1^\top A_1 + \lambda I)^{-1}(A_1^\top y + \lambda z)$ 
                \Comment{\emph{data consistency}}
            \EndFor
            \State \textbf{return} $x$ 
    \end{algorithmic}    
\end{algorithm}
\end{minipage}
\par
}\vspace{1em}

For simplicity, we focus on a RED approach with proximal gradient descent (see, e.g., \cite{parikh2014proximal}) as the base algorithm with stepsize $\tau > 0$. This results in an alternating scheme:
\begin{align*}
    z^{(k)} & = (1-\tau)x^{(k)} + \tau g(x^{(k)})\\
    x^{(k+1)} & = \argmin_{x} \frac{1}{2\lambda}\|A_1x - y\|_2^2 + \frac{1}{2\tau} \|x- z^{(k)}\|_2^2
\end{align*}
The $x$-update above has the closed-form expression
\begin{equation}
    x^{(k+1)} = \left(A_1^\top A_1 + \tfrac{\lambda}{\tau} I\right)^{-1}\left(A_1^\top y + \tfrac{\lambda}{\tau} z^{(k)}\right) \label{eq:xup}
\end{equation}
For simplicity of implementation and to reduce the number of tuning parameters, we fix the stepsize to $\tau = 1$ in all our experiments. We summarize these steps in Algorithm \ref{alg:rnr}.

Note that in the limit as $\lambda\rightarrow \infty$, Algorithm \ref{alg:rnr} reduces to the fixed-point scheme \eqref{eq:fp}, and in the limit as $\lambda \rightarrow 0$ Algorithm \ref{alg:rnr} will return the initialization $x = A_1^\dagger y$. In general, the output from Algorithm \ref{alg:rnr} will interpolate between these two extremes: $x$ will be an approximate fixed point of $g$ and will approximately satisfy data consistency, \ie $y \approx A_1 x$.

For certain types of forward models the $x$-update in \eqref{eq:xup} can be computed efficiently (e.g., if $A_1$ corresponds to a 2-D discrete convolution with circular boundary conditions, then $A_1^\T A_1$ diagonalizes under the 2-D discrete Fourier transform). However, in general, the matrix inverse $(A_1^\top A_1 + \lambda I)^{-1}$ may be expensive to apply. Therefore, in practice we propose approximating \eqref{eq:xup} with a fixed number of conjugate gradient iterations.

A notable aspect of the \rnr\ approach is that it has potential to improve the accuracy of network-based reconstructions \emph{even in the absence of model drift}, \ie even if $A_1 = A_0$. This is because data-consistency is not guaranteed by certain reconstruction networks (e.g., U-Nets). However, we are less likely to see a benefit in the case where the reconstruction network already incorporates data-consistency layers, such as networks inspired by unrolling iterative optimization algorithms. We explore this aspect empirically in Section \ref{sec:mrisamplingexp} in the context of MRI reconstruction.

{\centering
\begin{minipage}{.6\linewidth}
\centering
\begin{algorithm}[H]
    \caption{Reuse \& Regularize with fine-tuning (\rnr$+$)}
    \label{alg:rnrp}
    \centering
    \begin{algorithmic}[1]
        \Require{Pretrained reconstruction network $f_0(\cdot) = f(\cdot; \theta_0)$, original forward model $A_0$, new forward model $A_1$, regularization parameter $\lambda$, new measurements $y$.}
            \State Construct an estimator $\xhat(y;\theta_0)$ by unrolling $K$ iterations of Algorithm \ref{alg:rnr}
            \State Fine-tune the network weights as  $\theta_1 = \theta_0+\delta$ where $\delta$ is a perturbation learned by approximately minimizing the cost \eqref{eq:rnrpcost} via SGD.
            \State \textbf{return} $x = \xhat(y;\theta_1)$
    \end{algorithmic}    
\end{algorithm}
\end{minipage}
\par
}\vspace{1em}

The \rnr\ approach can also be extended the case where the new forward model $A_1$ depends on unknown parameters. First, we define an estimator $\hat{x}(y;A_1)$ by unrolling of a fixed number of iterates of Algorithm \ref{alg:rnr}, \ie we take $\hat{x}(y;A_1) = x^{(K)}$ where $x^{(K)}$ is the $K$th iterate of Algorithm \ref{alg:rnr} with input $y$ for some small fixed value of $K$ (e.g., $K=5$). Supposing $A_1$ belongs to a parameterized class of forward models $A(\sigma)$, i.e., $A_1 = A(\sigma_1)$ for some set of parameters $\sigma_1$, we propose estimating $\sigma_1$ by minimizing data-consistency of the estimator:
\begin{equation}\label{eq:B2}
\tilde{\sigma}_1 = \argmin_\sigma  \|A(\sigma)\hat{x}(y;A(\sigma)) - y\|_2^2
\end{equation}
The resulting image estimate is then taken to be $\hat{x}(y;\tilde{A}_1)$ where $\tilde{A}_1 = A(\tilde{\sigma}_1)$.

Finally, we also consider a combination of the \pnp\ and \rnr\ approaches where we additionally fine-tune the weights $\theta_0$ of the reconstruction network $f_0$ embedded in the unrolled \rnr\ estimator.  Writing this estimator as $\xhat(y;\theta_0)$, similar to the \pnp\ approach we propose ``fine-tuning'' the weights $\theta_0$ by approximately minimizing the cost function
\begin{equation}\label{eq:rnrpcost}
\min_\theta \|A_1 \xhat(y;\theta) -y\|_2^2,
\end{equation}
to obtain the updated network parameters $\theta_1 = \theta + \delta$ where $\delta$ is some small perturbation. The estimated image is then given by $x = \xhat(y;\theta_1)$. We call this approach \rnr$+$. Empirically we see consistent improvement in reconstruction accuracy from \rnr$+$ over \rnr\ without any fine-tuning (see Figures \ref{fig:megafigmotblur} and \ref{fig:megafigmri}). However, this comes at the additional computational cost of having to retrain the reconstruction network parameters at test time.
}

\section{Experiments}\label{sec:exp}

In this section we empirically demonstrate our approach to model adaptation on three types of inverse problems with two example reconstruction network architectures. We have chosen these comparison points for their simplicity and to illustrate the broad applicability of our proposed approaches. In particular, our approaches to model adaptation are not tied to a specific architectural design.

\subsection{Methods and datasets used}

We demonstrate our approaches on three inverse problems: motion deblurring, superresolution, and undersampled single-coil MRI reconstruction. 

For motion deblurring, our initial model $A_0$ corresponds to a $10^\circ$ motion blur with a $7\times7$ kernel, and $A_1$ is a \change{$20^\circ$} motion blur with a $7\times7$ kernel, with angle given with respect to the horizontal axis. In superresolution, our initial model is a bilinear downsampling with rate $2\times$, and $A_1$ corresponds to $2\times$ bicubic downsampling. 

\begin{figure}[ht]
\centering
\begin{tabular}{cc}
\includegraphics[width=0.25\columnwidth, trim= 20 20 20 20, clip]{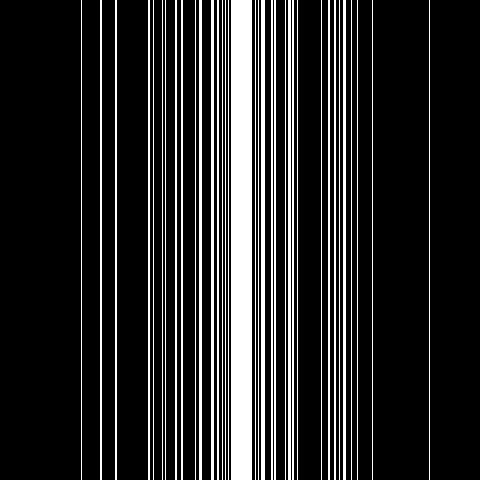} & \includegraphics[width=0.25\columnwidth,trim= 20 20 20 20, clip]{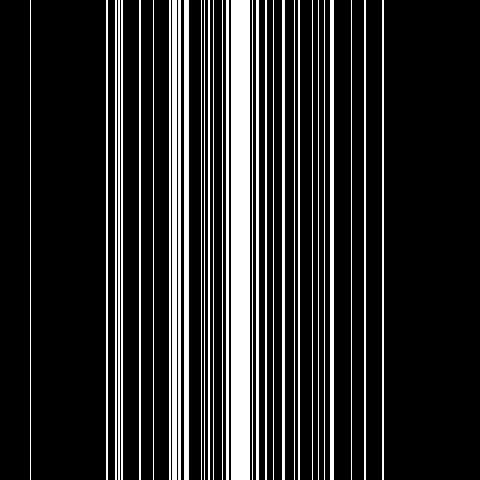}\\[0.1em]
\begin{minipage}{0.25\columnwidth}
(a) Original k-space sampling pattern ($A_0$)
\end{minipage} &
\begin{minipage}{0.25\columnwidth}
(b) Resampled k-space sampling pattern ($A_1$)
\end{minipage}
\end{tabular}
\caption{Visualization of k-space masks used for MRI experiments. Each mask represents a \change{6-fold Cartesian undersampling} with 4\% of the center k-space lines fully sampled, and the remaining lines sampled according to a Gaussian variable density scheme. \change{The $A_1$ mask contains the same center lines, but the higher frequency k-space lines are sampled separately.}}\label{fig:mrimasks}
\end{figure}

MRI reconstruction is performed with a \change{$6\times$} undersampling of k-space in the phase encoding direction for both $A_0$ and $A_1$. The sampling maps are shown in Fig \ref{fig:mrimasks}.

We use two datasets in our experiments. First, for motion deblurring and superresolution, we train and test on 128x128-pixel aligned photos of human faces from the CelebA dataset \cite{liu2015faceattributes}. 

The data used in the undersampled MRI experiments were obtained from the NYU fastMRI Initiative \cite{zbontar2018fastMRI}. The primary goal of the fastMRI dataset is to test whether machine learning can aid in the reconstruction of medical images. We trained and tested on a subset of the single-coil knee dataset\change{, which consist of simulated single-coil measurements. In all tests, we use complex-valued data, which interfaces with our deep networks by treating the real and imaginary parts of the images as separate channels. We measure reconstruction accuracy with respect to the center 320$\times$320 pixels of the complex IFFT of the fully-sampled k-space data. For the purpose of visualization, we display only the magnitude images in the following sections.}

Learning rates and regularization parameters (\ie $\mu$ in Algorithm \ref{alg:pnp} and $\lambda$ in Algorithm \ref{alg:rnr})
were tuned via cross-validation on a hold out validation set of 512 images for CelebA, and 64 MR images for fastMRI. Batch sizes were fixed in advance to be 128 for the motion blur and superresolution settings, and 8 for the MRI setting. Hyperparameters were tuned via grid search on a log scale. \change{For \rnr, we use $K=5$ iterations in the main loop of Algorithm \ref{alg:rnr}. During training, we add Gaussian noise with $\sigma=0.01$ to all measurements, as suggested by \cite{genzel2020solving} to improve robustness.} 

We compare the performance of two reconstruction network architectures across all datasets. First, we utilize the U-Net architecture \cite{ronneberger2015u}. Our U-Net implementation takes as input the adjoint of the measurements under the forward model $A_0^\top y$ or $A_1^\top y$, which is then passed through several CNN layers before obtaining a reconstructed image $\xhat$.

We also utilize the MoDL architecture \cite{aggarwal2018modl}, a learned architecture designed for solving inverse problems with known forward models. MoDL is an iterative or ``unrolled'' architecture, which alternates between a data-consistency step and a trained CNN denoiser, with weights tied across unrolled iterations. We use a U-Net architecture as the denoiser in our implementation of MoDL, ensuring that the overall number of parameters (except for a learned scaling factor in MoDL) is the same in both architectures.

\change{To compare to deep learning-based approaches which do not require training on particular forward models, we compare to the Image-Adaptive GAN (IAGAN) \cite{hussein2020image} and to Regularization by Denoising (RED) \cite{RED}. IAGAN leverages a pretrained GAN to reconstruct arbitrary linear measurements by fitting the latent code input to the GAN, while also tuning the GAN parameters in a way similar to our proposed \pnp\ approach and the Deep Image Prior approach \cite{dip}.

RED requires only a pretrained denoiser, which we implement by pretraining a set of residual U-Net denoisers on the fastMRI and CelebA training sets, with a variety of different Gaussian noise levels. Specifically, we train 15 denoisers for each problem setting, with $\sigma$ ranging from $10^{-4}$ to $10^1$ on a logarithmic scale. All results shown are tuned on the validation set to ensure the optimal denoisers are used.

We also compare to a penalized least squares approach with total variation regularization \cite{rudin1992nonlinear}, a classical approach that does not use any learned elements. While more complex regularizers are possible, total variation (TV) is used because of its status as a simple, widely-used conventional baseline.}

\subsection{Parametrizing forward models}

Both of our proposed model adaptation methods permit the new forward model to be unknown during training, provided it has a known parametrization. In this case, the parameters describing the forward model are learned along with the reconstruction. Here we describe the parametrizations of the forward models that are used.

For the deblurring task, the unknown blur kernel is parametrized as a 7x7 blur kernel, initialized with the weights used for the ground-truth kernel during the initial stage of training. Practically, this is identical to a standard convolutional layer with a fixed initialization and only one learned kernel.

A similar approach is used for superresolution. The forward model can be efficiently represented by strided convolution, and the adjoint is represented by a standard "convolution transpose" layer, again with the weights initialized to match the forward operator in the initial pre-training phase.

\change{In the case of MRI, we use two choices of $A_1$, depending on whether we assume $A_1$ is fully known or not. In the case $A_1$ is fully known, we utilize another $6\times$ undersampled k-space mask, but with resampled high-frequency lines. We display the original and new k-space sampling masks in Figure \ref{fig:mrimasks}. To illustrate the utility of our approach under miscalibration of the forward model in an MRI reconstruction setting, we also consider a unknown random perturbation of the original k-space lines, which we attempt to learn during reconstruction. The vertical k-space lines are still fully sampled, as are the center 4$\%$ of frequencies, but all high frequency lines are perturbed uniformly at random with a continuous value from -2 to 2. We wish to emphasize that this experiment is not meant to reflect clinical practice, since such miscalibration of k-space sampling locations is not typically encountered in anatomical imaging with Cartesian k-space sampling trajectories. However, we include this experiment simply to illustrate that our approach could be extended to unknown parametric changes in the forward model in an MR reconstruction setting.
}

\subsection{Main results}

\begin{table*}	
\centering
	\adjustbox{max width=\columnwidth}{
	\begin{tabular}{cc|c|c|c|c|c|c|}
		\hline
         & & \multicolumn{6}{c|}{Baselines} \\
         & & \multirow{2}{*}{TV} & \multirow{2}{*}{RED} & & Train w/$A_0$ & Train w/$A_0$ & Train w/$A_1$ \\
		 & & &  & & Test w/$A_0$ & Test w/$A_1$ & Test w/$A_1$ \\ \hline \hline
	 \multicolumn{2}{c|}{\multirow{2}{*}{Blur}} & \multirow{2}{*}{27.61}  & \multirow{2}{*}{30.23} & U-Net & 34.15   & 25.42  & 33.98 \\
	  & & & & MoDL & 36.25  & 23.91 & 36.13 \\ \hline
	 \multicolumn{2}{c|}{\multirow{2}{*}{SR}} & \multirow{2}{*}{28.33} & \multirow{2}{*}{28.59} &  U-Net & 30.74  & 26.3 & 31.22 \\
	  & & & & MoDL & 31.32  & 22.27 & 31.98 \\ \hline
	 \multicolumn{2}{c|}{\multirow{2}{*}{MRI}} & \multirow{2}{*}{25.09} & \multirow{2}{*}{27.76} & U-Net & 31.51  & 27.47 & 32.33 \\
		 & & & & MoDL & 31.88 & 22.82 & 31.79 \\ \hline \hline
		 \multicolumn{8}{c}{Proposed Model Adaptation Methods} \\
	 &	 & \multicolumn{3}{c|}{Known $A_1$} &  \multicolumn{3}{c}{Unknown $A_1$} \\
	 &	 & P\&P (Alg. \ref{alg:pnp}) & R\&R (Alg. \ref{alg:rnr}) & R\&R+ (Alg. \ref{alg:rnrp}) & P\&P (Alg. \ref{alg:pnp}) & R\&R (Alg. \ref{alg:rnr}) & R\&R+ (Alg. \ref{alg:rnrp}) \\ \hline
		 \multirow{2}{*}{Blur}  & U-Net & 33.01 & 32.11 & \bf  33.50 & 29.18  & 27.67 & 30.05 \\
		 & MoDL & 30.08 & 33.82 & \bf 34.73 & 29.89  & 27.81 & 27.94 \\ \hline
		 \multirow{2}{*}{SR}  & U-Net & 28.00 & 29.95 & \bf  29.99  & 27.77  & 26.98 & 29.35 \\
		 & MoDL & 24.59 & 28.18 & \bf  29.83 & 23.14 & 24.93 & 25.29 \\ \hline
		 \multirow{2}{*}{MRI}  & U-Net & 29.07 & 29.71 & \bf  31.43 & 28.92  & 28.06 & 29.54 \\
		 & MoDL & 30.63 & 30.25 & \bf 31.44 & 26.64 & 23.46 & 27.67 \\
	\end{tabular}
	}
	\vspace{0.5em}
	\caption{\change{Comparison of performance of various baseline methods for inverse problems across a variety of datasets and forward models. The metric presented is the mean PSNR. SSIM values can be found in Table \ref{table:ssimtable}}}
	\label{table:mainadaptationtable}
\end{table*}

In Table \ref{table:mainadaptationtable} we present our main results. We present sample reconstructions for the deblurring problem and MRI reconstruction problem in Figs. \ref{fig:megafigmotblur} and \ref{fig:megafigmri}. For reference, the ground truth, inputs to the networks, a total variation regularized reconstruction, and a RED reconstruction are presented in Figs. \ref{fig:deblurinitialmethods} and \ref{fig:mriinitialmethods}. We also provide in the Appendix a table of SSIM values as well as the full version of Table \ref{table:mainadaptationtable}, which contains the standard deviations of PSNR.

While the magnitude of the improvements vary across domains and problems, we find that retraining the network with the proposed model adaptation techniques significantly improve performance by several dBs in the new setting. This effect is particularly striking in the case of MRI reconstruction with MoDL, where the ``naive'' approach of replacing $A_0$ with $A_1$ in the network gives catastrophic results (a roughly 9 dB drop in reconstruction PSNR), while the proposed model adaptation approaches give reconstruction PSNRs within 1-2 dB of the baseline approach of training and testing with the same forward model in the case where $A_1$ is known.

\begin{figure}
\centering
\adjustbox{max width=0.5\columnwidth}{
\renewcommand*{\arraystretch}{0}
\begin{tabular}{c@{}c@{}c@{}c@{}}
  \small Ground & \small Blurred & \small TV-Regularized & \small RED  \\
  \small Truth &                    & \small Reconstruction & \\ \vspace{2pt} && \\
 \includegraphics[align=c, width = 0.15\columnwidth]{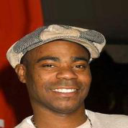} &
\includegraphics[align=c,width = 0.15\columnwidth]{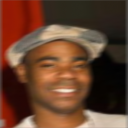} &
 \includegraphics[align=c, width = 0.15\columnwidth]{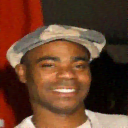} &
\includegraphics[align=c, width = 0.15\columnwidth]{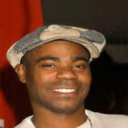} \\ 
 \includegraphics[align=c, width = 0.15\columnwidth]{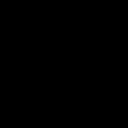}
& \includegraphics[align=c, width = 0.15\columnwidth]{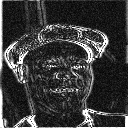} &
\includegraphics[align=c, width = 0.15\columnwidth]{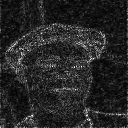} &
\includegraphics[align=c, width = 0.15\columnwidth]{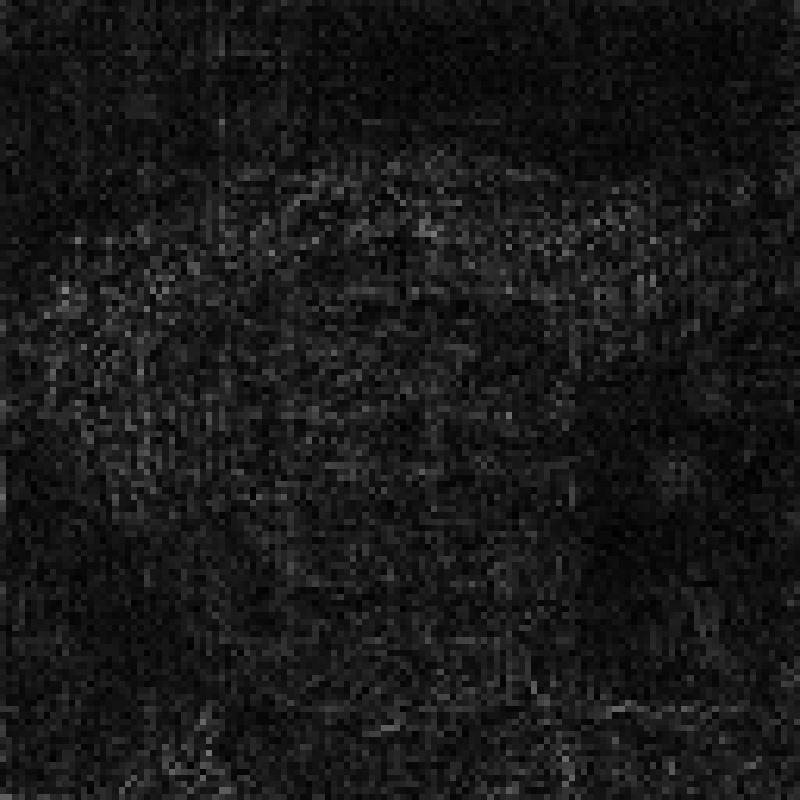}
\end{tabular}
}
\caption{Comparison figures for the deblurring methods in Figure \ref{fig:megafigmotblur}. We present the ground truth, the blurred image (with Gaussian noise with $\sigma=0.01$ added), a total variation (TV) regularized reconstruction, \change{and a comparison to Regularization by Denoising (RED), a model-agnostic method leveraging a deep denoiser}. Below each of the above is the residual image, multiplied by 5$\times$ for ease of visualization.}
\label{fig:deblurinitialmethods}
\end{figure}

\begin{figure}
\centering
\adjustbox{max width=0.5\columnwidth}{
\renewcommand*{\arraystretch}{0}
\begin{tabular}{@{}c@{}c@{}c@{}c@{}}
 \small Ground & \small IFFT & \small TV-Regularized & \small RED \\
 \small Truth & \small Reconstruction & \small Reconstruction & \\ \vspace{1pt} && \\
 \includegraphics[align=c,width = 0.15\columnwidth]{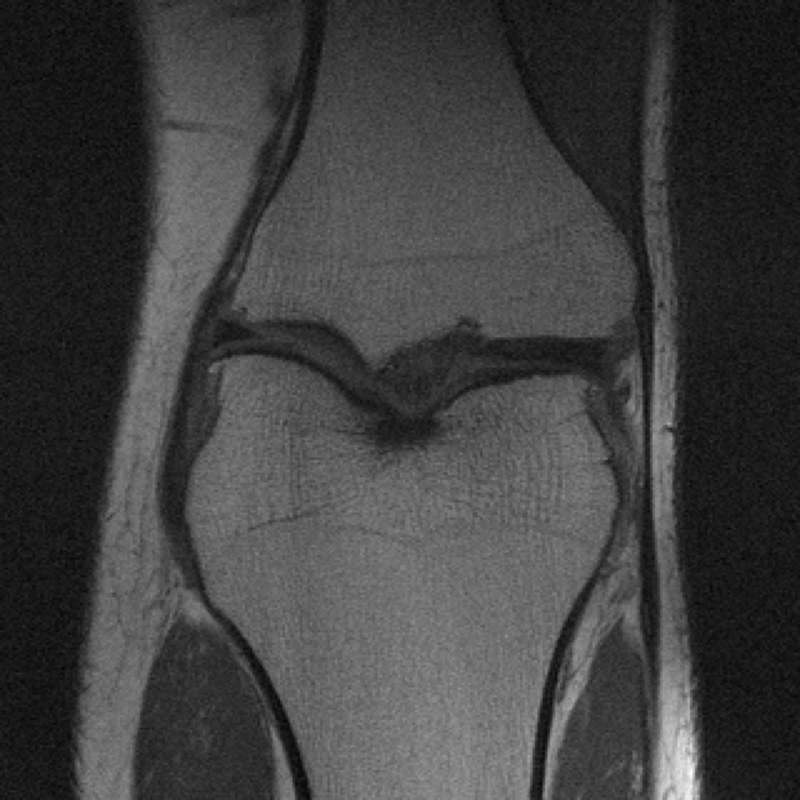} &
\includegraphics[align=c,width = 0.15\columnwidth]{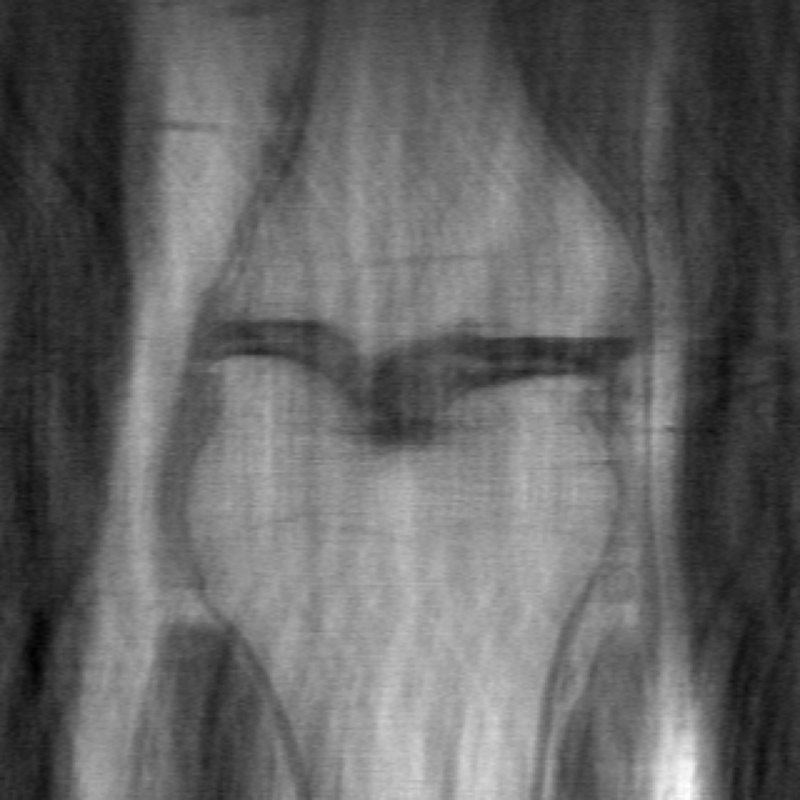} &
\includegraphics[align=c,width = 0.15\columnwidth]{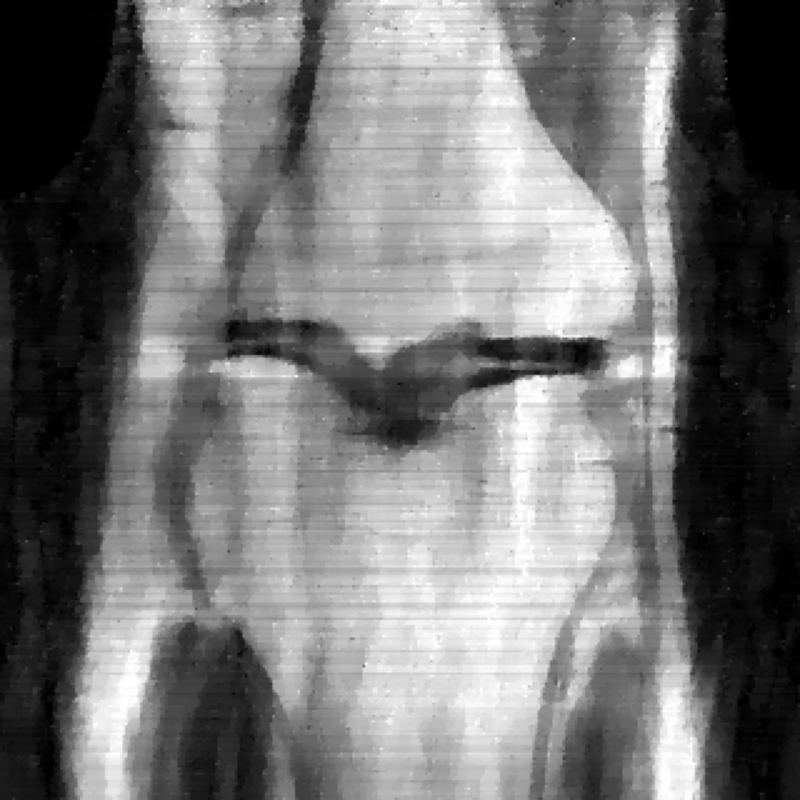} &
\includegraphics[align=c,width = 0.15\columnwidth]{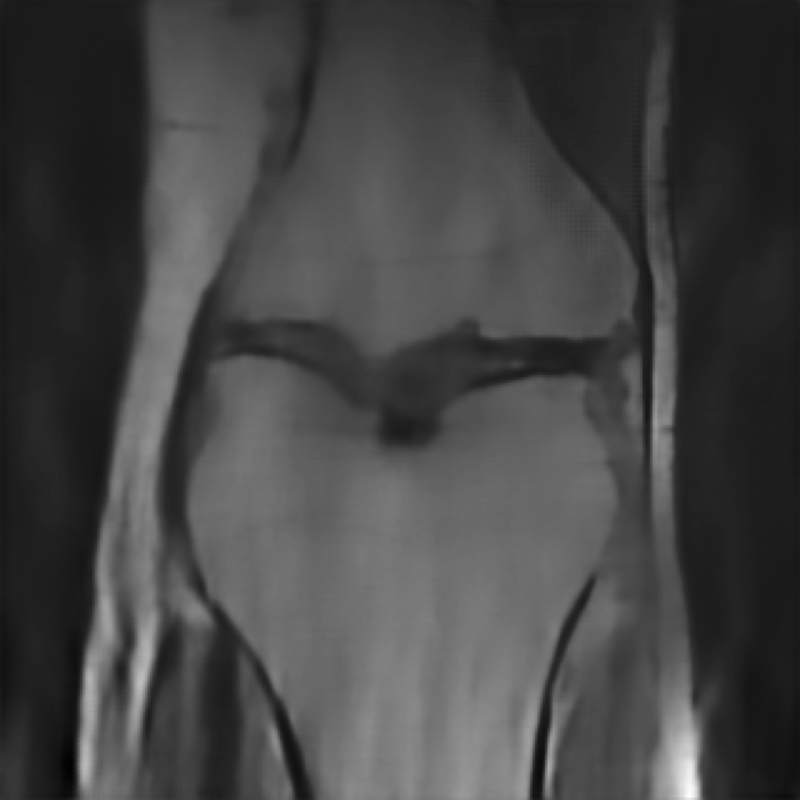} \\
 \includegraphics[align=c,width = 0.15\columnwidth]{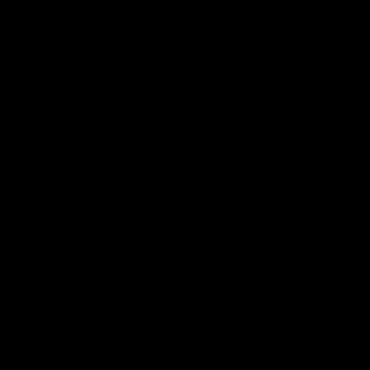} &
\includegraphics[align=c,width = 0.15\columnwidth]{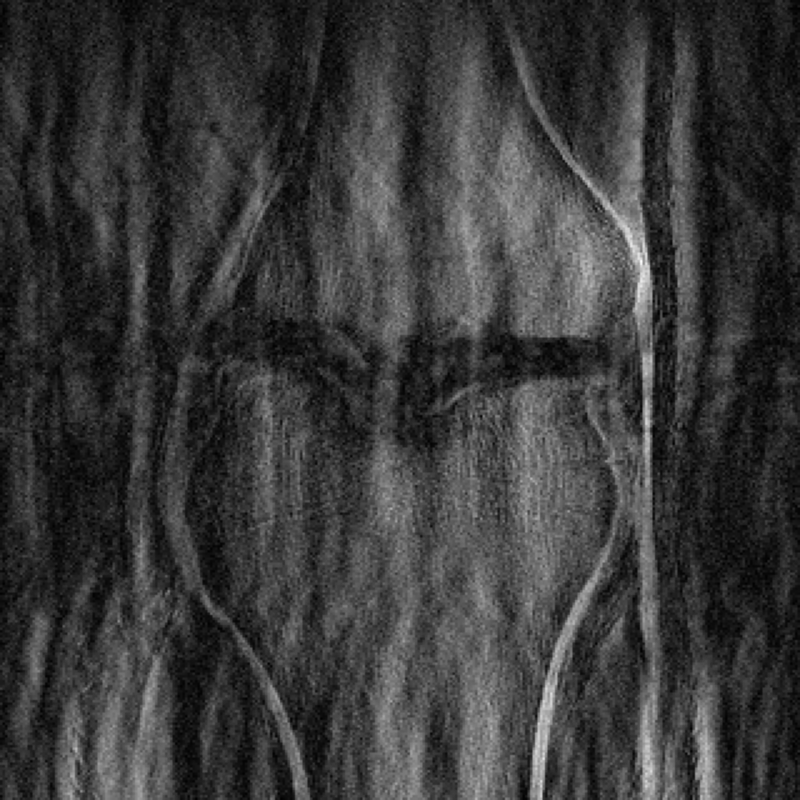} &
\includegraphics[align=c,width = 0.15\columnwidth]{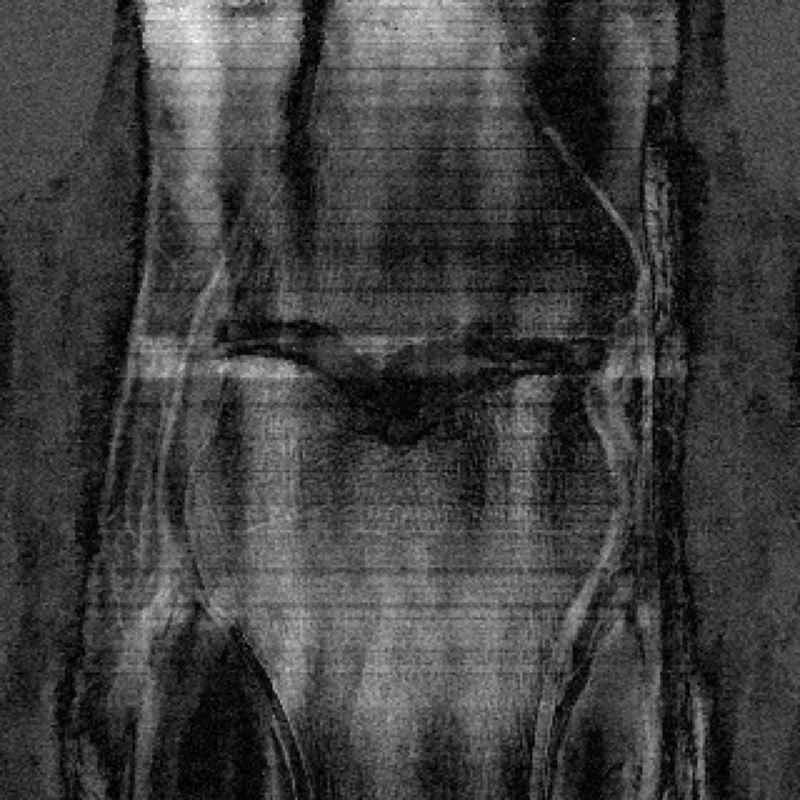} &
\includegraphics[align=c,width = 0.15\columnwidth]{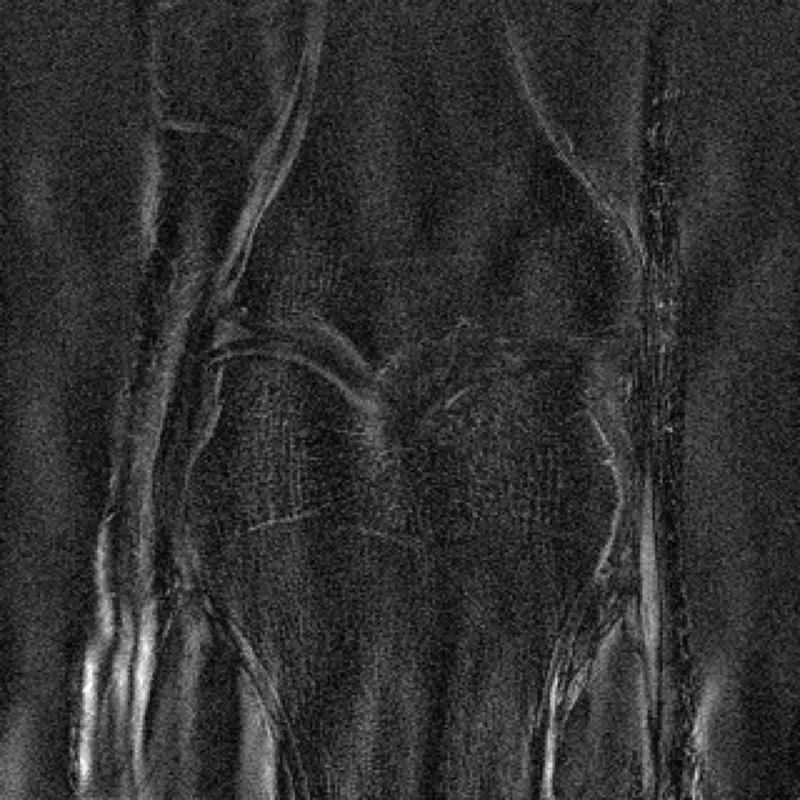}
\end{tabular}
}
\caption{Comparison figures for the MRI reconstruction methods in Figure \ref{fig:megafigmri}. We present the IFFT with all k-space data maintained, the naïve IFFT reconstruction after k-space masking, a total variation (TV) regularized reconstruction (with PSNR 27.3 dB), and a RED reconstruction (with PSNR 28.4 dB). We also present the residuals relative to the fully-sampled IFFT, multiplied by 5$\times$ for ease of visualization.}

\label{fig:mriinitialmethods}
\end{figure}

\begin{figure*}
\centering
\adjustbox{max width=\columnwidth}{
\renewcommand*{\arraystretch}{0}
\begin{tabular}{@{}c@{}c@{}c@{}c@{}c@{}c@{}c@{}c@{}c@{}}
& \small Train w/$A_0$  & \small Train w/$A_0$ & \small \pnp & \small \rnr & \small \rnr+ & \small \pnp & \small \rnr & \small \rnr+ \\
 & \small Test w/$A_0$ & \small Test w/$A_1$ & \small Known $A_1$ & \small Known $A_1$ & \small Known $A_1$ & \small Unknown $A_1$ & \small Unknown $A_1$ & \small Unknown $A_1$ \\ \vspace{2pt} \\
\begin{minipage}{0.08\linewidth} U-Net \end{minipage} &
\includegraphics[align=c, width = 0.12\linewidth]{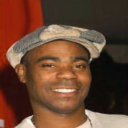} & 
\includegraphics[align=c, width = 0.12\linewidth]{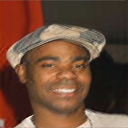} & 
\includegraphics[align=c, width = 0.12\linewidth]{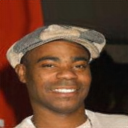} &
\includegraphics[align=c, width = 0.12\linewidth]{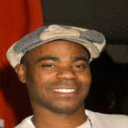} & 
\includegraphics[align=c, width = 0.12\linewidth]{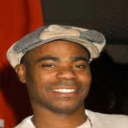} & 
\includegraphics[align=c, width = 0.12\linewidth]{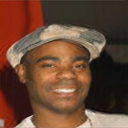} & 
\includegraphics[align=c, width = 0.12\linewidth]{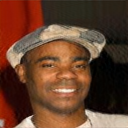} &
\includegraphics[align=c, width = 0.12\linewidth]{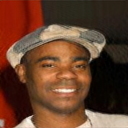} \\
\vspace{2pt}\begin{minipage}{0.08\linewidth} U-Net \\ Residual \end{minipage} &
\includegraphics[align=c, width = 0.12\linewidth]{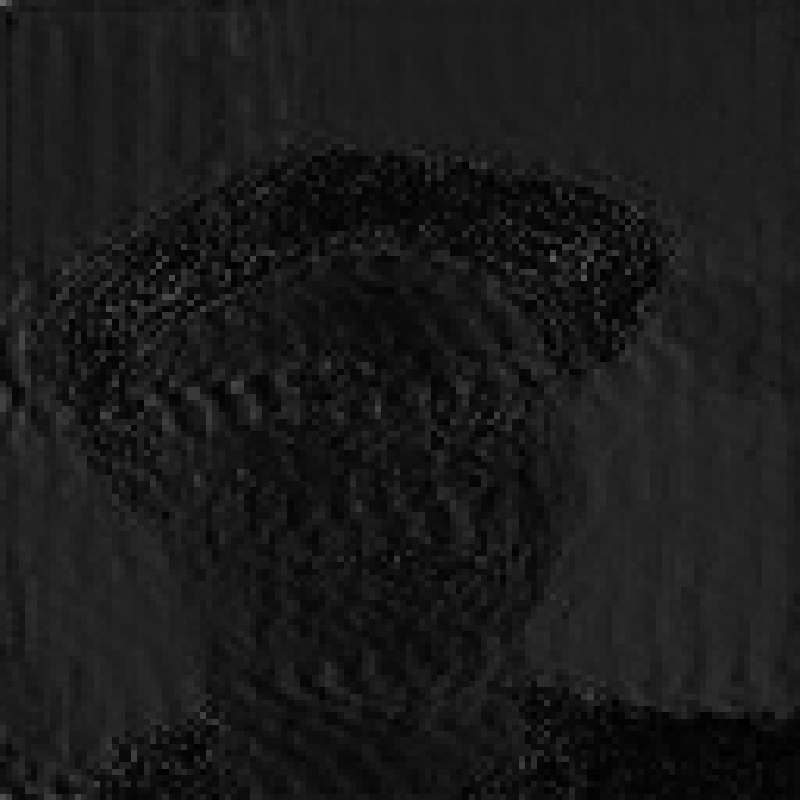} & 
\includegraphics[align=c, width = 0.12\linewidth]{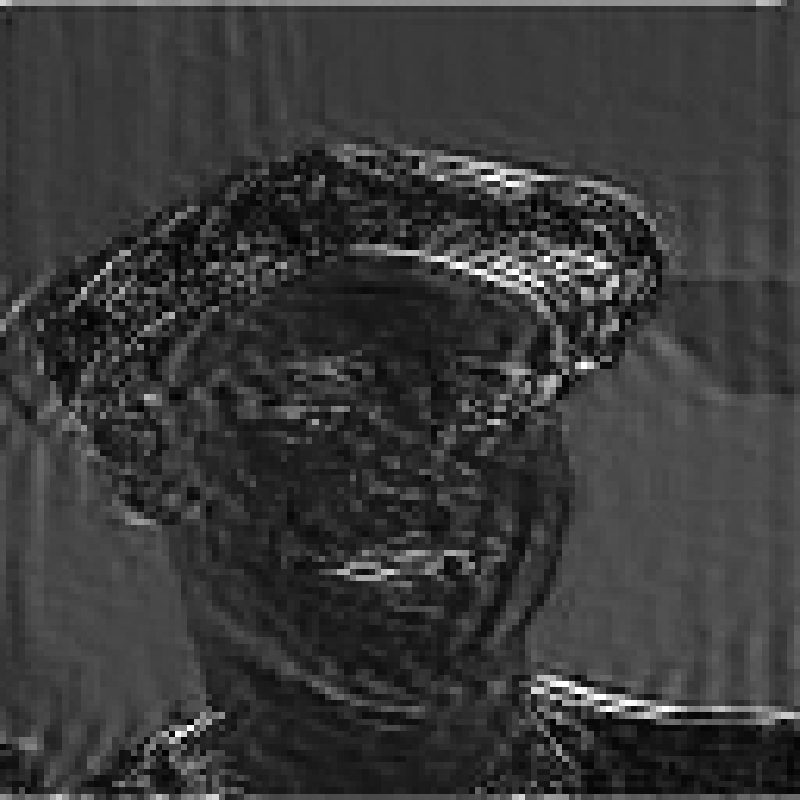} & 
\includegraphics[align=c, width = 0.12\linewidth]{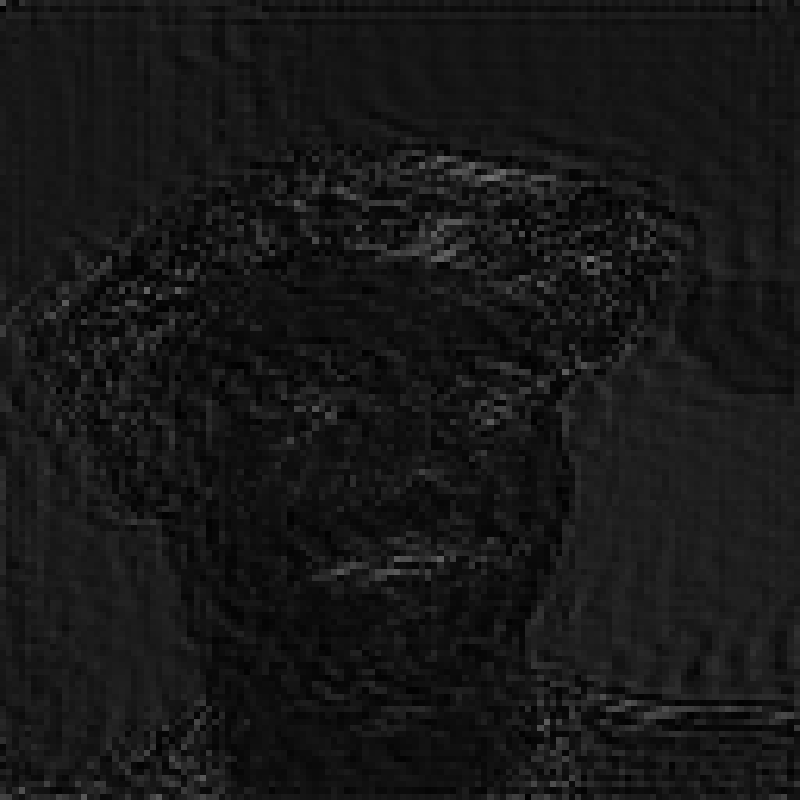} &
\includegraphics[align=c, width = 0.12\linewidth]{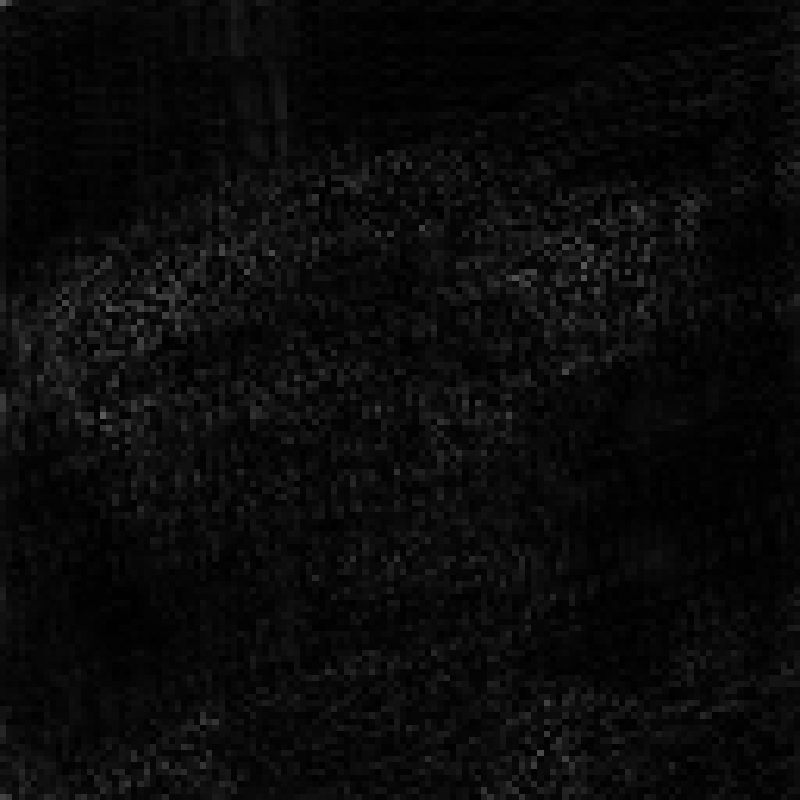} & 
\includegraphics[align=c, width = 0.12\linewidth]{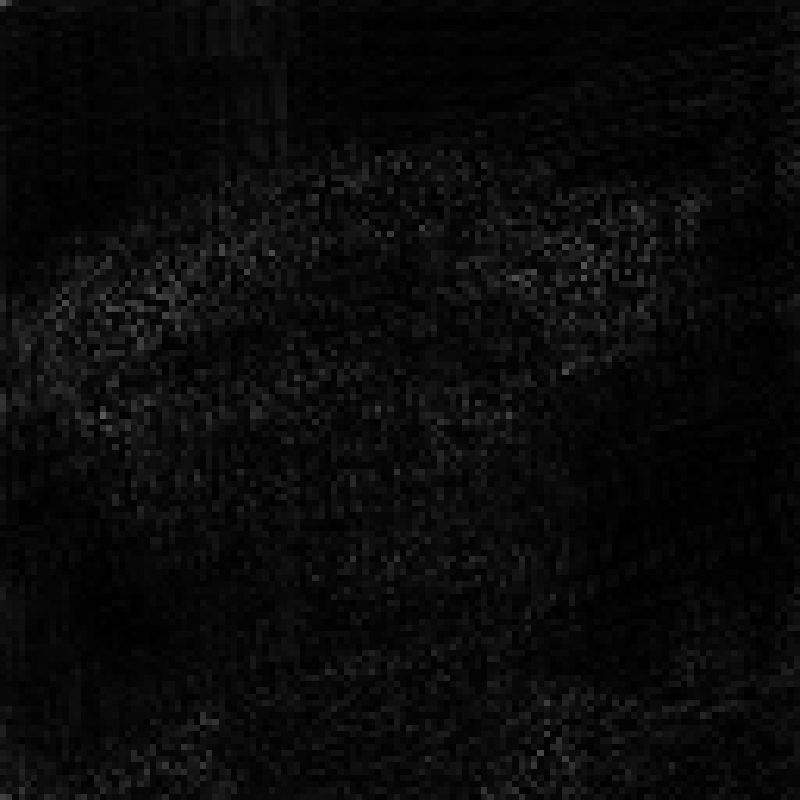} & 
\includegraphics[align=c, width = 0.12\linewidth]{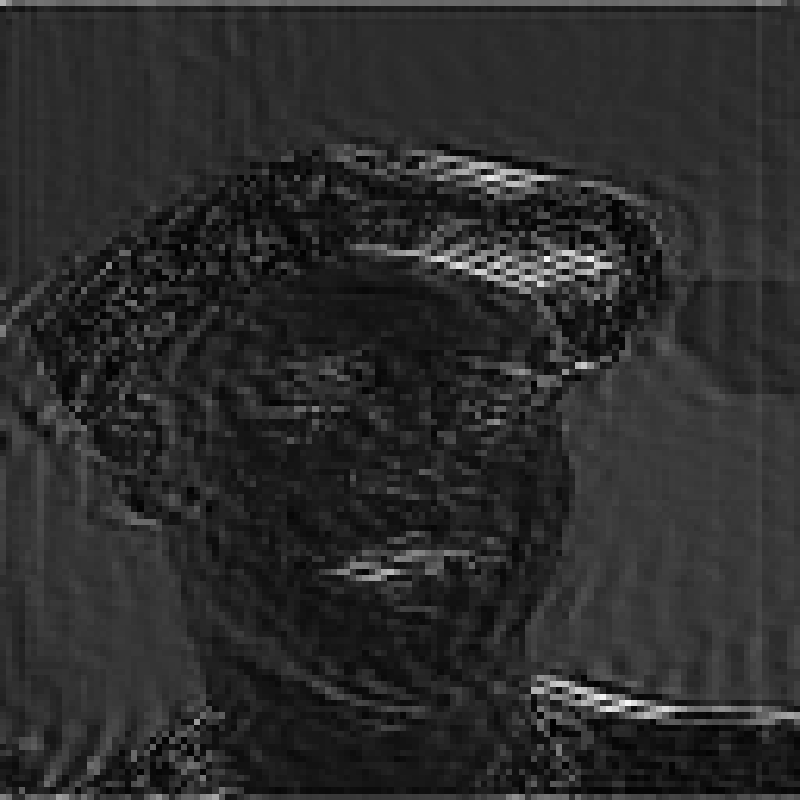} &
\includegraphics[align=c, width = 0.12\linewidth]{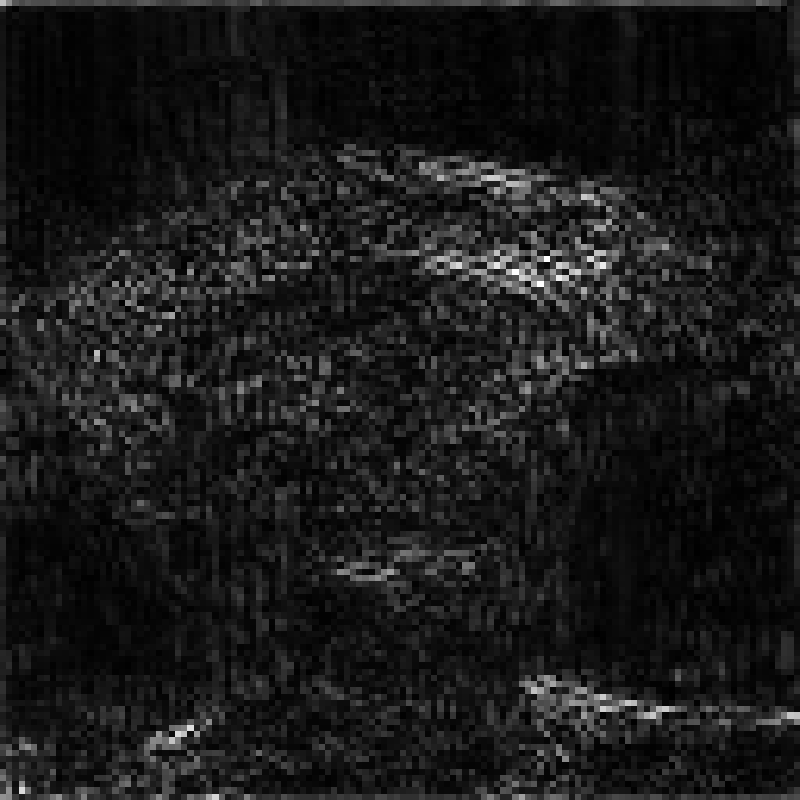} & 
\includegraphics[align=c, width = 0.12\linewidth]{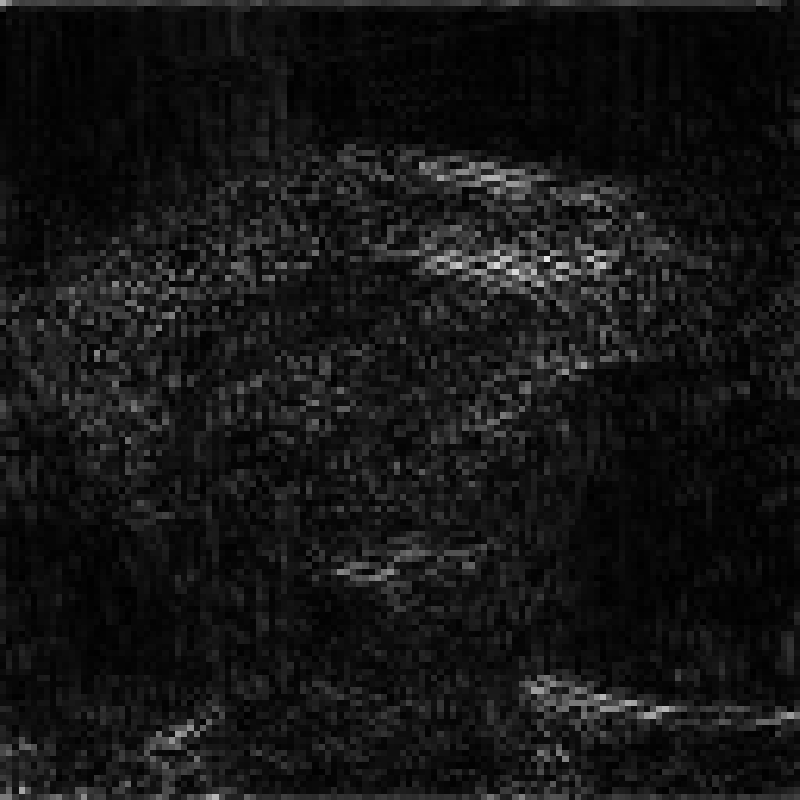} \\
\begin{minipage}{0.08\linewidth} MoDL \end{minipage} &
\includegraphics[align=c, width = 0.12\linewidth]{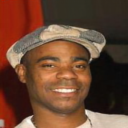} & 
\includegraphics[align=c, width = 0.12\linewidth]{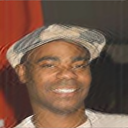} & 
\includegraphics[align=c, width = 0.12\linewidth]{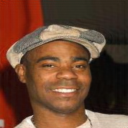} &
\includegraphics[align=c, width = 0.12\linewidth]{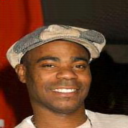} & 
\includegraphics[align=c, width = 0.12\linewidth]{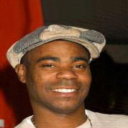} & 
\includegraphics[align=c, width = 0.12\linewidth]{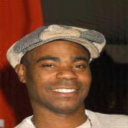} & 
\includegraphics[align=c, width = 0.12\linewidth]{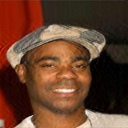} & 
\includegraphics[align=c, width = 0.12\linewidth]{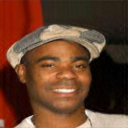} \\
\begin{minipage}{0.08\linewidth} MoDL \\ Residual \end{minipage} &
\includegraphics[align=c, width = 0.12\linewidth]{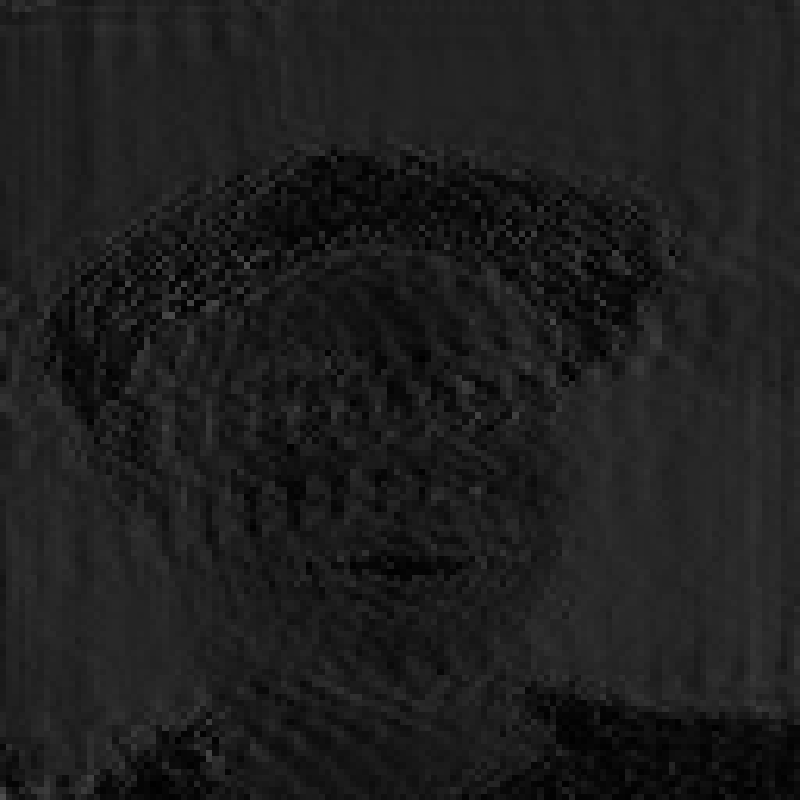} & 
\includegraphics[align=c, width = 0.12\linewidth]{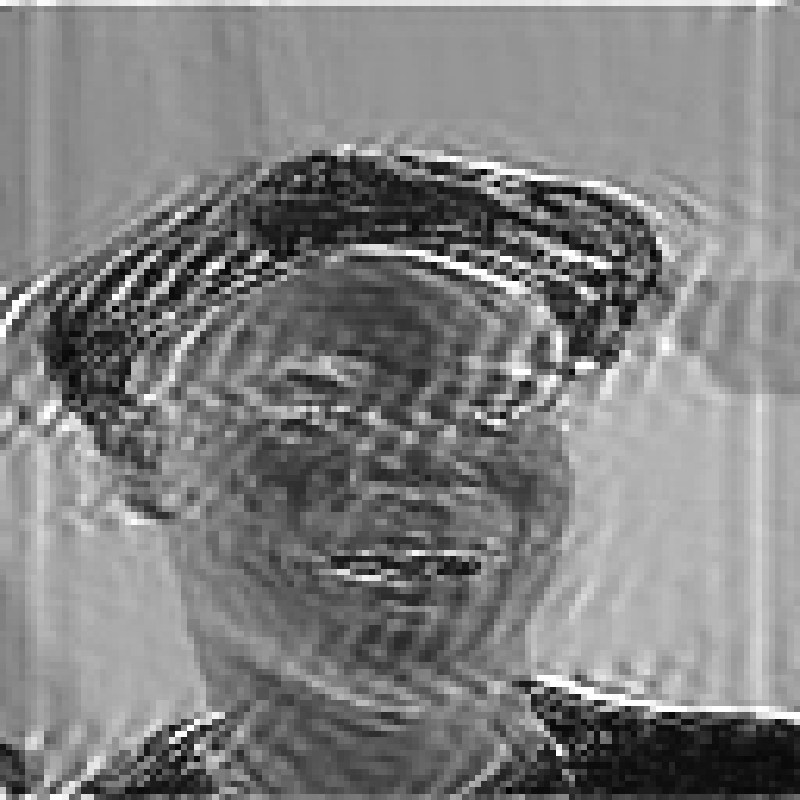} & 
\includegraphics[align=c, width = 0.12\linewidth]{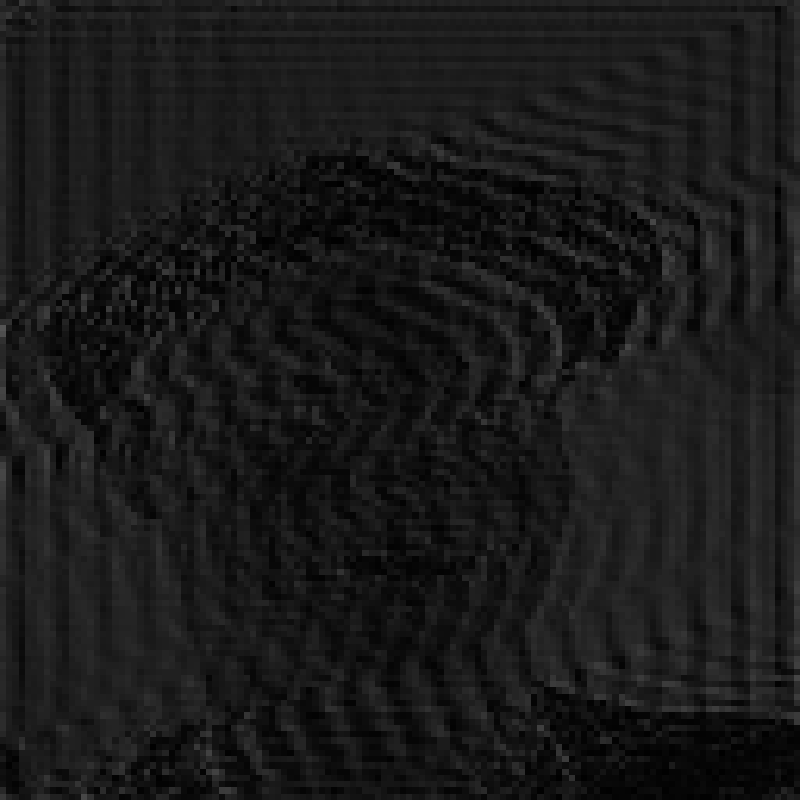} &
\includegraphics[align=c, width = 0.12\linewidth]{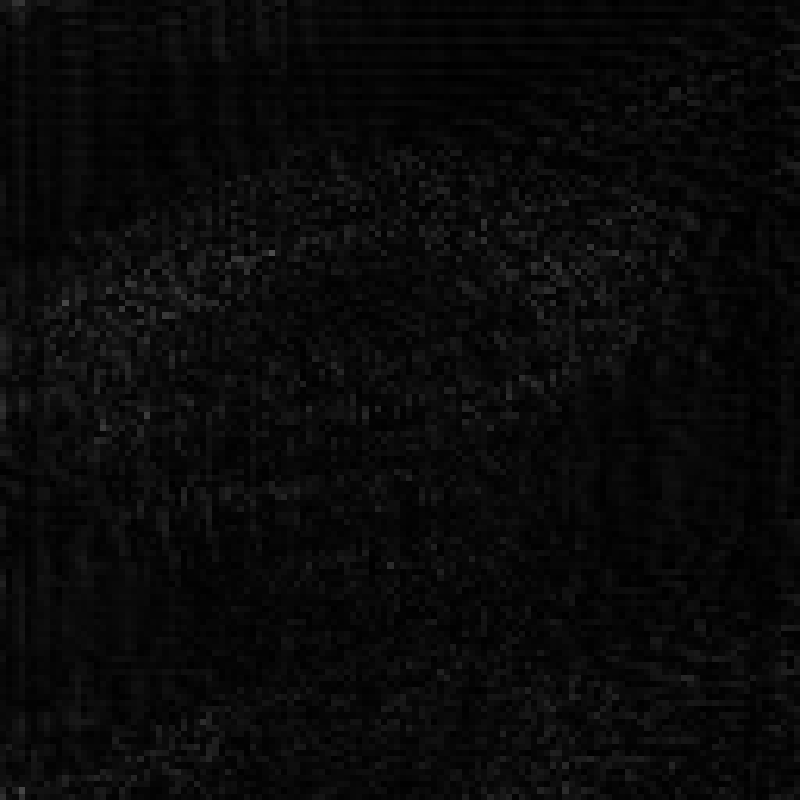} & 
\includegraphics[align=c, width = 0.12\linewidth]{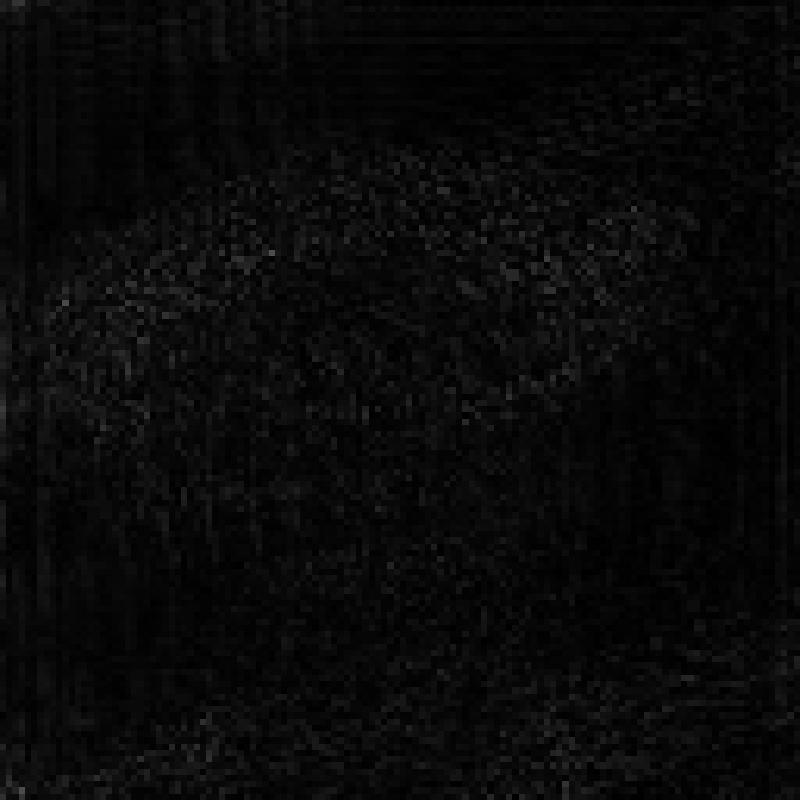} & 
\includegraphics[align=c, width = 0.12\linewidth]{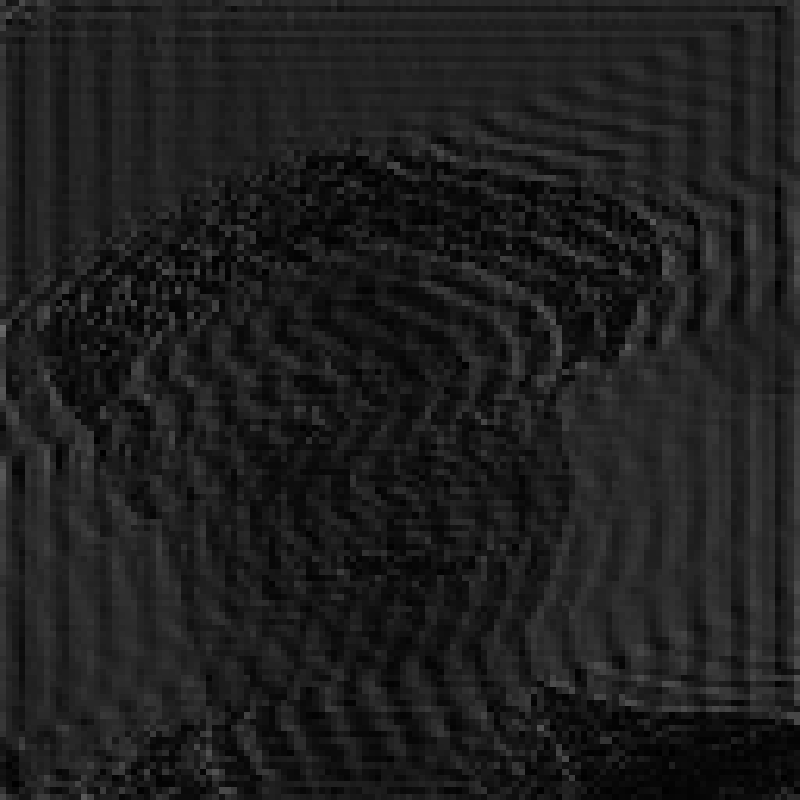} & 
\includegraphics[align=c, width = 0.12\linewidth]{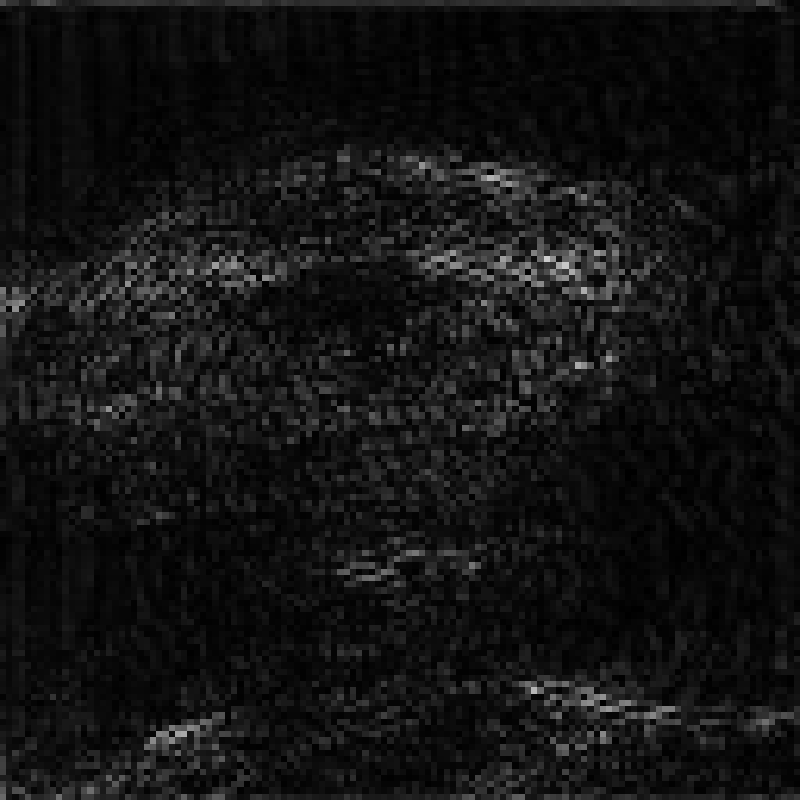} & 
\includegraphics[align=c, width = 0.12\linewidth]{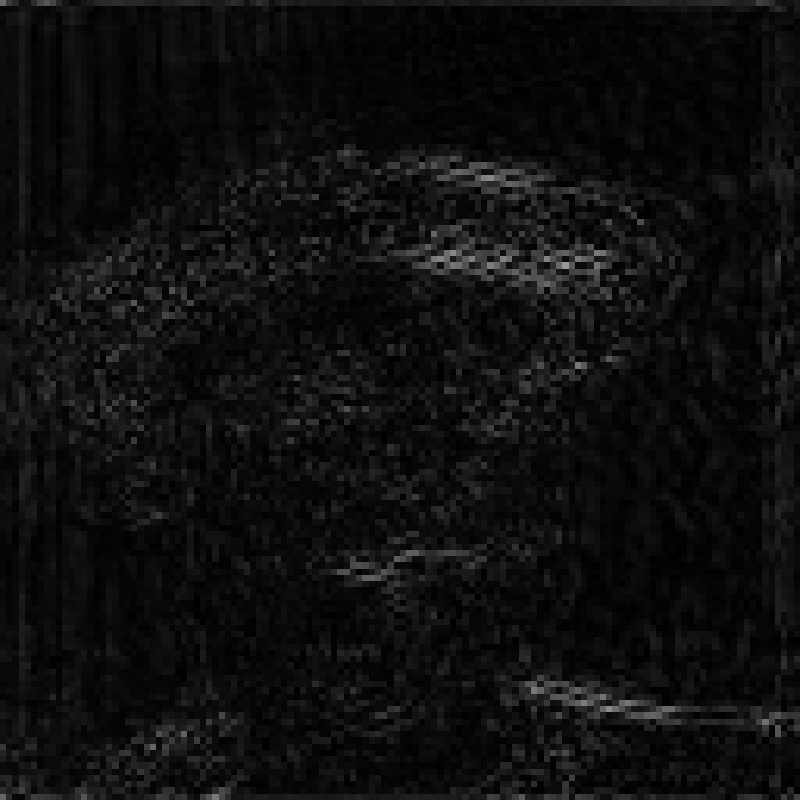}
\end{tabular}
}
\caption{Visual examples of reconstruction quality for the motion deblurring inverse problem solved by U-Net and MoDL, as well as the associated residuals. Each residual is multiplied by 5 for ease of inspection. The initial forward model $A_0$ is a 7x7 motion blur with angle $10^\circ$, and the $A_1$ model has a 7x7 motion blur kernel with angle $20^\circ$. The analogous figure for the superresolution problem, and further examples, are available in the Supplement. Best viewed electronically.}
\label{fig:megafigmotblur}
\end{figure*}

\begin{figure*}
\centering
\renewcommand*{\arraystretch}{0}
\begin{tabular}{@{}c@{}c@{}c@{}c@{}c@{}c@{}}
 & \small Train w/$A_0$  & \small Train w/$A_0$ & \small \pnp & \small \rnr & \small \rnr+ \\
& \small Test w/$A_0$ & \small Test w/$A_1$ & & & \\ 
\begin{minipage}{0.08\linewidth} \small U-Net \end{minipage} &
\includegraphics[align=c,width = 0.11\linewidth]{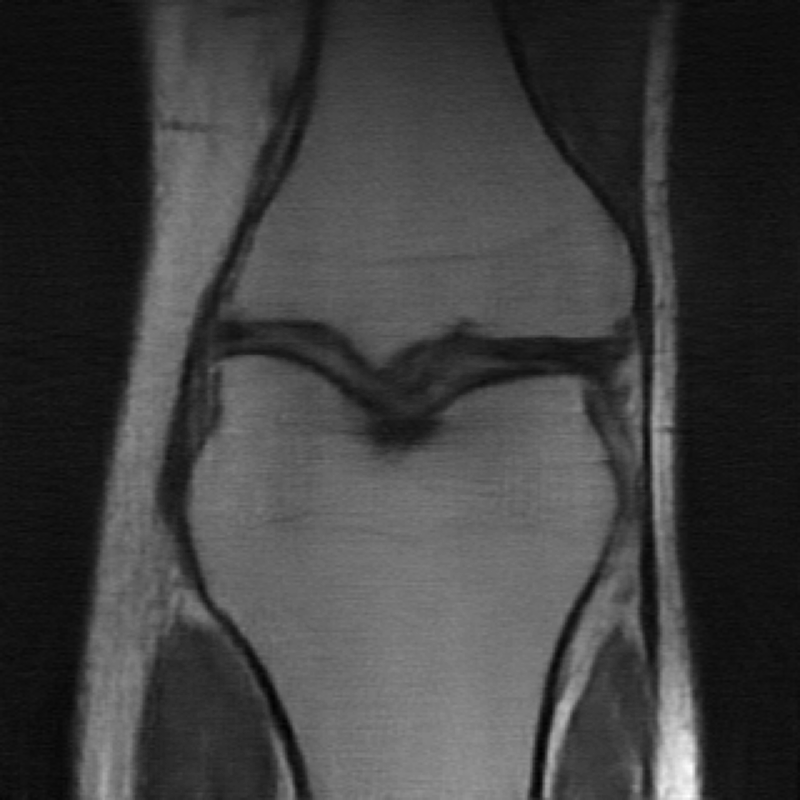} & 
\includegraphics[align=c,width = 0.11\linewidth]{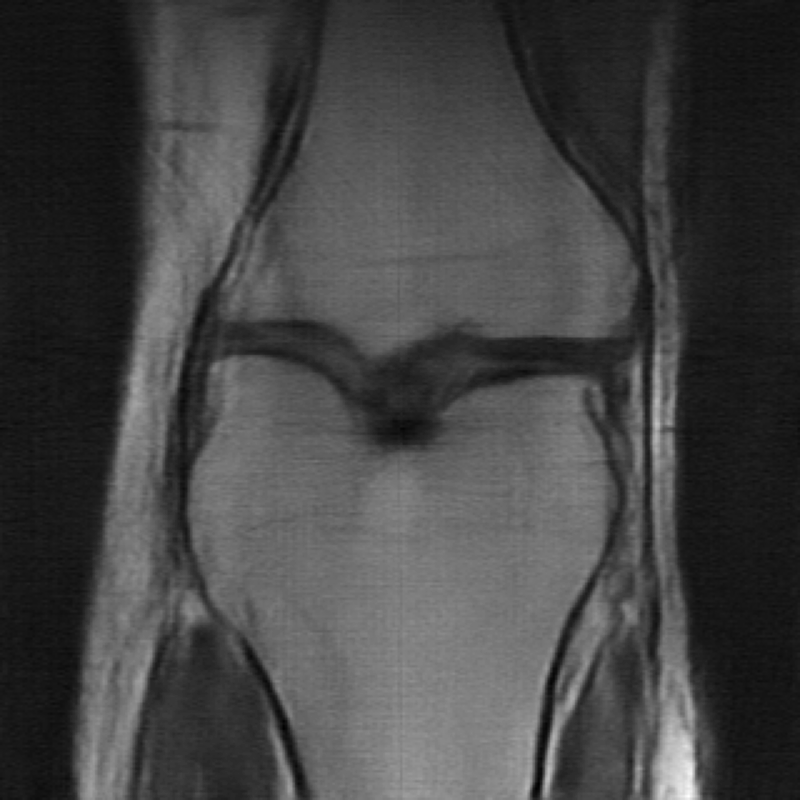} & 
\includegraphics[align=c,width = 0.11\linewidth]{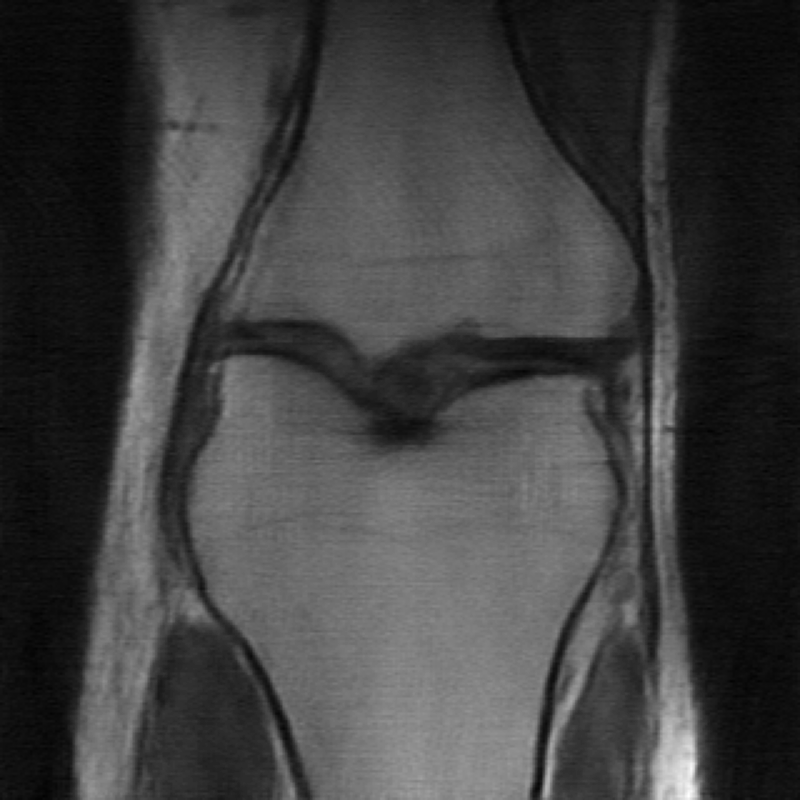} &
\includegraphics[align=c,width = 0.11\linewidth]{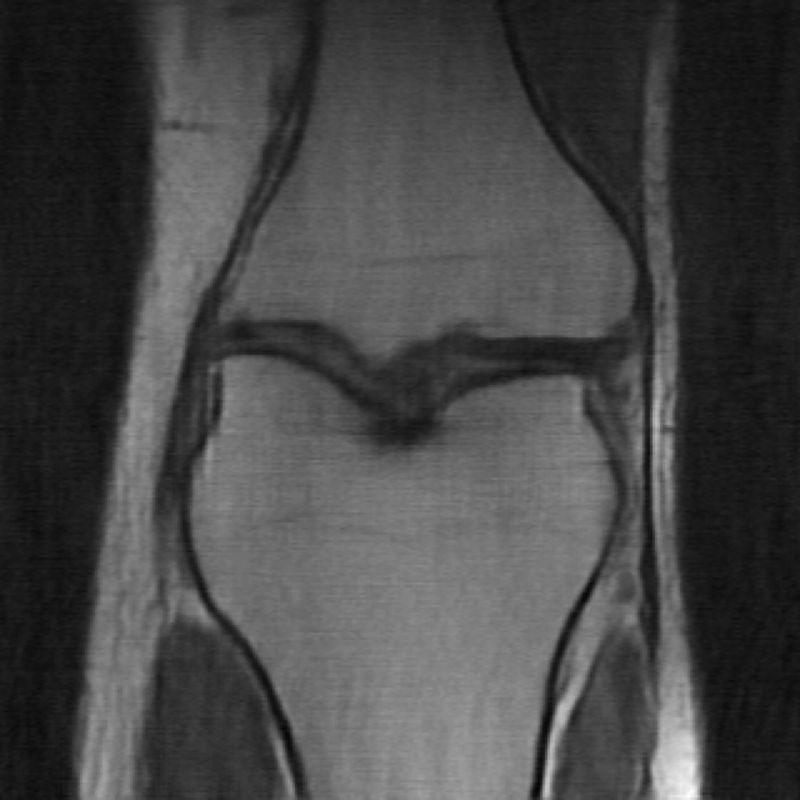} &
\includegraphics[align=c,width = 0.11\linewidth]{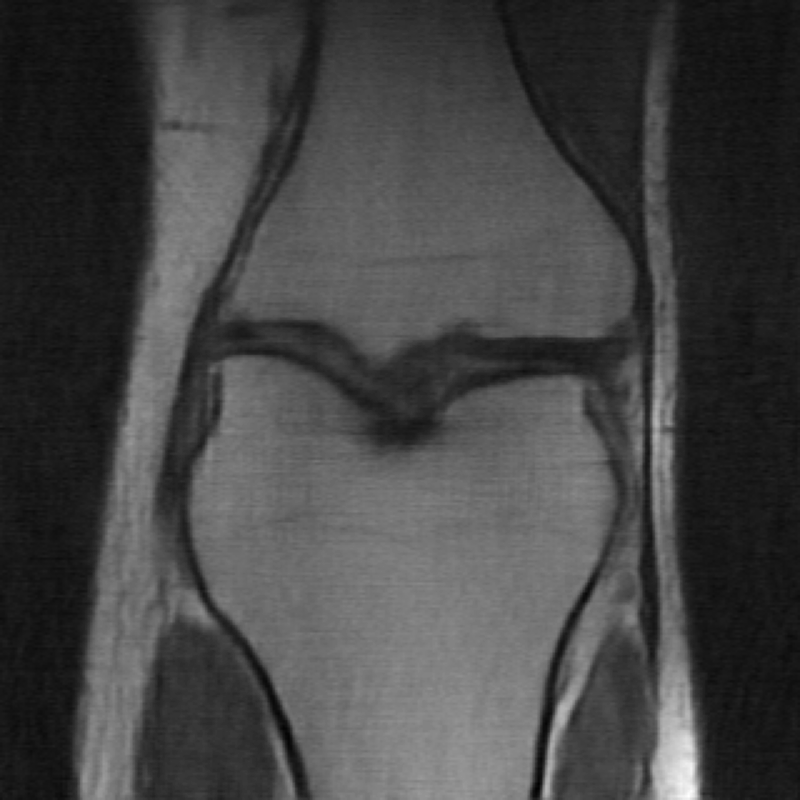} \\
\vspace{2pt}\begin{minipage}{0.08\linewidth}\small U-Net \\ Residual \end{minipage} & 
\includegraphics[align=c,width = 0.11\linewidth]{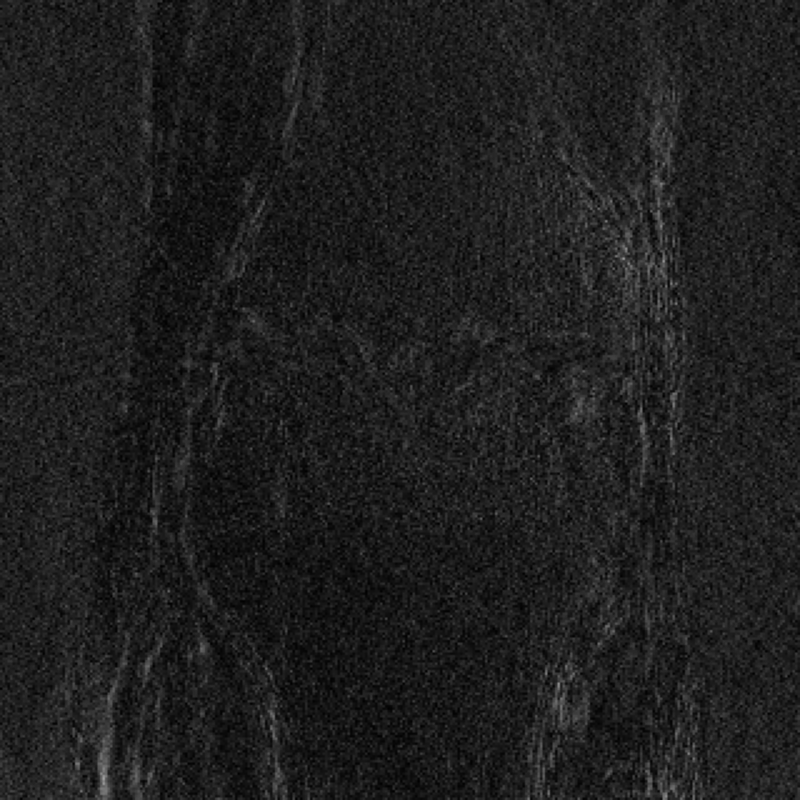} & 
\includegraphics[align=c,width = 0.11\linewidth]{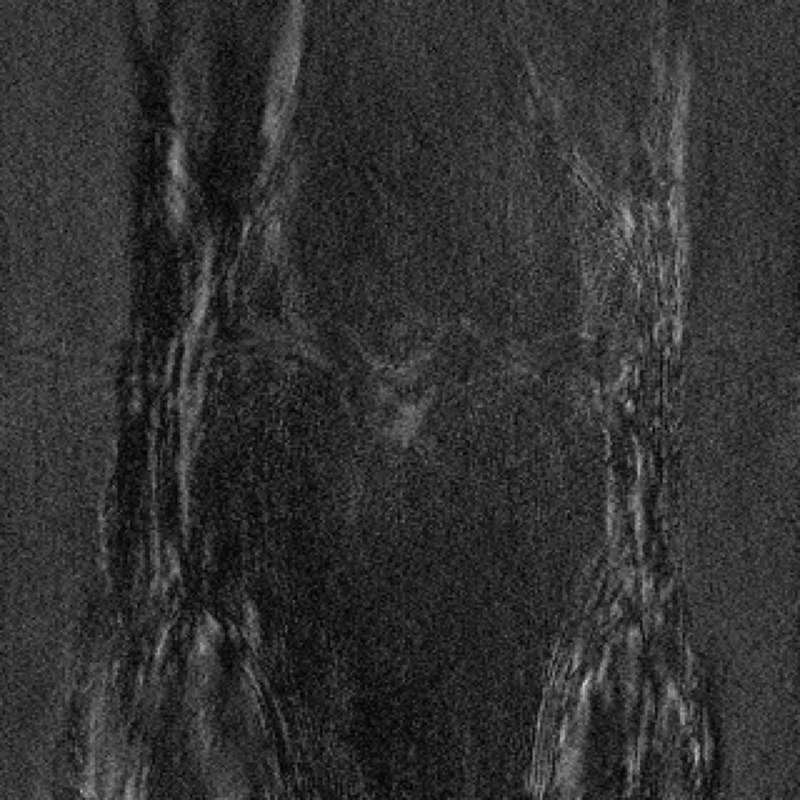} & 
\includegraphics[align=c,width = 0.11\linewidth]{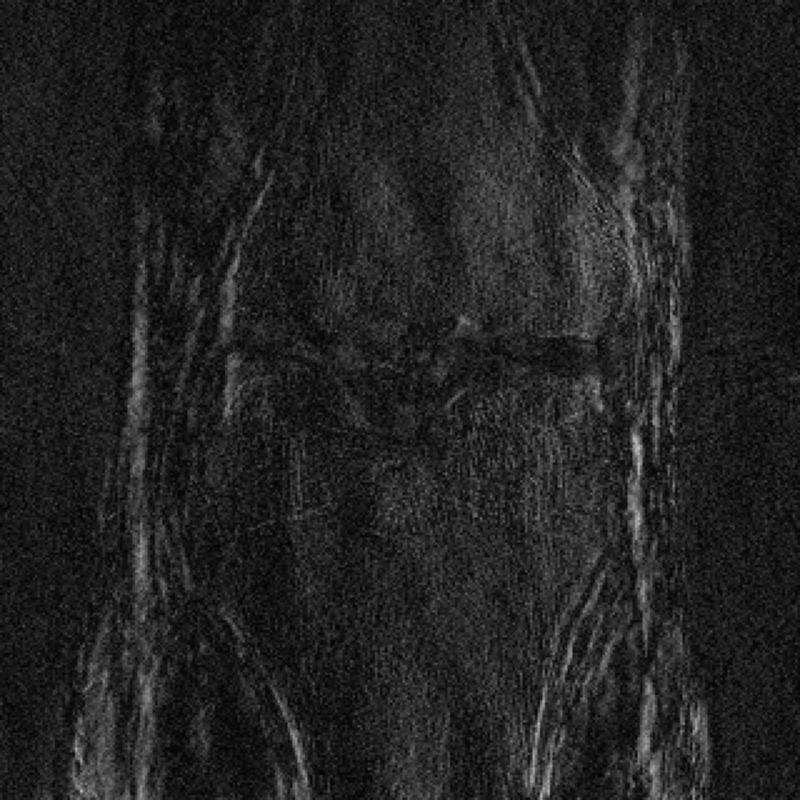} &
\includegraphics[align=c,width = 0.11\linewidth]{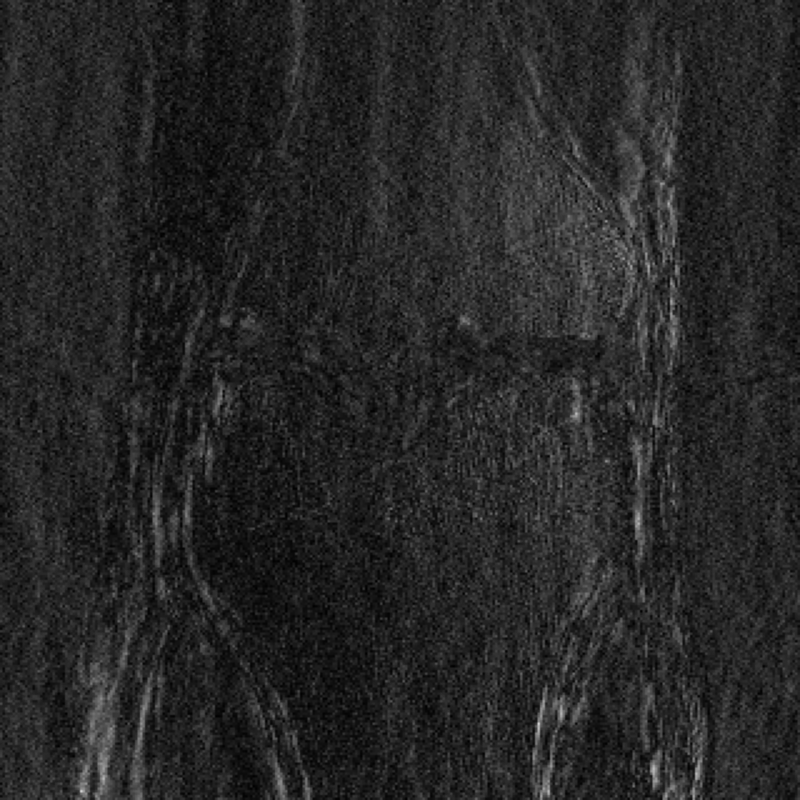} &
\includegraphics[align=c,width = 0.11\linewidth]{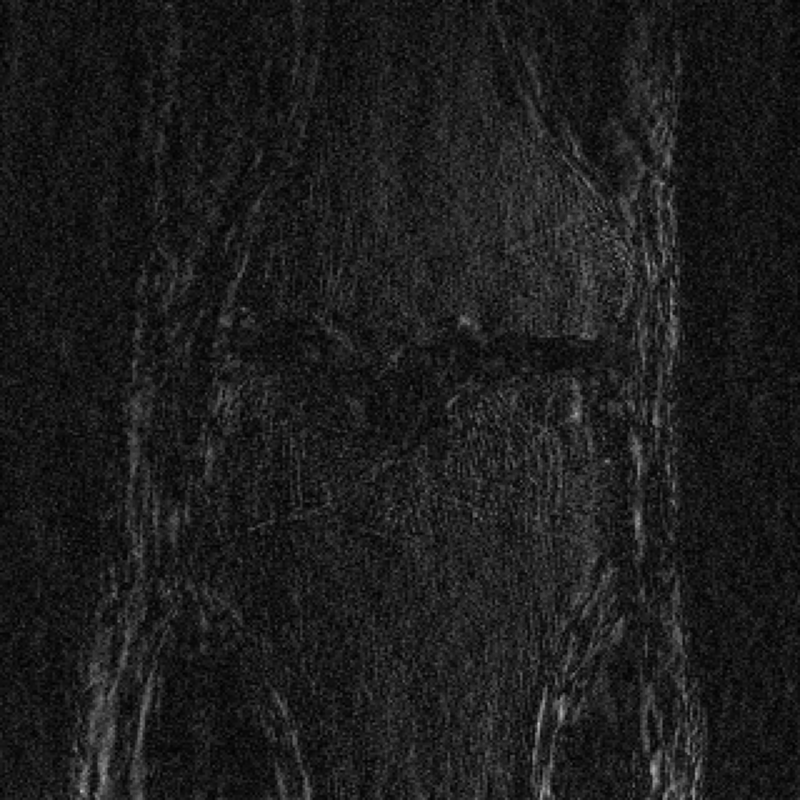} \\ 
\vspace{2pt} \begin{minipage}{0.08\linewidth} \small  MoDL \end{minipage} &
\includegraphics[align=c,width = 0.11\linewidth]{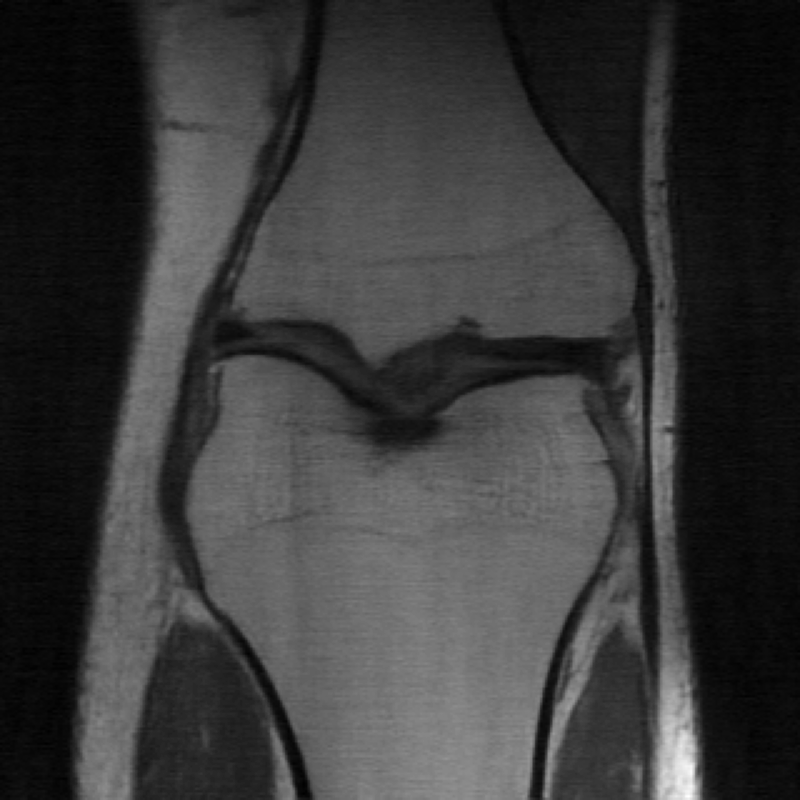} & 
\includegraphics[align=c,width = 0.11\linewidth]{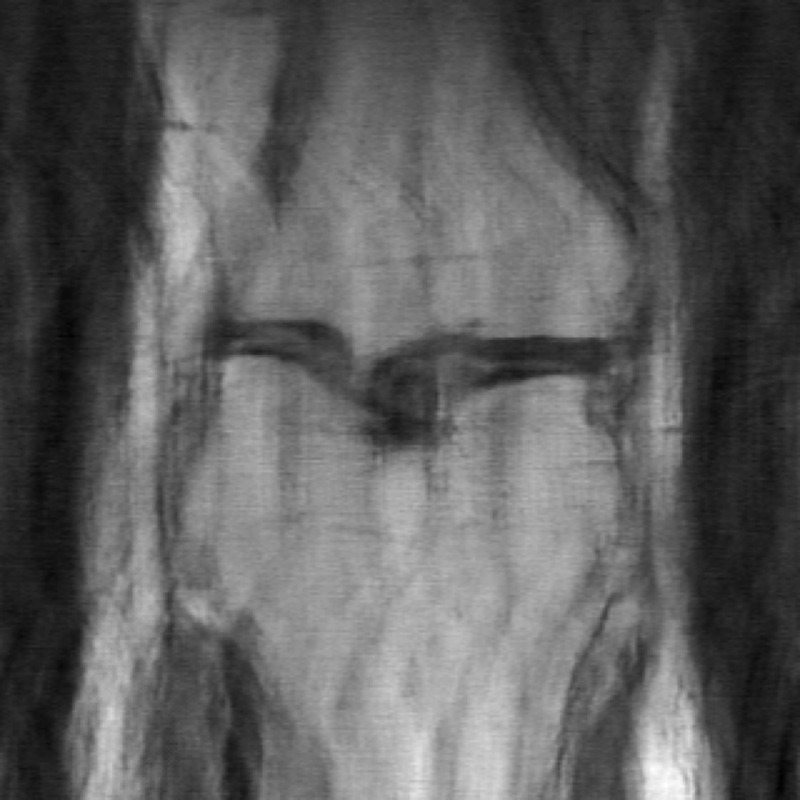} & 
\includegraphics[align=c,width = 0.11\linewidth]{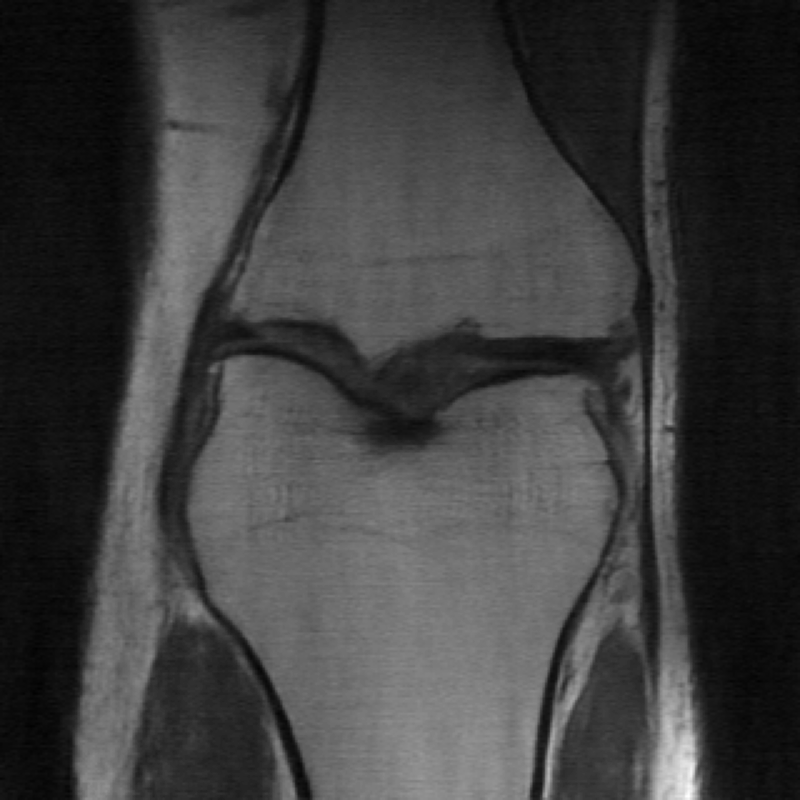} &
\includegraphics[align=c,width = 0.11\linewidth,align=c]{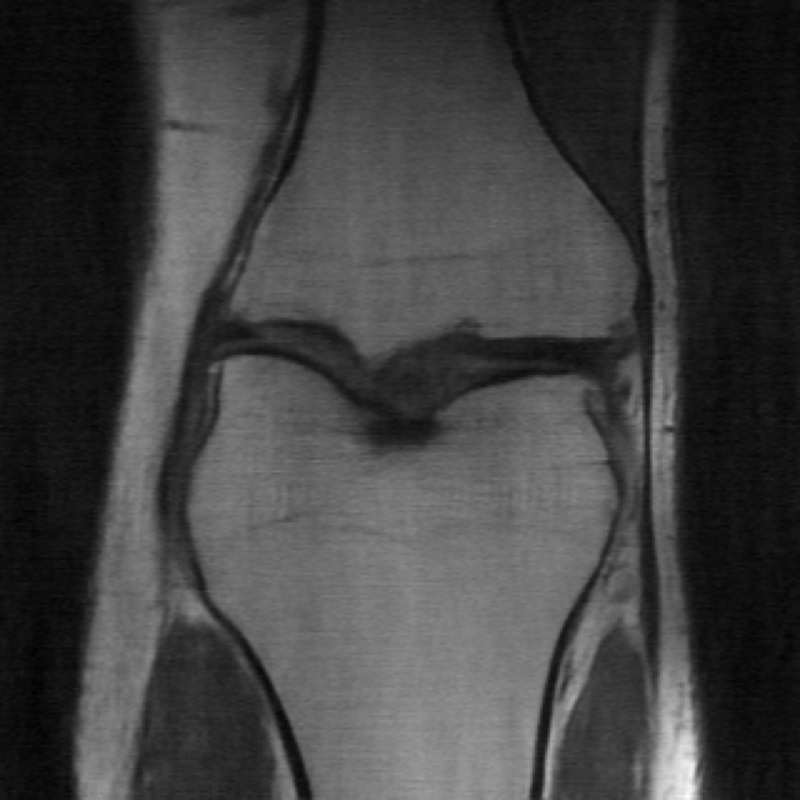} &
\includegraphics[align=c,width = 0.11\linewidth,align=c]{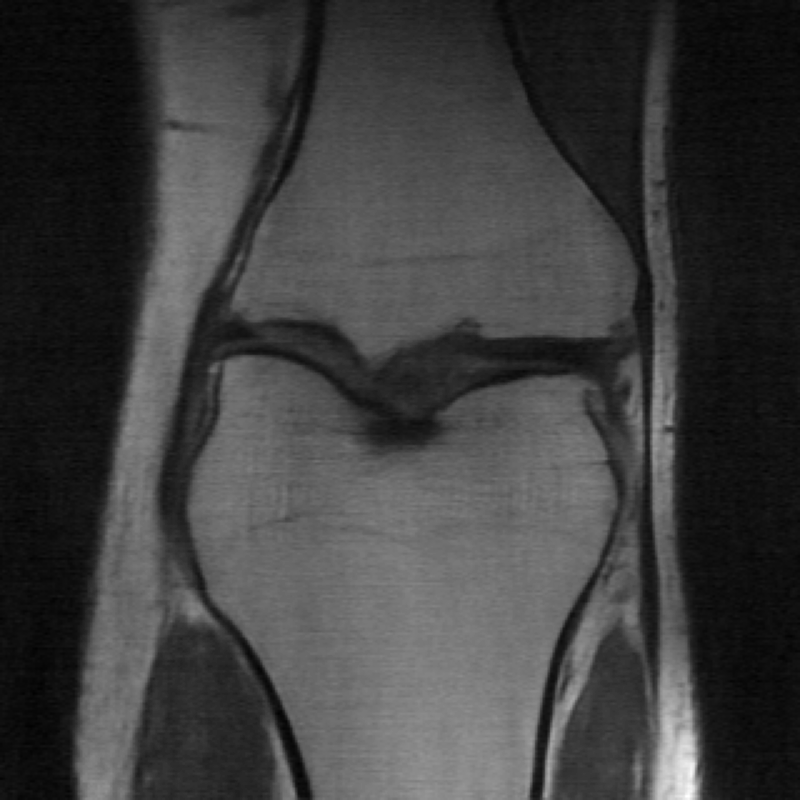} \\
\vspace{2pt}\begin{minipage}{0.08\linewidth}\small MoDL \\ Residual \end{minipage} & 
\includegraphics[align=c,width = 0.11\linewidth]{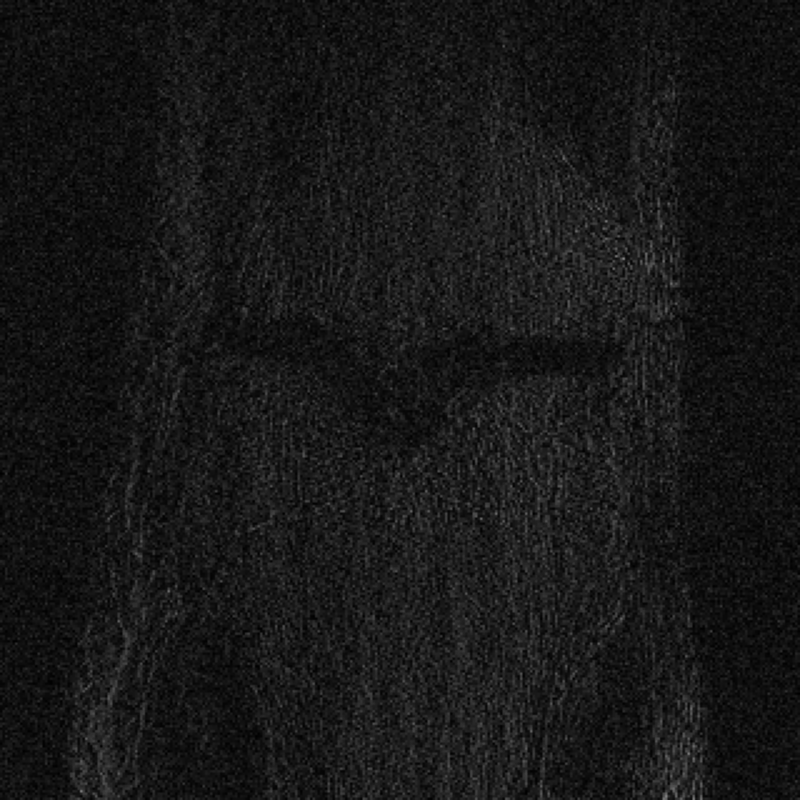} & 
\includegraphics[align=c,width = 0.11\linewidth]{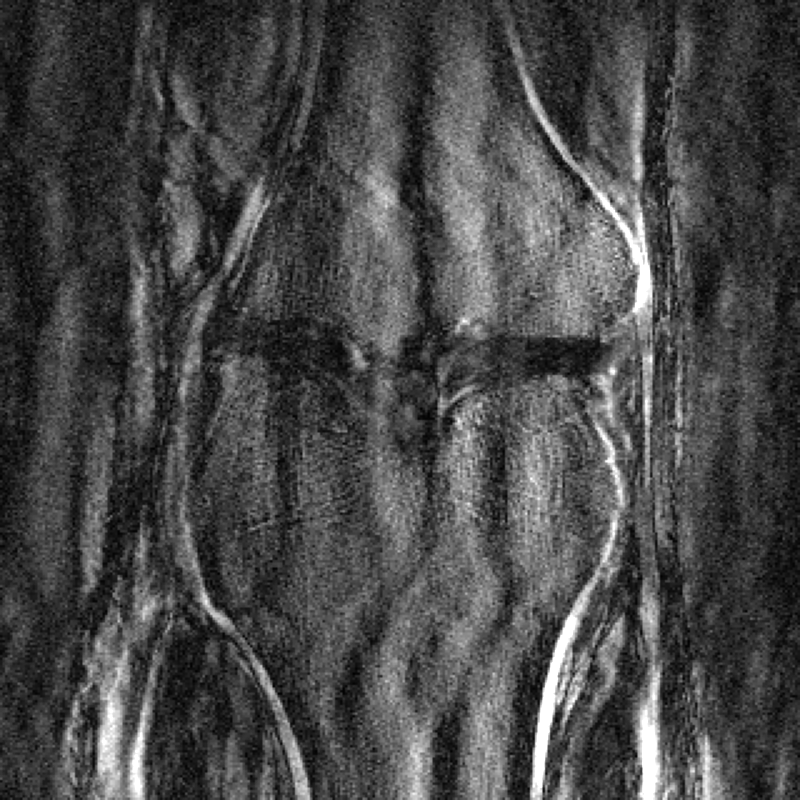} & 
\includegraphics[align=c,width = 0.11\linewidth]{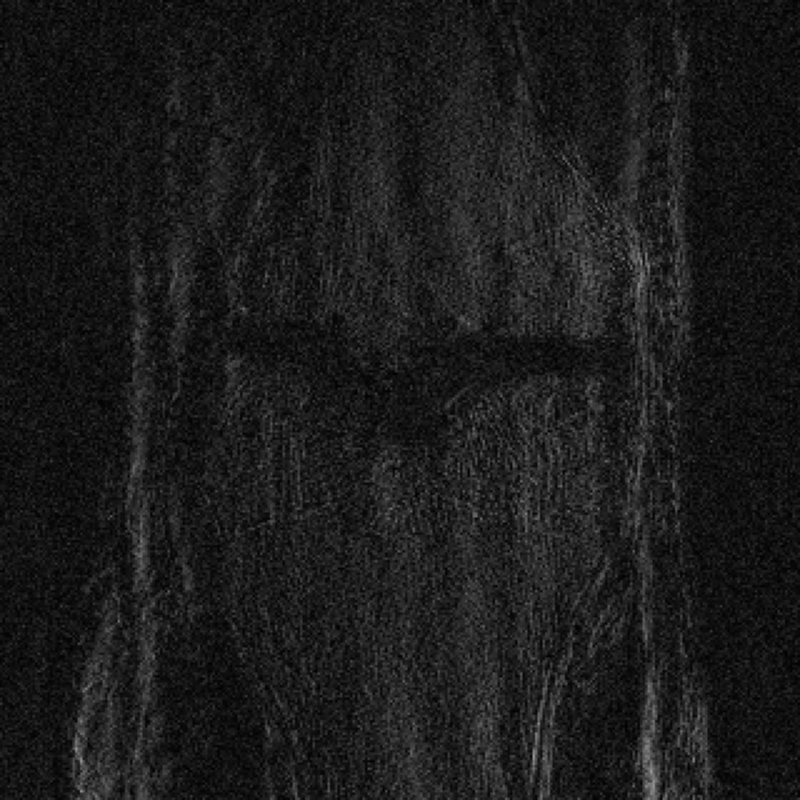} &
\includegraphics[align=c,width = 0.11\linewidth,align=c]{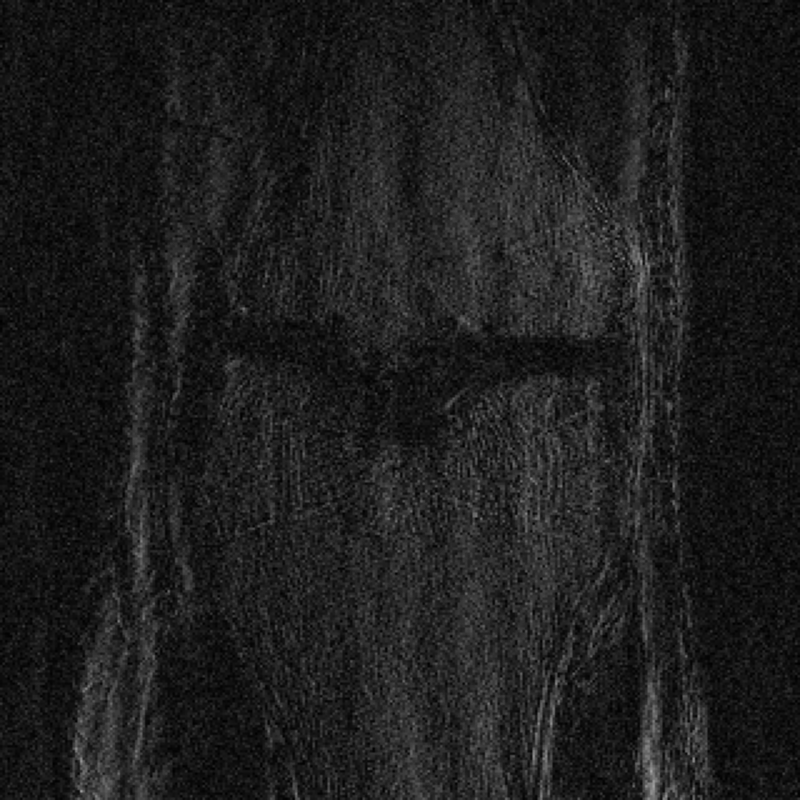} &
\includegraphics[align=c,width = 0.11\linewidth,align=c]{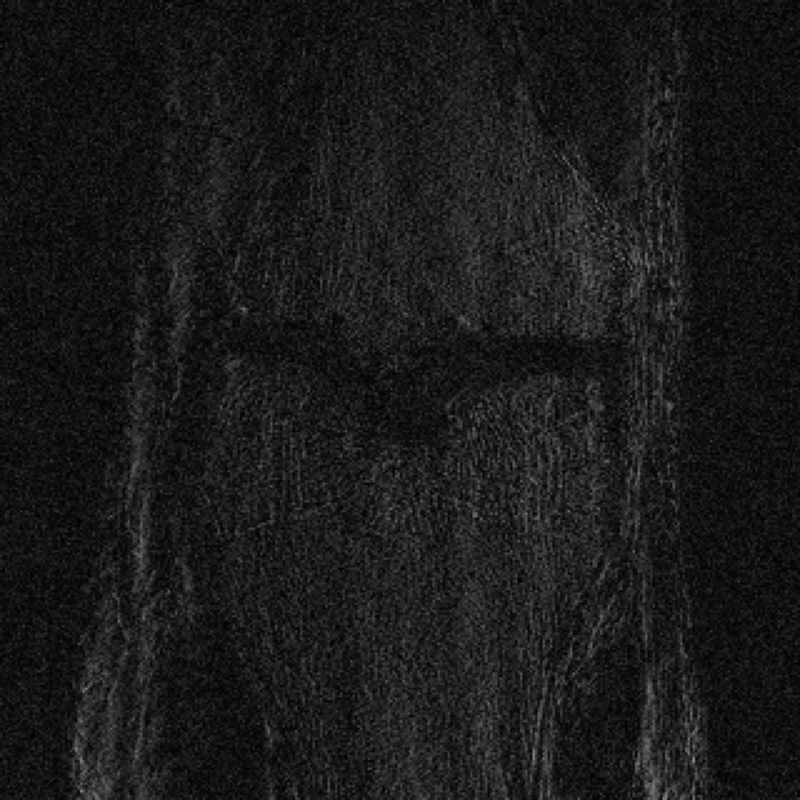}
\end{tabular}
\caption{Visual examples of different reconstruction approaches for the MRI inverse problem under model drift, along with associated residuals. All residual images are scaled by 5x for ease of inspection. Best viewed electronically.} 
\label{fig:megafigmri}
\end{figure*}

\subsection{Learning multiple forward models}

\definecolor{mygreen}{HTML}{01B051}
\definecolor{myblue}{HTML}{5B9BD5}
\definecolor{myred}{HTML}{FF0001}
\definecolor{myorange}{HTML}{ED7D31}
\definecolor{myyellow}{HTML}{FFC000}

\begin{figure}
\centering
\includegraphics[width=0.5\linewidth]{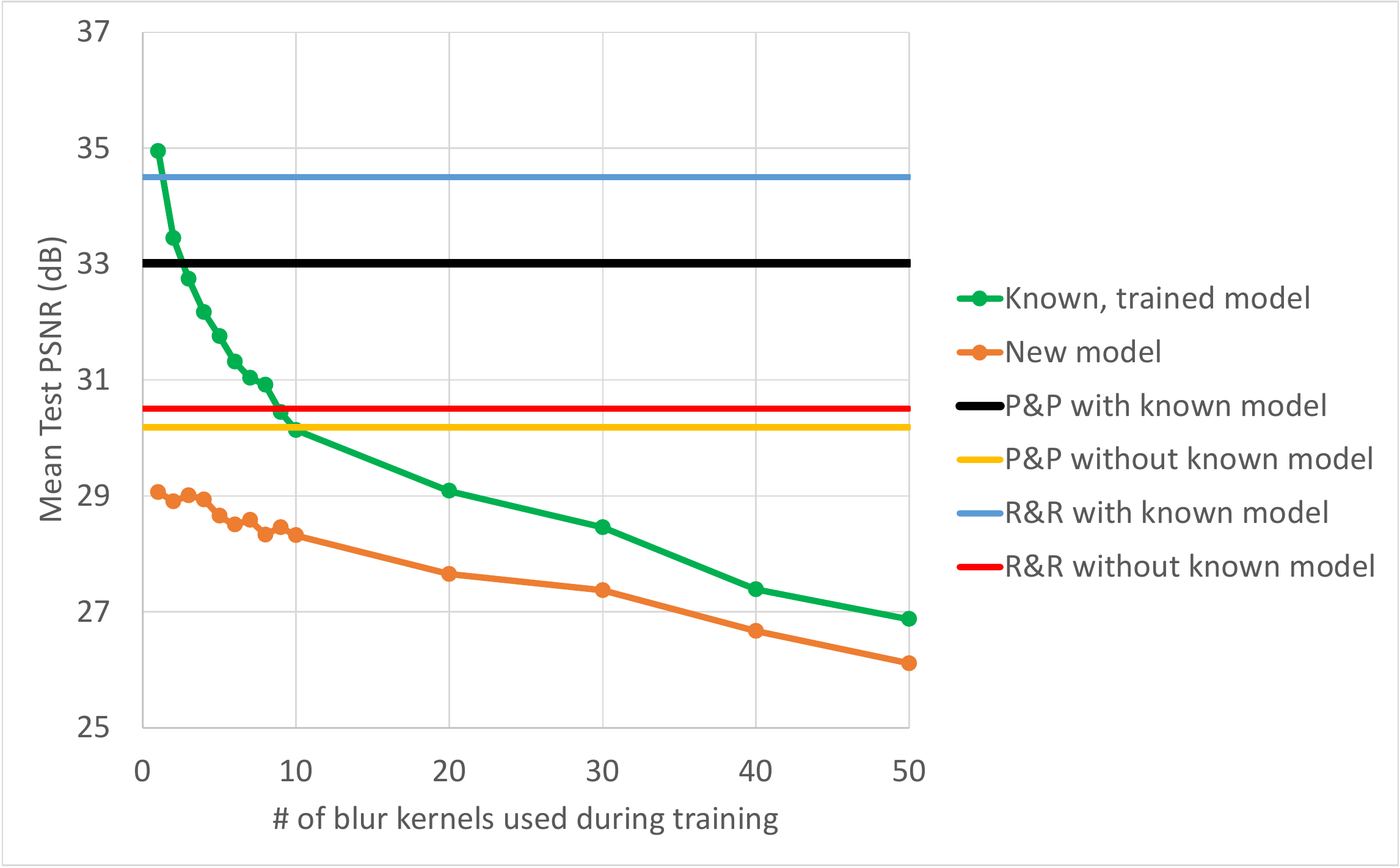}
\caption{Na\"ively learning to deblur with a single network and multiple blur kernels sacrifices performance on all blurs. 
In \textcolor{mygreen}{\bf green}, the test-time accuracy of a network trained to deblur multiple blurs, tested on a known kernel. In \textcolor{myorange}{\bf orange}, the same network, but tested on a new blur that was not used during training. In \textcolor{black}{\bf black}, our proposed P\&P approach (Alg. \ref{alg:pnp}), with a known model, and in \textcolor{myyellow}{\bf yellow} the same with a learned forward model. \textcolor{myblue}{\bf Blue} and \textcolor{myred}{\bf red} show the performance of our \change{R\&R} approach (Alg. \ref{alg:rnr}), with and without a known forward model.}
\label{fig:multimodelexp}
\end{figure}

In this section we explore an alternative approach to model adaptation. In this setting, we assume that a set of candidate forward models are known during training time. During test time, a single forward model is used for measurement, but the test-time forward model is not known during training. This case represents the setting where the forward model might be parametrized, and so a reasonable approach may be to train the learned network using a number of different forward models to improve robustness.

In simple settings, training on multiple models might be reasonable. However, when the forward model parameterization is high-dimensional, learning to invert all possible forward models may be difficult. 

We demonstrate this setting with a deblurring example, in which the same network is trained using a number of blur kernels. The blur kernels are the same kernels used for comparisons in \cite{hradivs2015convolutional}. For consistency, we resize all 50 blur kernels to 7x7, and normalize the weights to sum to 1. We compare reconstruction accuracy when the ground truth blur kernel is included in the set of kernels used for training, as well as when the reconstruction network has never seen data blurred with the testing kernel.

The results are shown in Fig \ref{fig:multimodelexp}. Experimentally, we find that training on multiple blur kernels simultaneously incurs a  performance penalty as the number of blur kernels used in training increases. In this setting, where the forward model has many degrees of freedom and data is limited, attempting to learn to solve all models simultaneously is worse than transferring a single learned model, even in the absence of further ground truth data for calibration.

\begin{figure*}[ht]
  \centering
  \includegraphics[width=\linewidth]{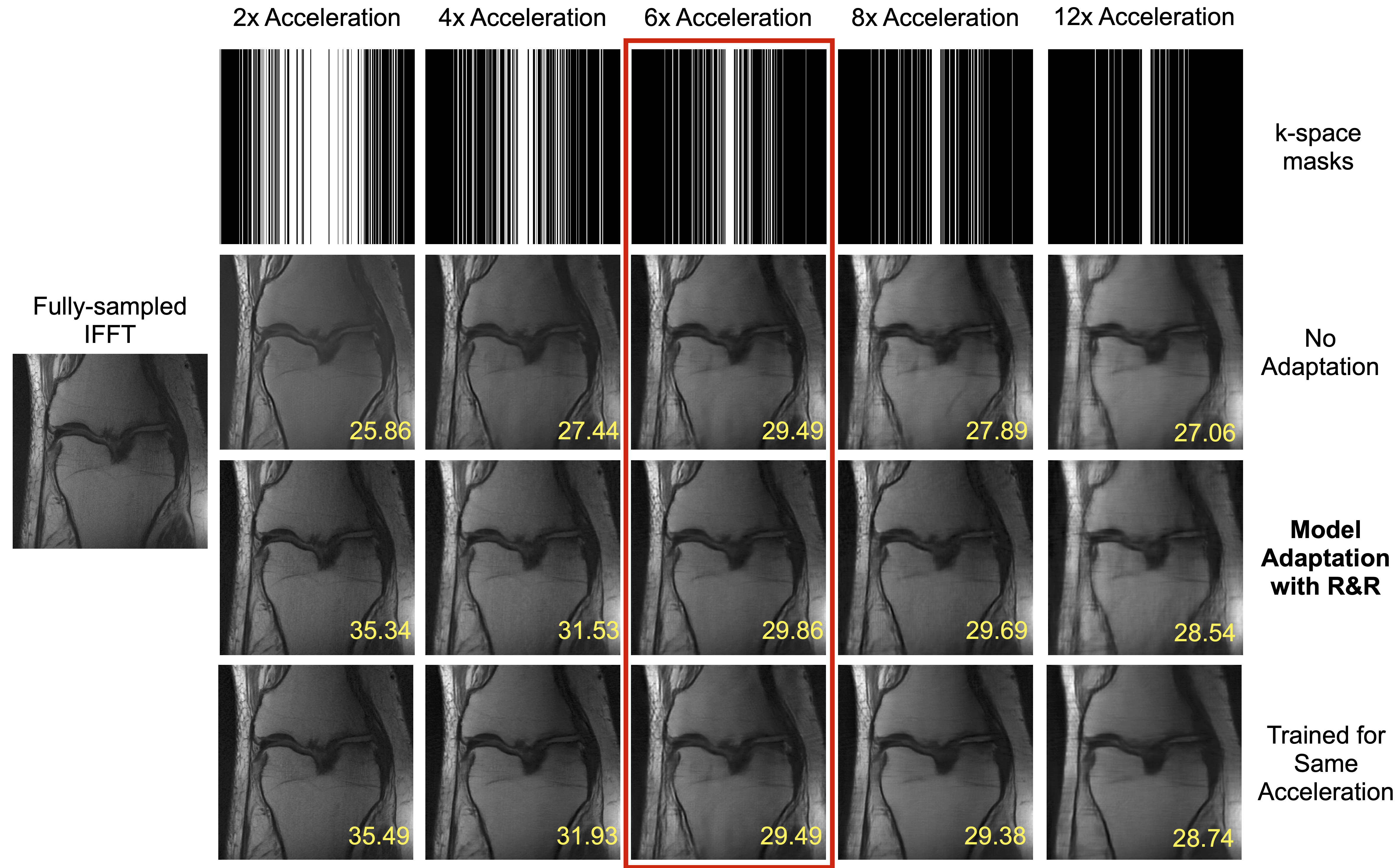}
\caption{\change{Using the \rnr\ model adaptation approach permits using a U-Net trained for $6\times$ acceleration on MRI reconstruction across a range of acceleration parameters. The various k-space sampling patterns used in these experiments are shown in the top row. Without adaptation (second row), the reconstruction quality decreases when changing the acceleration factor, \emph{even when more k-space measurements are taken}, as originally observed in \cite{antun2020instabilities}. The \rnr\ reconstructions (third row) compare favorably to the performance of networks trained on each particular k-space sampling pattern (bottom row). The PSNR of each image is presented in dB in yellow on each image.}}
\label{fig:samplerate}
\end{figure*}

\change{\subsection{Adapting to variable sampling rates in single-coil MRI}\label{sec:mrisamplingexp}

A particular concern raised in \cite{antun2020instabilities} is related to the stability of a learned solver with respect to the level of undersampling at measurement time. In particular, the authors of that work observe that an image reconstruction system trained to recover images sampled at a particular rate would experience a degradation in reconstruction accuracy for \emph{higher} sampling rates than the one the system was trained on.

In Fig. \ref{fig:samplerate} we explore this problem in the MRI setting using a U-Net as the reconstruction method, and demonstrate that our \rnr\ method can adapt to this setting as well. By using \rnr\ during inference, the learned network was trained at a $6\times$ accceleration acquisition setting, but was safely deployed for other accelerations \emph{without significant degradation in reconstruction quality}, and \emph{comparing favorably to networks trained explicitly for other sampling rates}.

For comparison purposes, we also train a U-Net using multiple sampling masks. During training, the multiple-model solver is trained to reconstruct MRIs that are measured using the five different sampling patterns demonstrated in Fig. \ref{fig:samplerate}. We present the mean PSNR on the test set in Table \ref{table:samplingratetable}, along with the mean test PSNR for applying \rnr\ to the multiple-model solver, assuming at test time that the sampling pattern is known. Reconstructions from the multiple-model solver can be found in the Supplement. We observe that training with multiple models means that at test time all models produce reasonable reconstructions, but at the cost of reconstruction quality compared to networks trained for single sampling patterns.

In this experiment, we also observe an interesting side-effect of \rnr: when \rnr\ is used to ``adapt'' to a forward model $A_0$ that the original network was trained on, we tend to see an improvement in reconstruction quality. This effect is most pronounced for the U-Net trained to reconstruct multiple sampling patterns, but is also true for the ``dedicated'' solver, demonstrated in Fig. \ref{fig:samplerate}.
}

\begin{table}	
\centering
	\begin{tabular}{c|c|c|c|c|c}
		\hline
	 & 2$\times$ & 4$\times$ & 6$\times$ & 8$\times$ & 12$\times$ \\ \hline
		 Single-Model & 35.74 & 32.53 & 31.51 & \bf 30.69 & \bf 29.48 \\
		 6$\times$ No adaptation & 27.02 & 30.20 & 31.51 & 27.76 & 26.15 \\
		 6$\times$ \rnr & 35.11 & \bf 32.61 & \bf 31.73 & 29.40 & 27.34 \\
		 Multi-Model & 33.99 & 31.62 & 30.48 & 29.25 & 28.35 \\
		 Multi-Model \rnr & \bf 35.80 & 32.35 & 30.81 & 29.60 & 28.61 \\
	\end{tabular}
	\vspace{0.5em}
	\caption{\change{Comparison of reconstruction PSNR for a variety of MRI acceleration factors for several different approaches. ``Multi-model'' refers to a U-Net trained for reconstruction on all shown sampling rates, whereas each column of the ``Single-Model'' results represents a network trained for that particular sampling pattern. $6\times$ refers to the network shown in other experiments. Training a single-sample model consistently performs well for that particular forward model, but at the cost of lower performance on other accelerations, even for higher sampling rates. The multi-model approach sacrifices performance on any one forward model, but most of the difference can be removed by augmenting the multi-model network with our \rnr\ method.}}
	\label{table:samplingratetable}
\end{table}

\change{\subsection{Model adaptation under variable model overlap}\label{sec:exp:nullspaceshift}

\begin{figure}
\centering
\includegraphics[width=0.5\linewidth]{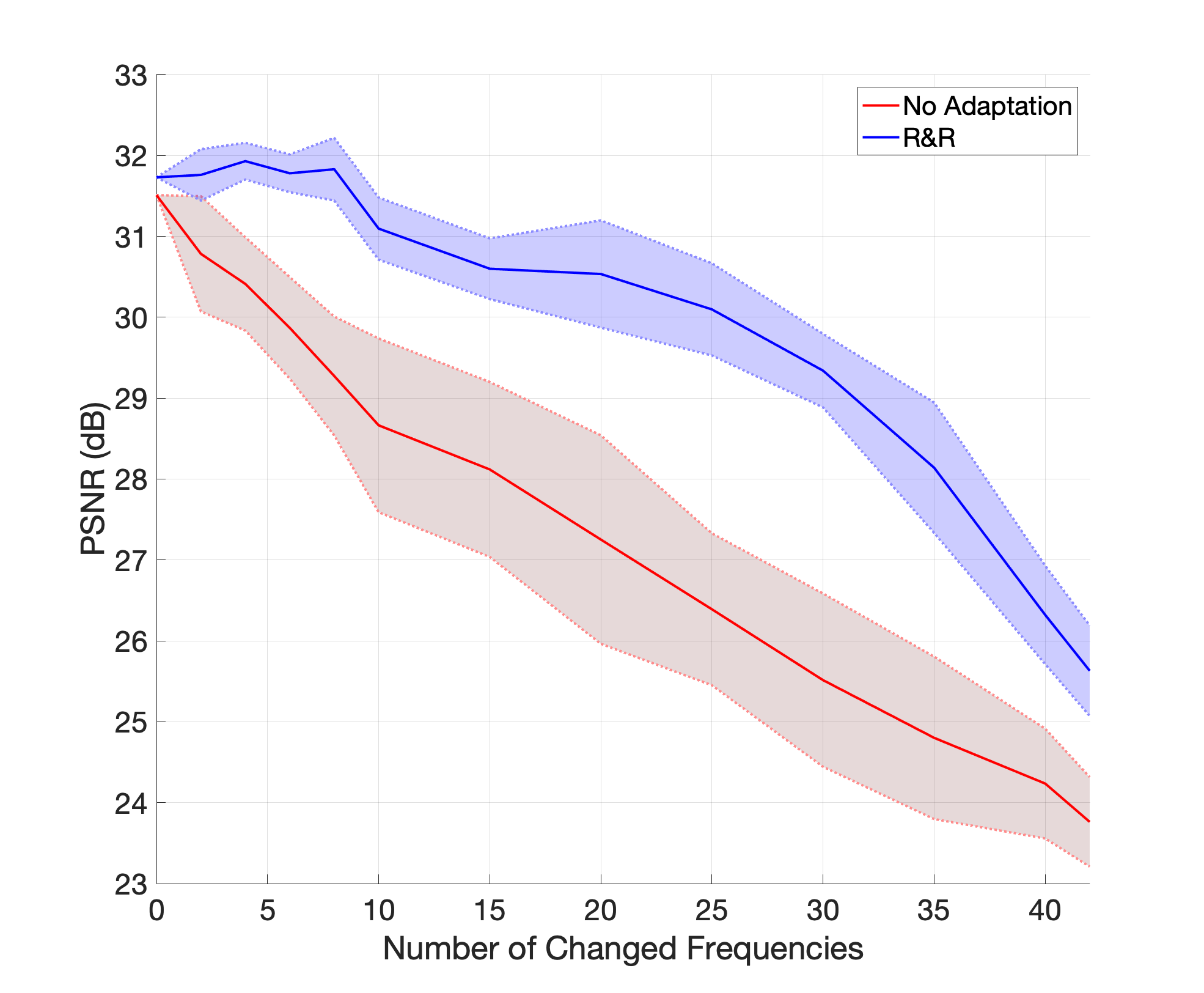}
\caption{\change{Comparison of the mean PSNR for the \rnr\ method and no adaptation for single-sample MRI reconstruction vs. the number of frequencies that differ between $A_0$ and $A_1$. The shaded areas represent the standard deviation of mean test PSNR over 10 runs, since frequencies are replaced randomly.}}
\label{fig:nullspaceoverlap}
\end{figure}

In this section we explore how varying the distance between the forward models $A_0$ and $A_1$ affects reconstruction quality, and how our proposed \rnr\ method deals with different amounts of overlap between $A_0$ and $A_1$. The forward model under investigation is $6\times$ single-coil MRI reconstruction.

To explore variable levels of model drift in the single-coil MRI reconstruction case, we vary which k-space frequencies are sampled in a Cartesian pattern. Specifically, we construct a list of ``non-sampled'' frequencies and a list of ``sampled'' frequencies under $A_0$. We create $A_1$ by swapping $n$ ``sampled'' frequencies for $n$ frequencies in the original ``non-sampled'' list, to ensure that the new $A_1$ contains exactly $n$ frequencies that were not sampled under $A_0$. We do not swap the 4$\%$ center frequencies in any test.

In Fig. \ref{fig:nullspaceoverlap} we plot $n$ vs the mean PSNR over 10 separate instantiations of the above experiment for a no-adaptation approach as well as our \rnr\ method. We run 10 separate instances since the frequencies that are swapped, as well as what frequencies they are swapped to, is random, introducing some variance to the process. We also visually represent the maximum and minimum PSNR across all instances with shading.

Note that the setup in this subsection is different than the experiments used in Fig. \ref{fig:samplecompfig}, Table \ref{table:psnrtable}, or Fig. \ref{fig:megafigmri}. In those experiments, $A_1$ resamples using a Gaussian distribution, biasing the frequencies towards central frequencies, and does not check for overlaps. Since $A_0$ is also biased toward central frequencies, and all frequencies are uniformly swapped, the $A_1$'s in this section may be ``harder'' to reconstruct from than the original $A_1$. In this experiment, the number of changed frequencies acts as a proxy for the difference between $N(A_0)$ and $N(A_1)$, the null spaces of $A_0$ and $A_1$. This experiment also therefore illustrates the extent to which $f_0$ contains information about image components in the null space of $A_1$ for different $A_1$, since if $f_0$ did not contain any further information about the null space of $A_1$, we would expect the \rnr\ curve in Fig. \ref{fig:nullspaceoverlap} to overlap with the ``No Adaptation'' curve.
}

\subsection{Sample complexity}\label{sec:exp:samplecomp}

\change{Our other experiments assume that model adaptation is performed at the level of individual test samples. However, in the case where we have access to a \emph{calibration set} of measurements under the new forward model $A_1$ that we can leverage to retrain the network using the \pnp\ approach.}

In the transfer learning setting, a key concern is the size of the transfer learning set necessary to achieve high-quality results. In this section we compare the performance of \pnp across different calibration set sizes.

\begin{figure}
\centering
\includegraphics[width=0.5\linewidth]{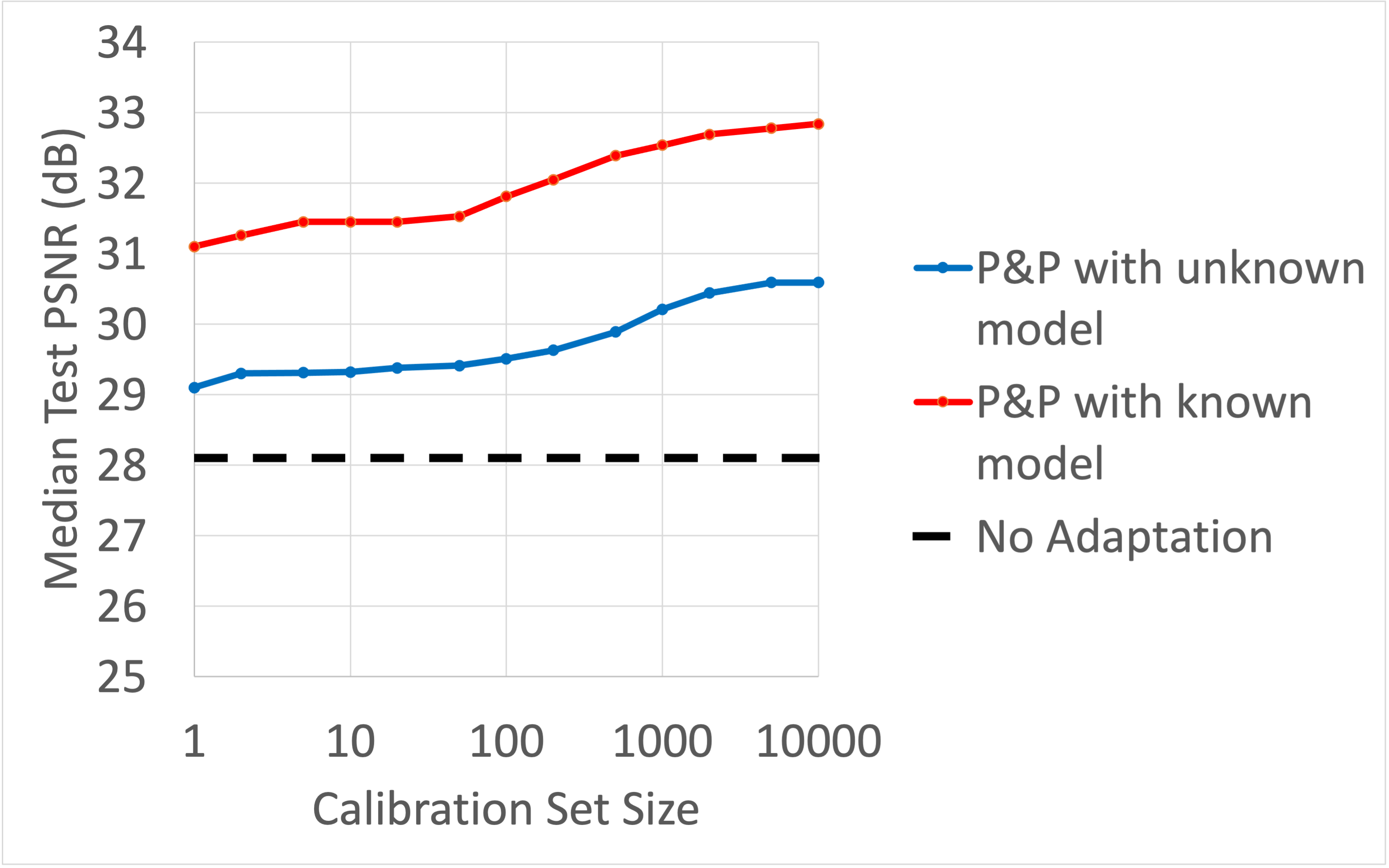}
\caption{Performance of the P\&P model adaptation approach for motion deblurring as a function of the number of calibration samples (blurred images) under the new forward model. Both of our approaches outperform a naive approach (``No Adaptation''), even without exact knowledge of the new forward model.}
\label{fig:samplecompfig}
\end{figure}

In Fig. \ref{fig:samplecompfig} we explore the effect of the number of  samples observed under the new forward model on the adapted model. We observe that even without knowing the forward model, a single calibration sample is sufficient to give improvement over the ``naive'' method that replaces $A_0$ with $A_1$ without further retraining. 
When the forward model is known during calibration and testing, a single example image can result in a 2 dB improvement in PSNR.

\subsection{Model-blind reconstruction with generative networks}\label{sec:exp:gan}

Recent work \cite{bora2017compressed, anirudh2018unsupervised, basioti2020image, hussein2020image} has explored solving inverse problems using generative networks, which permit reconstruction under arbitrary forward models assuming an expressive enough generative network. In particular, \cite{anirudh2018unsupervised} and \cite{basioti2020image} consider the case where the forward model is either partially or entirely unknown, and hence may be learned by parameterizing and jointly optimizing over both the forward model and the latent code for the generative network.

\begin{figure}
\centering
\renewcommand*{\arraystretch}{0}
\begin{tabular}{@{}c@{}c@{}c@{}}
\small Truth & \small \change{\rnr$+$} & \small IAGAN \\ \vspace{2pt} \\
\subfloat{\includegraphics[width = 0.15\linewidth, height = 0.15\linewidth]{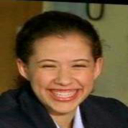}} \hfill &
\subfloat{\includegraphics[width = 0.15\linewidth]{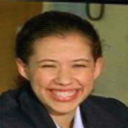}} &  
\subfloat{\includegraphics[width = 0.15\linewidth]{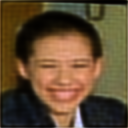}} \\
\subfloat{\includegraphics[width = 0.15\linewidth]{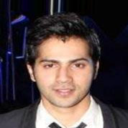}} &
\subfloat{\includegraphics[width = 0.15\linewidth]{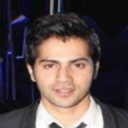}} &  
\subfloat{\includegraphics[width = 0.15\linewidth]{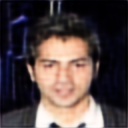}} \\
\subfloat{\includegraphics[width = 0.15\linewidth]{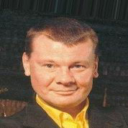}} & 
\subfloat{\includegraphics[width = 0.15\linewidth]{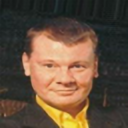}} & 
\subfloat{\includegraphics[width = 0.15\linewidth]{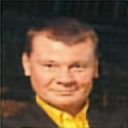}}\\
\end{tabular}
\caption{Comparison of model adaptation (\change{\rnr$+$}) with a model-blind GAN-based reconstruction approach \change{(IAGAN \cite{hussein2020image})} for 2$\times$ super-resolution. While a GAN-based approach only requires learning a single generative network for all forward models, our results suggest that a network trained for a specific forward model with the same number of training samples gives better reconstructions. Best viewed electronically.}
\label{fig:ganfig}
\end{figure}

\change{In Fig. \ref{fig:ganfig} we provide an illustration of reconstructions obtained by the method of \cite{hussein2020image}, compared to our proposed \rnr\ approach. In our demonstration, as in \cite{hussein2020image}, the generative network under consideration is a pretrained Boundary Equilibrium GAN (BEGAN) \cite{berthelot2017began}.
The reconstruction quality is higher when a model-specific network is used, especially when examining fine details and textures.}

In the absence of $(x_i, y_i)$ pairs, a generative approach may be reasonable. However, learning the data manifold in its entirety requires a great deal of data at minimum, along with a sufficiently large and well-tuned generator. The authors of \cite{yeh2017semantic} also note this fundamental limitation: for smaller or simpler applications, learning a high-quality GAN is straightforward, but for more complex applications it is difficult to train GAN models that are sufficiently accurate to rely on for high-quality reconstructions.

\section{Discussion and Conclusion}

This paper explores solutions to the fragility of learned inverse problem solvers in the face of model drift. We demonstrate across a range of simple, practical applications that using a learned image reconstruction network in settings even slightly different than they were trained in results in significant reconstruction errors, both subtle and obvious. We propose two model adaptation procedures: the first is based on a transfer learning approach that attempts to learn a perturbation to the pre-trained network parameters, which we call Parametrize and Perturb (\pnp); \change{the second reuses the network as an implicitly defined regularizer in an iterative model-based reconstruction scheme, which we call Reuse and Regularize (\rnr). We also look at a hybrid approach combining these techniques we call \rnr+.}

We show that our model adaptation techniques enable retraining/reuse of learned solvers under a change in the forward model, even when the change in forward model is not known. In addition, we demonstrate that just learning to invert a variety of forward models at once is not necessarily the solution to the problem of model drift: directly training on many forward models empirically appears to cause reconstruction quality to fall across all learned models. We also show that our approach is superior to one that requires learning a model of the entire image space via a generative model.

\change{The proposed \pnp, \rnr, and \rnr+ model adaptation approaches each have different trade-offs, and may be useful in different scenarios. In general, we observe that \rnr+\ produces superior reconstructions over \rnr\ and \pnp, but incurs significant computation and time costs associated with network retraining specified in  \eqref{eq:rnrpcost}. The \pnp\ approach also incurs similar costs associated with network retraining. However, when a calibration set is available (as in Section \ref{sec:exp:samplecomp}), the \pnp\ approach only needs to be retrained once, and computation cost at deployment matches the original solver. However, we observe two significant benefits of the \rnr\ approach. First, empirically we observe that only few iterations of \rnr\ (see Algorithm \ref{alg:rnr}) tend to be required to give accurate results (namely, $5$ iterations in all our experiments), which increases computational cost by only a constant factor relative to the original reconstruction network. In addition, in the \rnr\ approach only one new parameter is introduced, in contrast to several parameters related to the optimization required for \pnp\ and \rnr+. Finally, our experiments suggest that the improvement offered by \rnr+ tends to be marginal relative to the improvement seen by going from no adaptation to \rnr. Therefore, in situations where reconstruction time is crucial, model adaptation by \rnr\ may be preferred over \rnr+.}

\change{One surprising benefit of the \rnr\ approach is that even in the absence of model drift (i.e., $A_0=A_1$) the reconstruction accuracy improves relative to the output from the reconstruction network. This is because \rnr\ iteratively modifies the output of the network to enforce data-consistency at test time. This may potentially resolve the issue raised in \cite{sidky2020cnns} about whether learned image reconstruction networks are truly ``solving'' a given inverse problem, \ie give a well-defined inverse map of the measurement model. However, to show this would require a much more detailed analysis of the estimator defined by the \rnr\ approach that is beyond the scope of this work.}

Adapting learned inverse problem solvers to function under new forward models is just one step towards robustifying these powerful approaches. Our approach for unknown $A_1$ assumes an explicit parametrization of the forward model, but such a parametrization is not always straightforward or realistic. How best to adapt to complex changes in the forward model that are not easily parametrized is an important open question for future work; \change{see \cite{lunz2021learned} for one recent approach to learning non-parametric (and potentially non-linear) changes to forward models in an iterative reconstruction framework.}

\change{While this work focused on linear inverse problem, many of the principles introduced in this work extend also to non-linear inverse problems. For example, the \rnr\ approach, which is based on the regularization-by-denoising technique (RED), is readily adapted to non-linear problems amenable to a RED approach, which includes phase retrieval \cite{metzler2018prdeep} among others.}

\change{Our empirical evidence suggests that successful model adaptation is possible provided the nullspace (or approximate nullspaces) of $A_0$ and $A_1$ are close in some sense. However, in settings where nullspaces of $A_0$ and $A_1$ are far apart, model adaptation may lead to  artifacts or hallucinated details in the reconstructions. In order to understand these limitations of model adaptation, recent methods introduced to quantify hallucinations induced by neural-network based reconstructions may prove to be useful \cite{bhadra2020hallucinations}.}

Finally, while we focused our attention on model drift, an important open problem is how to adapt to simultaneous model and data distribution drift, and the extent to which these effects can be treated independently. We hope to address these questions in future work.

\bibliographystyle{plain}
\bibliography{refs}
\newpage

\section{Appendix}

In this Appendix we present companion tables to Table \ref{table:mainadaptationtable} which contain further information, including corresponding mean SSIM over the test set, as well as the standard deviations of PSNR and SSIM.

\begin{table*}	
\centering
	\adjustbox{max width=\columnwidth}{
	\begin{tabular}{cc|c|c|c|c|c|c|}
		\hline
         & & \multicolumn{6}{c|}{Baselines} \\
         & & \multirow{2}{*}{TV} & \multirow{2}{*}{RED} & & Train w/$A_0$ & Train w/$A_0$ & Train w/$A_1$ \\
		 & & &  & & Test w/$A_0$ & Test w/$A_1$ & Test w/$A_1$ \\ \hline \hline
	 \multicolumn{2}{c|}{\multirow{2}{*}{Blur}} & \multirow{2}{*}{27.61 $\pm$ 2.57}  & \multirow{2}{*}{30.23 $\pm$ 2.98 } & U-Net & 34.15 $\pm$ 2.33 & 25.42 $\pm$ 1.74 & 33.98 $\pm$ 2.15 \\
	  & & & & MoDL & 36.25 $\pm$ 2.25 & 23.91 $\pm$ 2.02 & 36.13 $\pm$ 2.19 \\ \hline
	 \multicolumn{2}{c|}{\multirow{2}{*}{SR}} & \multirow{2}{*}{28.33 $\pm$ 2.42} & \multirow{2}{*}{28.59 $\pm$ 2.09}  &  U-Net & 30.74 $\pm$ 2.59 & 26.3 $\pm$ 1.65 & 31.22 $\pm$ 2.71 \\
	  & & & & MoDL & 31.32 $\pm$ 2.65 & 22.27 $\pm$ 2.04 & 31.98 $\pm$ 2.61 \\ \hline
	 \multicolumn{2}{c|}{\multirow{2}{*}{MRI}} & \multirow{2}{*}{25.09 $\pm$ 2.50} & \multirow{2}{*}{27.76 $\pm$ 3.37} & U-Net & 31.51 $\pm$ 2.83 & 27.47 $\pm$ 2.47 & 32.33 $\pm$ 2.64 \\
		 & & & & MoDL & 31.88 $\pm$ 2.85 & 22.82 $\pm$ 2.75 & 31.79 $\pm$ 2.81 \\ \hline \hline
		 \multicolumn{8}{c}{Proposed Model Adaptation Methods} \\
	 &	 & \multicolumn{3}{c|}{Known $A_1$} &  \multicolumn{3}{c}{Unknown $A_1$} \\
	 &	 & P\&P (Alg. \ref{alg:pnp}) & R\&R (Alg. \ref{alg:rnr}) & R\&R+ (Alg. \ref{alg:rnrp}) & P\&P (Alg. \ref{alg:pnp}) & R\&R (Alg. \ref{alg:rnr}) & R\&R+ (Alg. \ref{alg:rnrp}) \\ \hline
		 \multirow{2}{*}{Blur}  & U-Net & 33.01 $\pm$ 1.85 & 32.11 $\pm$ 2.64 & 33.50 $\pm$ 2.47 & 29.18 $\pm$ 1.81 & 27.67 $\pm$ 2.23 & 30.05 $\pm$ 2.73 \\
		 & MoDL & 30.08 $\pm$ 1.59 & 33.82 $\pm$ 1.73 & 34.73 $\pm$ 1.82 & 29.89 $\pm$ 1.66 & 27.81 $\pm$ 2.3 & 27.94 $\pm$ 2.4 \\ \hline
		 \multirow{2}{*}{SR}  & U-Net & 28.00 $\pm$ 1.83 & 29.95 $\pm$ 2.49 & 29.99 $\pm$ 2.48 & 27.77 $\pm$ 2.15 & 26.98 $\pm$ 2.39 & 29.35 $\pm$ 2.36 \\
		 & MoDL & 24.59 $\pm$ 2.31 & 28.18 $\pm$ 1.64 & 29.83 $\pm$ 2.00 & 23.14 $\pm$ 2.01 & 24.93 $\pm$ 2.04 & 25.29 $\pm$ 2.33 \\ \hline
		 \multirow{2}{*}{MRI}  & U-Net & 29.07 $\pm$ 2.72 & 29.71 $\pm$ 2.75 & 31.43 $\pm$ 2.99 & 28.92 $\pm$ 3.04 & 28.06 $\pm$ 2.63 & 29.54 $\pm$ 2.53 \\
		 & MoDL & 30.63 $\pm$ 2.85 & 30.25 $\pm$ 3.10 & 31.44 $\pm$ 2.75 & 26.64 $\pm$ 2.60 & 23.46 $\pm$ 2.54 & 27.67 $\pm$ 2.62 \\
	\end{tabular}
	}
	\vspace{0.5em}
	\caption{Comparison of performance of various baseline methods for inverse problems across a variety of datasets and forward models. The metric presented is the mean PSNR $\pm$ the standard deviation.}
	\label{table:psnrtable}
\end{table*}

\begin{table*}	
\centering
	\adjustbox{max width=\columnwidth}{
	\begin{tabular}{cc|c|c|c|c|c|c|}
		\hline
         & & \multicolumn{6}{c|}{Baselines} \\
         & & \multirow{2}{*}{TV} & \multirow{2}{*}{RED} & & Train w/$A_0$ & Train w/$A_0$ & Train w/$A_1$ \\
		 & & &  & & Test w/$A_0$ & Test w/$A_1$ & Test w/$A_1$ \\ \hline \hline
	 \multicolumn{2}{c|}{\multirow{2}{*}{Blur}} & \multirow{2}{*}{0.94 $\pm$ 0.06}  & \multirow{2}{*}{0.96 $\pm$ 0.05} & U-Net & 0.98 $\pm$ 0.05   & 0.89 $\pm$ 0.09 & 0.98 $\pm$ 0.05 \\
	  & & & & MoDL & 0.98 $\pm$ 0.03  & 0.84 $\pm$ 0.08 & 0.98 $\pm$ 0.04 \\ \hline
	 \multicolumn{2}{c|}{\multirow{2}{*}{SR}} & \multirow{2}{*}{0.95 $\pm$ 0.06} & \multirow{2}{*}{0.96 $\pm$ 0.03} &  U-Net & 0.97  $\pm$ 0.03  & 0.92 $\pm$ 0.09 & 0.97 $\pm$ 0.02 \\
	  & & & & MoDL & 0.97 $\pm$ 0.04 & 0.89 $\pm$ 0.06 & 0.98 $\pm$ 0.04 \\ \hline
	 \multicolumn{2}{c|}{\multirow{2}{*}{MRI}} & \multirow{2}{*}{0.79 $\pm$ 0.04} & \multirow{2}{*}{0.80 $\pm$ 0.05} & U-Net & 0.82 $\pm$ 0.06  & 0.74 $\pm$ 0.06 & 0.82 $\pm$  0.06 \\
		 & & & & MoDL & 0.83 $\pm$ 0.06 & 0.65 $\pm$ 0.08 & 0.83 $\pm$ 0.06  \\ \hline \hline
		 \multicolumn{8}{c}{Proposed Model Adaptation Methods} \\
	 &	 & \multicolumn{3}{c|}{Known $A_1$} &  \multicolumn{3}{c}{Unknown $A_1$} \\
	 &	 & P\&P (Alg. \ref{alg:pnp}) & R\&R (Alg. \ref{alg:rnr}) & R\&R+ (Alg. \ref{alg:rnrp}) & P\&P (Alg. \ref{alg:pnp}) & R\&R (Alg. \ref{alg:rnr}) & R\&R+ (Alg. \ref{alg:rnrp}) \\ \hline
		 \multirow{2}{*}{Blur}  & U-Net & 0.81 $\pm$ 0.09 & 0.81 $\pm$ 0.07 & 0.82 $\pm$ 0.07 & 0.75 $\pm$ 0.05  & 0.79 $\pm$ 0.09 & 0.77 $\pm$ 0.08 \\
		 & MoDL & 0.82 $\pm$ 0.07 & 0.81 $\pm$ 0.09 & 0.83 $\pm$ 0.07 & 0.72 $\pm$ 0.09 & 0.68 $\pm$ 0.07 & 0.75 $\pm$ 0.10 \\ \hline
		 \multirow{2}{*}{SR}  & U-Net & 0.94 $\pm$ 0.09 & 0.96 $\pm$ 0.04 & 0.96 $\pm$ 0.04 & 0.94 $\pm$ 0.09 & 0.94 $\pm$ 0.08  & 0.96 $\pm$ 0.06 \\
		 & MoDL & 0.92 $\pm$ 0.06 & 0.94 $\pm$ 0.02 & 0.95 $\pm$ 0.02 & 0.91 $\pm$ 0.04 & 0.92 $\pm$ 0.02  & 0.94 $\pm$ 0.01 \\ \hline
		 \multirow{2}{*}{MRI}  & U-Net & 0.81 $\pm$ 0.09 & 0.81 $\pm$ 0.07 & 0.82 $\pm$ 0.07 & 0.75 $\pm$ 0.05 & 0.79 $\pm$ 0.09 & 0.77 $\pm$ 0.08 \\
		 & MoDL & 0.82 $\pm$ 0.07 & 0.81 $\pm$ 0.09 & 0.83 $\pm$ 0.07 & 0.72 $\pm$ 0.09  & 0.68 $\pm$ 0.07  & 0.75 $\pm$ 0.10 \\
	\end{tabular}
	}
	\vspace{0.5em}
	\caption{Comparison of performance of various baseline methods for inverse problems across a variety of datasets and forward models. The metric presented is the mean SSIM $\pm$ the standard deviation.}
	\label{table:ssimtable}
\end{table*}

\subsection{Supplemental Figures}

Here we include visual figures to illustrate some sample reconstructions from the methods outlined in the main work. The figures illustrated are chosen to be close to the median performance while maintaining a range of subjects. The three test images used in the figures are within the inter-quartile range of reconstruction PSNR for all reconstruction methods.

We also include a brief ablation study indicating that the proximity term in (5) of the main text is necessary for high-quality reconstructions, as well as providing an additional row of Figure 10 for a U-Net trained on multiple forward models.

\begin{figure}
\centering
\begin{tabular}{ccc}
\subfloat{\includegraphics[width = 0.125\textwidth]{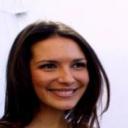}} &
\subfloat{\includegraphics[width = 0.125\textwidth]{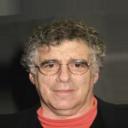}} &
\subfloat{\includegraphics[width = 0.125\textwidth]{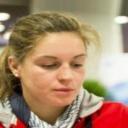}} \\
\subfloat{\includegraphics[width = 0.125\textwidth]{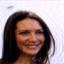}} &
\subfloat{\includegraphics[width = 0.125\textwidth]{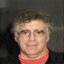}} &
\subfloat{\includegraphics[width = 0.125\textwidth]{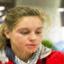}} \\
\subfloat{\includegraphics[width = 0.125\textwidth]{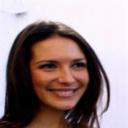}} &
\subfloat{\includegraphics[width = 0.125\textwidth]{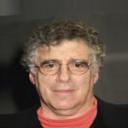}} &
\subfloat{\includegraphics[width = 0.125\textwidth]{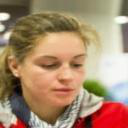}}
\end{tabular}
\caption{Role of proximity term in model adaptation for motion deblurring with \pnp. When ablating the term that promotes proximity to the initial network weights in the \pnp\ approach (\ie setting $\mu = 0$), the retrained network outputs degenerate solutions that match the measurements (center row), but lack fidelity with the ground truth (top row). By including the proximity term ($\mu \neq 0$), we find a small corrective perturbation to the network weights that improves reconstruction accuracy significantly (bottom row).}
\label{fig:blurfig}
\end{figure}

\subsection{Motion Deblurring}

\begin{figure*}
\centering
\renewcommand*{\arraystretch}{0}
\begin{tabular}{@{}c@{}c@{}c@{}c}
\small Ground & \small Blurred & \small TV & \small RED \vspace{2pt} \\
\small Truth &                    & \small Reconstruction & \\ \vspace{2pt} \\
\includegraphics[align=c, width = 0.15\linewidth]{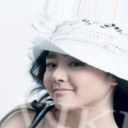} &
\includegraphics[align=c, width = 0.15\linewidth]{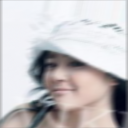} &
\includegraphics[align=c, width = 0.15\linewidth]{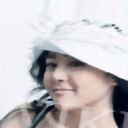} &
\includegraphics[align=c, width = 0.15\linewidth]{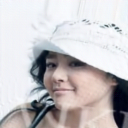} \\
\includegraphics[align=c, width = 0.15\linewidth]{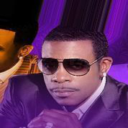} &
\includegraphics[align=c, width = 0.15\linewidth]{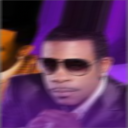} &
\includegraphics[align=c, width = 0.15\linewidth]{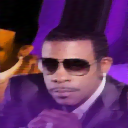} &
\includegraphics[align=c, width = 0.15\linewidth]{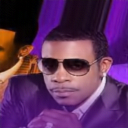} \\
\includegraphics[align=c, width = 0.15\linewidth]{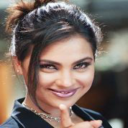} &
\includegraphics[align=c, width = 0.15\linewidth]{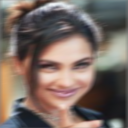} &
\includegraphics[align=c, width = 0.15\linewidth]{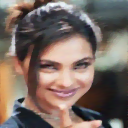} &
\includegraphics[align=c, width = 0.15\linewidth]{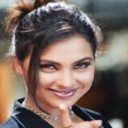}
\end{tabular}
\caption{Ground truth, blurred, and a classical TV-regularized reconstruction for the motion deblurring inverse problem. Compare the reconstructions to the learned reconstructions in Fig. \ref{fig:megafigdeblurunet} and \ref{fig:megafigdeblurmodl}.}
\label{fig:truthdeblur}
\end{figure*}

\begin{figure*}
\centering
\renewcommand*{\arraystretch}{0}
\begin{tabular}{@{}c@{}c@{}c@{}c@{}c@{}c@{}c@{}c@{}c@{}}
& \small Train $A_0$  & \small Train $A_0$ & \small \pnp & \small \rnr & \small \rnr+ & \small \pnp & \small \rnr & \small \rnr+ \\
 & \small Test $A_0$ & \small Test $A_1$ & Known $A_1$ & Known $A_1$ & Known $A_1$ & Unknown $A_1$ & Unknown $A_1$ & Unknown $A_1$ \\ \vspace{2pt} \\
 &
\includegraphics[align=c, width = 0.12\linewidth]{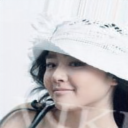} & 
\includegraphics[align=c, width = 0.12\linewidth]{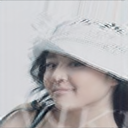} & 
\includegraphics[align=c, width = 0.12\linewidth]{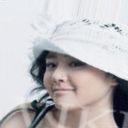} &
\includegraphics[align=c, width = 0.12\linewidth]{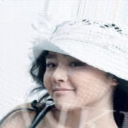} & 
\includegraphics[align=c, width = 0.12\linewidth]{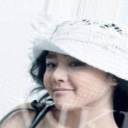} & 
\includegraphics[align=c, width = 0.12\linewidth]{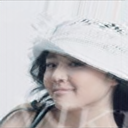} & 
\includegraphics[align=c, width = 0.12\linewidth]{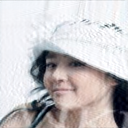} &
\includegraphics[align=c, width = 0.12\linewidth]{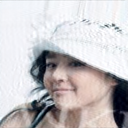} \\
 &
\includegraphics[align=c, width = 0.12\linewidth]{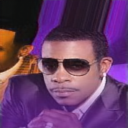} & 
\includegraphics[align=c, width = 0.12\linewidth]{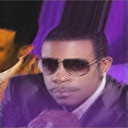} & 
\includegraphics[align=c, width = 0.12\linewidth]{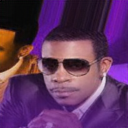} &
\includegraphics[align=c, width = 0.12\linewidth]{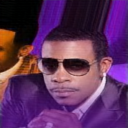} & 
\includegraphics[align=c, width = 0.12\linewidth]{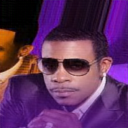} & 
\includegraphics[align=c, width = 0.12\linewidth]{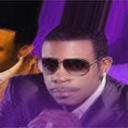} & 
\includegraphics[align=c, width = 0.12\linewidth]{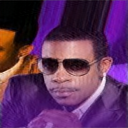} &
\includegraphics[align=c, width = 0.12\linewidth]{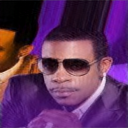} \\
 &
\includegraphics[align=c, width = 0.12\linewidth]{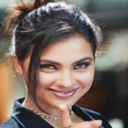} & 
\includegraphics[align=c, width = 0.12\linewidth]{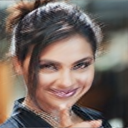} & 
\includegraphics[align=c, width = 0.12\linewidth]{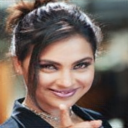} &
\includegraphics[align=c, width = 0.12\linewidth]{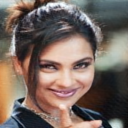} & 
\includegraphics[align=c, width = 0.12\linewidth]{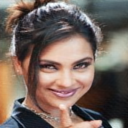} & 
\includegraphics[align=c, width = 0.12\linewidth]{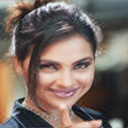} & 
\includegraphics[align=c, width = 0.12\linewidth]{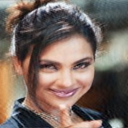} &
\includegraphics[align=c, width = 0.12\linewidth]{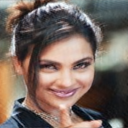} \\
\end{tabular}
\caption{Visual examples of reconstruction quality for the U-Net solver for the motion deblurring inverse problem. }
\label{fig:megafigdeblurunet}
\end{figure*}

\begin{figure*}
\centering
\renewcommand*{\arraystretch}{0}
\begin{tabular}{@{}c@{}c@{}c@{}c@{}c@{}c@{}c@{}c@{}c@{}}
& \small Train $A_0$  & \small Train $A_0$ & \small \pnp & \small \rnr  & \small \rnr+ & \small \pnp & \small \rnr & \small \rnr+ \\
 & \small Test $A_0$ & \small Test $A_1$ & Known $A_1$ & Known $A_1$ & Known $A_1$ & Unknown $A_1$ &  Unknown $A_1$ & Unknown $A_1$ \\ \vspace{2pt} \\
 &
\includegraphics[align=c, width = 0.12\linewidth]{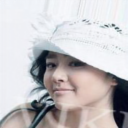} & 
\includegraphics[align=c, width = 0.12\linewidth]{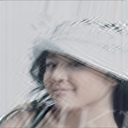} & 
\includegraphics[align=c, width = 0.12\linewidth]{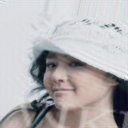} &
\includegraphics[align=c, width = 0.12\linewidth]{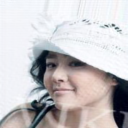} & 
\includegraphics[align=c, width = 0.12\linewidth]{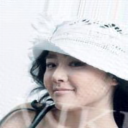} & 
\includegraphics[align=c, width = 0.12\linewidth]{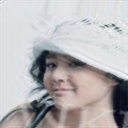} & 
\includegraphics[align=c, width = 0.12\linewidth]{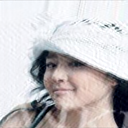} &
\includegraphics[align=c, width = 0.12\linewidth]{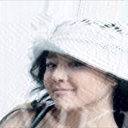} \\
 &
\includegraphics[align=c, width = 0.12\linewidth]{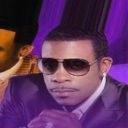} & 
\includegraphics[align=c, width = 0.12\linewidth]{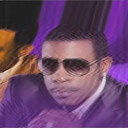} & 
\includegraphics[align=c, width = 0.12\linewidth]{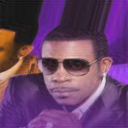} &
\includegraphics[align=c, width = 0.12\linewidth]{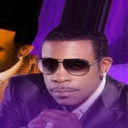} & 
\includegraphics[align=c, width = 0.12\linewidth]{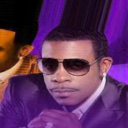} & 
\includegraphics[align=c, width = 0.12\linewidth]{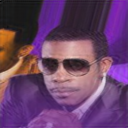} & 
\includegraphics[align=c, width = 0.12\linewidth]{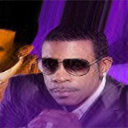} &
\includegraphics[align=c, width = 0.12\linewidth]{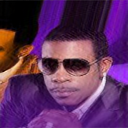} \\
 &
\includegraphics[align=c, width = 0.12\linewidth]{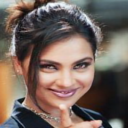} & 
\includegraphics[align=c, width = 0.12\linewidth]{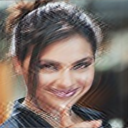} & 
\includegraphics[align=c, width = 0.12\linewidth]{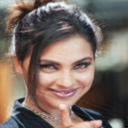} &
\includegraphics[align=c, width = 0.12\linewidth]{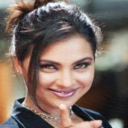} & 
\includegraphics[align=c, width = 0.12\linewidth]{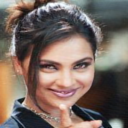} & 
\includegraphics[align=c, width = 0.12\linewidth]{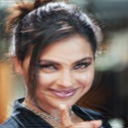} & 
\includegraphics[align=c, width = 0.12\linewidth]{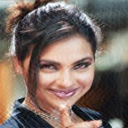} &
\includegraphics[align=c, width = 0.12\linewidth]{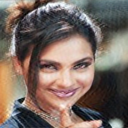} \\
\end{tabular}
\caption{Visual examples of reconstruction quality for the MoDL solver for the motion deblurring inverse problem.}
\label{fig:megafigdeblurmodl}
\end{figure*}

In Figures \ref{fig:truthdeblur}, \ref{fig:megafigdeblurunet}, and \ref{fig:megafigdeblurmodl} we present sample visualizations for the motion deblurring setting. Figure \ref{fig:truthdeblur} contains the original images, as well as the blurred images and a sample total-variation regularized reconstruction.

Figure \ref{fig:megafigdeblurunet} and \ref{fig:megafigdeblurmodl} present reconstruction results for the images shown in Figure \ref{fig:truthdeblur}. Digital viewing is recommended.

\subsection{Superresolution}

\begin{figure*}
\centering
\renewcommand*{\arraystretch}{0}
\begin{tabular}{@{}c@{}c@{}c@{}c}
\small Ground & \small $A_0^T y$ & \small TV & \small RED \vspace{2pt}  \\
\small Truth &                    & \small Reconstruction &  \\ \vspace{2pt} \\
\includegraphics[align=c, width = 0.15\linewidth]{figures/megafig/truth/celeba/41_8.458262.png} &
\includegraphics[align=c, width = 0.15\linewidth]{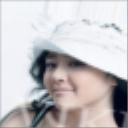} &
\includegraphics[align=c, width = 0.15\linewidth]{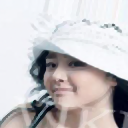} &
\includegraphics[align=c, width = 0.15\linewidth]{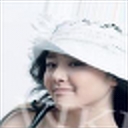} \\
\includegraphics[align=c, width = 0.15\linewidth]{figures/megafig/truth/celeba/23_9.048739.png} &
\includegraphics[align=c, width = 0.15\linewidth]{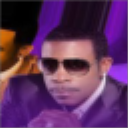} &
\includegraphics[align=c, width = 0.15\linewidth]{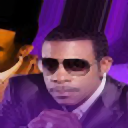} &
\includegraphics[align=c, width = 0.15\linewidth]{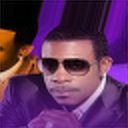} \\
\includegraphics[align=c, width = 0.15\linewidth]{figures/megafig/truth/celeba/24_7.981877.png} &
\includegraphics[align=c, width = 0.15\linewidth]{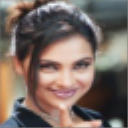} &
\includegraphics[align=c, width = 0.15\linewidth]{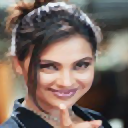} &
\includegraphics[align=c, width = 0.15\linewidth]{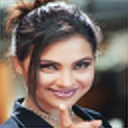}
\end{tabular}
\caption{Ground truth, $A_0^T y$, and a classical TV-regularized reconstruction for the superresolution inverse problem. Here, $A_0^T y$ is the input of the initial trained network, and corresponds to downsampling under $A_0$ and upsampling back to the original size with $A_0^T$.}
\label{fig:truthsuperres}
\end{figure*}

\begin{figure*}
\centering
\renewcommand*{\arraystretch}{0}

\begin{tabular}{@{}c@{}c@{}c@{}c@{}c@{}c@{}c@{}c@{}c@{}}
& \small Train $A_0$  & \small Train $A_0$ & \small \pnp & \small \rnr & \small \rnr+ & \small \pnp & \small \rnr & \small \rnr+ \\
& \small Test $A_0$ & \small Test $A_1$ & Known $A_1$ & Known $A_1$ & Known $A_1$ & Unknown $A_1$ & Unknown $A_1$ & Unknown $A_1$ \\ \vspace{2pt} \\
 &
\includegraphics[align=c, width = 0.12\linewidth]{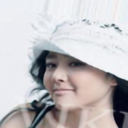} & 
\includegraphics[align=c, width = 0.12\linewidth]{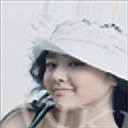} & 
\includegraphics[align=c, width = 0.12\linewidth]{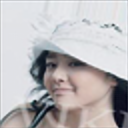} &
\includegraphics[align=c, width = 0.12\linewidth]{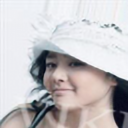} & 
\includegraphics[align=c, width = 0.12\linewidth]{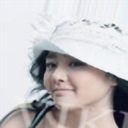} & 
\includegraphics[align=c, width = 0.12\linewidth]{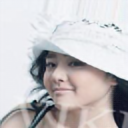} & 
\includegraphics[align=c, width = 0.12\linewidth]{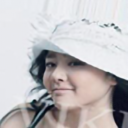} &
\includegraphics[align=c, width = 0.12\linewidth]{figures/revisionfigs/supp/superres/unet/41_rnrp__29.8809814453125_0.96674985.png} \\
 &
\includegraphics[align=c, width = 0.12\linewidth]{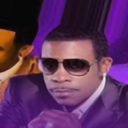} & 
\includegraphics[align=c, width = 0.12\linewidth]{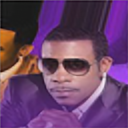} & 
\includegraphics[align=c, width = 0.12\linewidth]{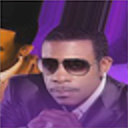} &
\includegraphics[align=c, width = 0.12\linewidth]{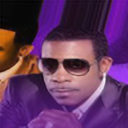} & 
\includegraphics[align=c, width = 0.12\linewidth]{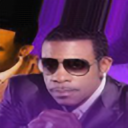} & 
\includegraphics[align=c, width = 0.12\linewidth]{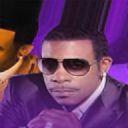} & 
\includegraphics[align=c, width = 0.12\linewidth]{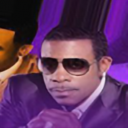} &
\includegraphics[align=c, width = 0.12\linewidth]{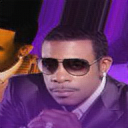} \\
 &
\includegraphics[align=c, width = 0.12\linewidth]{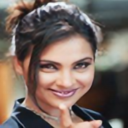} & 
\includegraphics[align=c, width = 0.12\linewidth]{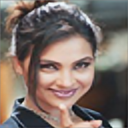} & 
\includegraphics[align=c, width = 0.12\linewidth]{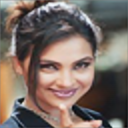} &
\includegraphics[align=c, width = 0.12\linewidth]{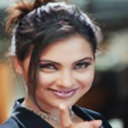} & 
\includegraphics[align=c, width = 0.12\linewidth]{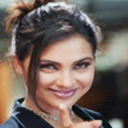} & 
\includegraphics[align=c, width = 0.12\linewidth]{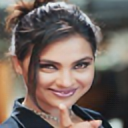} & 
\includegraphics[align=c, width = 0.12\linewidth]{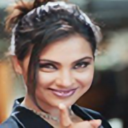} &
\includegraphics[align=c, width = 0.12\linewidth]{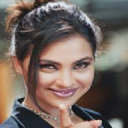} \end{tabular}
\caption{Visual examples of reconstruction quality for the U-Net network for the superresolution inverse problem.}
\label{fig:megafigsuperresunet}
\end{figure*}

\begin{figure*}
\centering
\renewcommand*{\arraystretch}{0}

\begin{tabular}{@{}c@{}c@{}c@{}c@{}c@{}c@{}c@{}c@{}c@{}}
& \small Train $A_0$  & \small Train $A_0$ & \small \pnp & \small \rnr & \small \rnr+ & \small \pnp & \small \rnr & \small \rnr+ \\
& \small Test $A_0$ & \small Test $A_1$ & Known $A_1$ & Known $A_1$ & Known $A_1$ & Unknown $A_1$ & Unknown $A_1$ & Unknown $A_1$ \\ \vspace{2pt} \\
 &
\includegraphics[align=c, width = 0.12\linewidth]{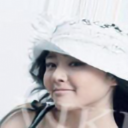} & 
\includegraphics[align=c, width = 0.12\linewidth]{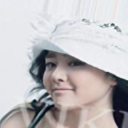} & 
\includegraphics[align=c, width = 0.12\linewidth]{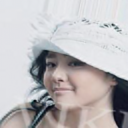} &
\includegraphics[align=c, width = 0.12\linewidth]{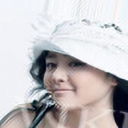} & 
\includegraphics[align=c, width = 0.12\linewidth]{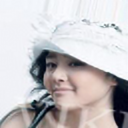} & 
\includegraphics[align=c, width = 0.12\linewidth]{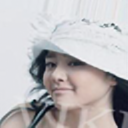} & 
\includegraphics[align=c, width = 0.12\linewidth]{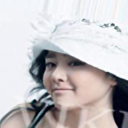} &
\includegraphics[align=c, width = 0.12\linewidth]{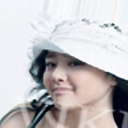} \\
 &
\includegraphics[align=c, width = 0.12\linewidth]{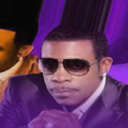} & 
\includegraphics[align=c, width = 0.12\linewidth]{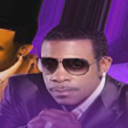} & 
\includegraphics[align=c, width = 0.12\linewidth]{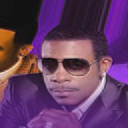} &
\includegraphics[align=c, width = 0.12\linewidth]{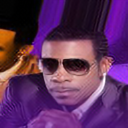} & 
\includegraphics[align=c, width = 0.12\linewidth]{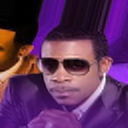} & 
\includegraphics[align=c, width = 0.12\linewidth]{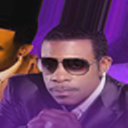} & 
\includegraphics[align=c, width = 0.12\linewidth]{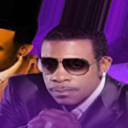} &
\includegraphics[align=c, width = 0.12\linewidth]{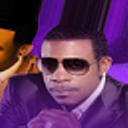} \\
 &
\includegraphics[align=c, width = 0.12\linewidth]{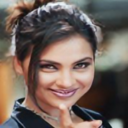} & 
\includegraphics[align=c, width = 0.12\linewidth]{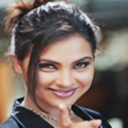} & 
\includegraphics[align=c, width = 0.12\linewidth]{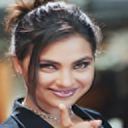} &
\includegraphics[align=c, width = 0.12\linewidth]{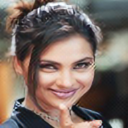} & 
\includegraphics[align=c, width = 0.12\linewidth]{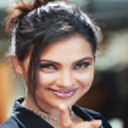} & 
\includegraphics[align=c, width = 0.12\linewidth]{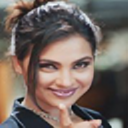} & 
\includegraphics[align=c, width = 0.12\linewidth]{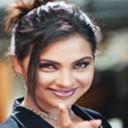} &
\includegraphics[align=c, width = 0.12\linewidth]{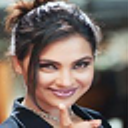} \end{tabular}
\caption{Visual examples of reconstruction quality for the MoDL network for the superresolution inverse problem.}
\label{fig:megafigsuperresmodl}
\end{figure*}

As in the previous subsection, Figures \ref{fig:truthsuperres}, \ref{fig:megafigsuperresunet}, and \ref{fig:megafigsuperresmodl} demonstrate sample visualizations for the superresolution setting. Figure \ref{fig:truthdeblur} contains the original images, as well as the upsampled $y$ and a sample total-variation regularized reconstruction.

Figure \ref{fig:megafigsuperresunet} and \ref{fig:megafigsuperresmodl} present reconstruction results for the images shown in Figure \ref{fig:truthsuperres}. Digital viewing is recommended.

\subsection{Proximity loss study}\label{sec:exp:proximityloss}

\change{In this section we examine the effect of training without the proximity term in the adaptation loss of Eq. 5, demonstrating that just refitting the data without staying close to the original model is not enough. 

In Figure \ref{fig:blurfig} we demonstrate the effect of ablating the proximity term in Eq. 5, which we believe are necessary to avoid degenerate solutions. Optimizing Eq. 5 without proximity terms
fails to leverage the network learned for $A_0$ {\em using ground truth images $x$}. Without the proximity term, our solution relies too heavily on $y$, and not enough on the information encoded in $f_0$.
We empirically find the proximity term is necessary to maintain good reconstructions.}

\begin{figure*}[ht]
  \centering
  \includegraphics[width=\linewidth]{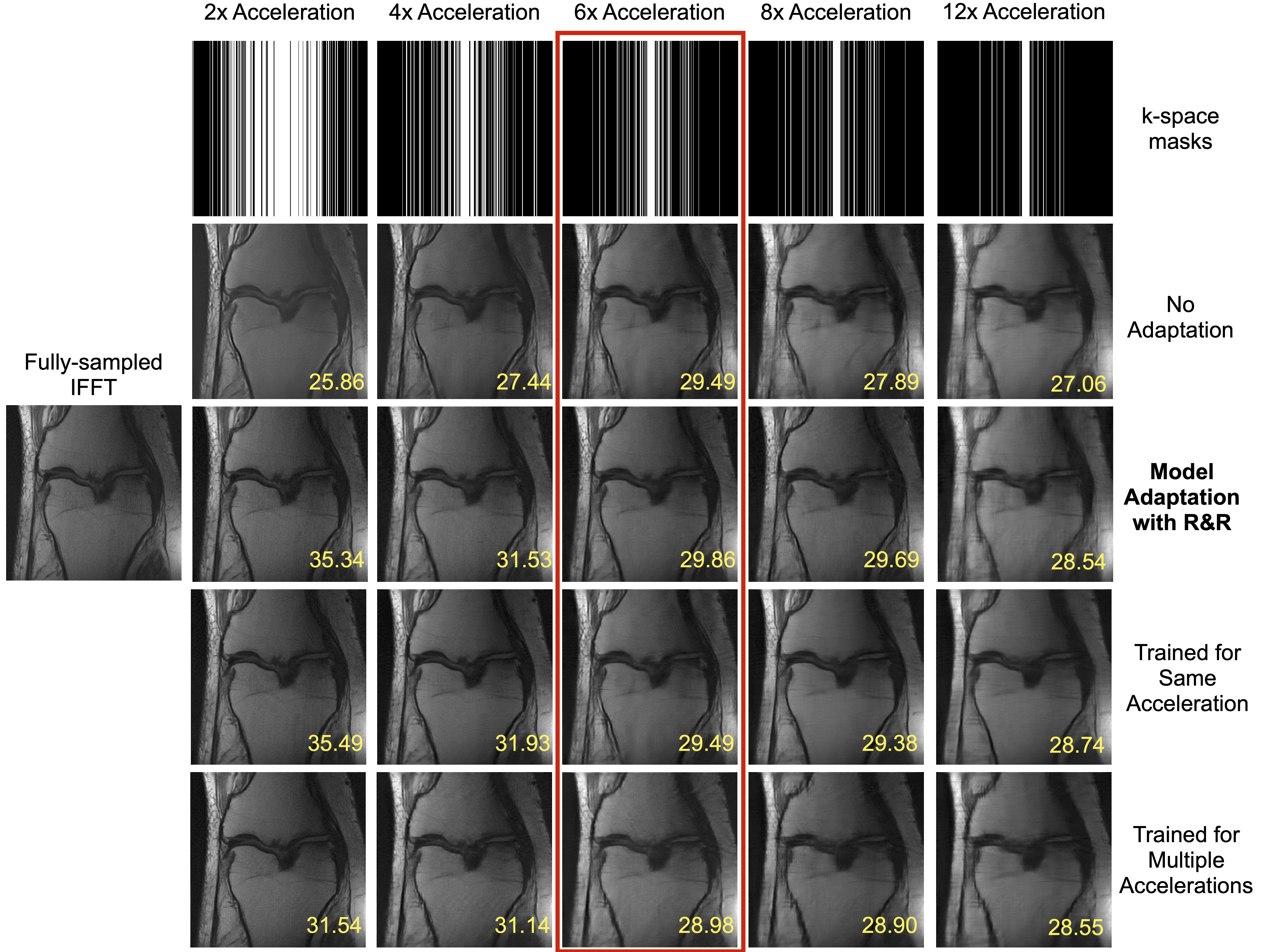}
\caption{\change{Using the \rnr\ model adaptation approach permits using a U-Net trained for $6\times$ acceleration on MRI reconstruction across a range of acceleration parameters. The various k-space sampling patterns used in these experiments are shown in the top row. Without adaptation (second row), the reconstruction quality decreases when changing the acceleration factor, \emph{even when more k-space measurements are taken}, as originally observed in \cite{antun2020instabilities}. The \rnr\ reconstructions (third row) compare favorably to the performance of networks trained on each particular k-space sampling pattern (second-to-bottom row). Training for multiple accelerations (bottom row) appears to be inferior in terms of reconstruction quality to dedicated training or adapting with \rnr. The PSNR of each image is presented in dB in yellow on each image.}}
\label{fig:sampleratesupp}
\end{figure*}

\change{\subsection{Adapting to variable sampling rates in single-coil MRI: Expanded figure}\label{sec:mrisamplingexpsupplement}

In Fig. \ref{fig:sampleratesupp} we present an expanded version of the original figure in the main body, including the reconstructions produced by the U-Net trained to reconstruct multiple forward models.
}

\end{document}